\documentclass[sigconf, authorversion]{acmart}
\AtBeginDocument{%
  }

\copyrightyear{2026}
\acmYear{2026}
\setcopyright{acmlicensed}
\acmConference[CHI '26]{Proceedings of the 2026 CHI Conference on Human Factors in Computing Systems}{April 13--17, 2026}{Barcelona, Spain}
\acmBooktitle{Proceedings of the 2026 CHI Conference on Human Factors in Computing Systems (CHI '26), April 13--17, 2026, Barcelona, Spain}
\acmDOI{10.1145/3772318.3791200}
\acmISBN{979-8-4007-2278-3/2026/04}

\usepackage[utf8]{inputenc}
\usepackage{colortbl}
\usepackage{subcaption}
\usepackage{array}
\usepackage{multirow}
\usepackage{makecell}
\usepackage{enumitem}
\usepackage{calc}
\usepackage{mwe}

\newcommand{\edit}[1]{{#1}}

\hypersetup{
  colorlinks=true,
  linkcolor=blue,
  filecolor=magenta,
  urlcolor=blue,
  citecolor=blue,
  pdftitle={Show It, Don't Just Say It: The Complementary Effects of Instruction Multimodality for Software Guidance},
  pdfauthor={Emran Poh; Yueyue Hou; Tianyi Zhang; Jiannan Li},
  pdfsubject={Multimodal instruction and adaptive AI tutoring for software learning},
  pdfkeywords={Multimodal Instruction, Software Learning, Learner Agency, Visual Annotations, Cognitive Load Theory, Adaptive Tutoring Systems, Digital Territoriality}
}

\begin{document}

\raggedbottom

\title[\texorpdfstring{Show It, Don't Just Say It: The Complementary Effects of Instruction Multimodality for Software Guidance}{Show It, Don't Just Say It: The Complementary Effects of Instruction Multimodality for Software Guidance}]{\texorpdfstring{`Show It, Don't Just Say It': The Complementary Effects of Instruction Multimodality for Software Guidance}{Show It, Don't Just Say It: The Complementary Effects of Instruction Multimodality for Software Guidance}}

\author{Emran Poh}
\orcid{0000-0003-4721-1916}
\email{emran.poh.2025@smu.edu.sg}
\affiliation{
  \institution{School of Computing and Information Systems\\Singapore Management University}
  \city{Singapore}
  \country{Singapore}
}

\author{Yueyue Hou}
\orcid{0000-0001-6020-4599}
\email{yueyue.hou.2024@smu.edu.sg}
\affiliation{
  \institution{School of Computing and Information Systems\\Singapore Management University}
  \city{Singapore}
  \country{Singapore}
}

\author{Tianyi Zhang}
\orcid{0009-0009-1318-3655}
\email{tianyizhang.2023@smu.edu.sg}
\affiliation{
  \institution{School of Computing and Information Systems\\Singapore Management University}
  \city{Singapore}
  \country{Singapore}
}

\author{Jiannan Li}
\orcid{0000-0001-8409-4910}
\email{jiannanli@smu.edu.sg}
\affiliation{
  \institution{School of Computing and Information Systems\\Singapore Management University}
  \city{Singapore}
  \country{Singapore}
}

\renewcommand{\shortauthors}{Poh et al.}


\begin{abstract}
Designing adaptive tutoring systems for software learning presents challenges in determining appropriate instructional modalities. To inform the design of such systems, we conducted an observational study of ten human teacher-student pairs \edit{(N=10)}, where experienced design software users taught novices two new graphic design software features through multi-step procedures. These lessons were limited to three communication channels (speech, visual annotations, and remote screen control) to mimic possible AI tutor modalities. We found that annotations complement speech with spatial precision and remote control complements it with spatial and temporal precision, but both cause intrusion to learner agency. Teachers adaptively select modalities to balance the need for instruction progress with students' cognitive engagement and sense of digital territory ownership. Our results provide further support to the contiguity principles and the value of agency in learning, while suggesting precision-agency trade-off and digital territoriality as new design constraints for adaptive software guidance.

\end{abstract}




\begin{teaserfigure}
  \includegraphics[width=\textwidth]{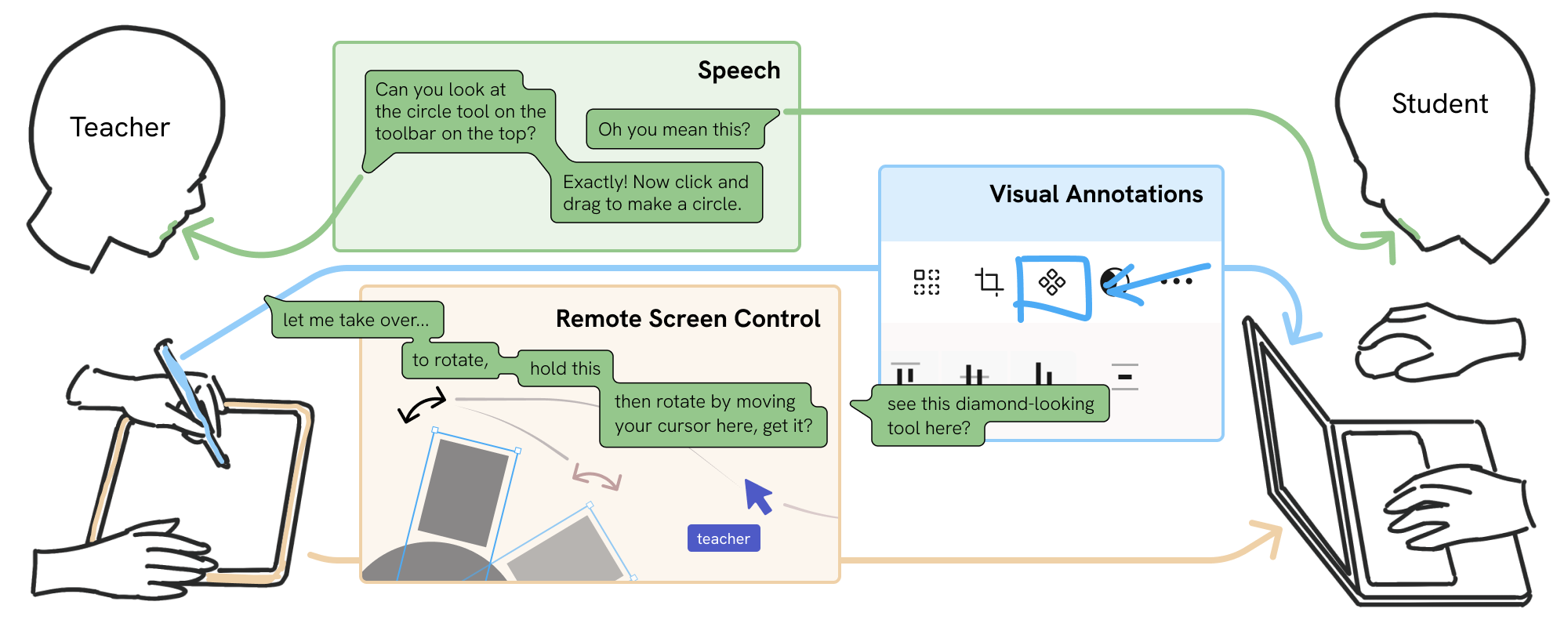}
  \caption{Human teachers coordinate multiple communication modalities in this observational study format—speech (bidirectional), visual annotations (unidirectional), and remote screen control (unidirectional)—to provide effective software instruction, demonstrating sophisticated strategies that can inform AI tutoring system design.}
  \Description{This diagram illustrates a remote learning scenario between a teacher and student, showing three primary communication modalities: Speech (bidirectional green arrows), Visual Annotations (unidirectional blue arrows), and Remote Screen Control (unidirectional orange arrows). The teacher, represented by a head outline on the left, uses speech to give instructions like "Can you look at the toolbar on the left? See this circle tool here?" while the student, shown as a head outline on the right, responds with "Oh, this one? The ellipse tool?" Visual annotations highlight specific UI elements like a diamond-shaped component tool in a toolbar, with speech bubbles indicating "see this diamond-looking tool here?" Remote screen control shows the teacher taking over the student's screen to demonstrate actions like rotating objects while explaining "let me take over... then hold `Shift' and rotate." The student interacts with both a laptop (with mouse and keyboard) and a drawing tablet, receiving guidance through all three modalities. Dashed arrows connect each modality between teacher and student, demonstrating how these communication channels work together to facilitate effective software instruction.}
  \label{fig:teaser}
\end{teaserfigure}


\maketitle

\section{Introduction}
With the increasing digitization of essential products and services, software proficiency has become critical not only for productivity, but for basic participation in everyday life \cite{knowles2018wisdom}. Despite advances in user experience design, many modern applications remain complex, fragmented, and difficult to master \cite{Kiani2020IWJ}. As a result, users frequently rely on tutorials and help systems, yet these resources are often incomplete, inconsistent, or prohibitively difficult to scale \cite{joshi2020micromentor,tanprasert2024helpcall}.
Recent advances in generative AI offer a compelling alternative: personalized, step-by-step software guidance at scale~\cite{hou2024effects}. However, purely text-based instructions from LLMs are often difficult to follow, as they are not grounded in actual software interfaces. Users may struggle to interpret unfamiliar terminology or connect abstract textual descriptions to concrete UI actions~\cite{khurana2024and}.  For example, the instruction ``Click the lock button in the tool panel'' presumes that the user can locate the tool panel and visually identify the lock icon among many possibilities.


Multimodal generative AI has the potential to mirror human teaching strategies when helping others use software, integrating visual annotations, spatial guidance, and demonstrations to create rich and effective learning experiences. Multimedia learning theories suggest that combining multiple modalities affords spatial and temporal contiguity in instruction content, that is, aligning words and pictures in space and time. Such contiguity enhances learning~\cite{mayer2001cognitive,mayer2023past}. Research in human-computer interaction shows that multimodal guidance, which augments text instructions with additional cues such as visual highlights~\cite{kelleher2005stencils} and demonstrations~\cite{chi2012mixt,jin2022synapse}, improves understanding and learnability. Furthermore, human teachers are highly \textit{adaptive} at choosing appropriate knowledge representations according to their functions~\cite{ainsworth2006deft} (e.g.\ diagrams for structure and text for rules), teaching goals, and the receptivity of students~\cite{mainali2021representation}.

However, a critical gap remains in understanding how such multimodal adaptivity can be effectively modeled and implemented in automated software guidance systems to create more expressive and contextually aware guidance.
Our research is thus guided by one primary question:

\textbf{How are the modalities used to complement each other in software instruction in response to tutorial contents and student states?}

Inspired by prior work that analyzed human practices to inform the design of digital systems~\cite{lanir2008observing,luo2018time,heath1991collaborative}, we conducted an observational study of ten teacher–student pairs \edit{(N=10)} where experienced graphic design software users taught novices two new software features by guiding them through multi-step task procedures.
The study involved experienced Figma users with six months or more experience, and they taught novices with minimal or no prior experience via videoconferencing.
Teachers could communicate through three distinct modalities: verbal communication (speech), visual annotations (drawing on the student's screen), and remote screen control (direct manipulation when permitted). This setup mimicked the capabilities that future on-screen AI guidance systems might have, allowing us to observe natural multimodal teaching patterns in a controlled yet realistic environment.

Our analysis of modality functions shows that speech forms the foundation of nearly all instructional activities, while visual annotation and remote control serve as complementary channels, each with distinct strengths. All three modalities support the basic functions of directing attention and suggesting actions or values, but they do so with different trade-offs. Annotation is particularly effective for illustrating spatial arrangements and visually clarifying new concepts. Remote control enables direct demonstration, which is especially valuable for conveying the dynamics and inter-dependencies of complex action sequences. At a high level, annotation complements speech with spatial precision, while remote control complements it with both spatial and temporal precision. 
\edit{This aligns with the spatial and temporal contiguity principle~\cite{mayer2001cognitive,mayer2023past}.}

Yet, greater precision often comes at a cost.
Our observations and interviews revealed that teachers were aware of the varying degrees of intrusion to student agency\edit{, defined as students' ability to initiate and control their own actions and shared visual focus,} associated with each modality.
While annotations and screen control could illustrate software operations with sharp clarity, they diminished the students' need to actively interact with the software and thus the opportunities for deeper cognitive engagement, as suggested by Cognitive Load Theory~\cite{sweller1988cognitive,sweller2011cognitive} and help regulation research~\cite{aleven2007assistance}, albeit to different extents.
Teachers were also mindful of the impact of visual annotation and screen control on students' ownership of the software interface as their digital territories~\cite{scott2004territoriality} as a form of agency intrusion.
Teachers’ adaptive modality choices reflected a continuous balancing act: they sought to minimize agency intrusion while ensuring their guidance provides students with sufficient support to progress. This balancing effort depended not only on the spatial and temporal precision required for the instructional content but also on the student’s receptivity at a given moment.
These findings suggest that for AI-based guidance systems to be both effective and respectful, they must account for the semantic, cognitive, and social factors. These include the intrusion to student agency of each modality, the precision demands of the instructional content, and the dynamic states of students. This research makes two key contributions to the design of adaptive software guidance:

\begin{enumerate}[leftmargin=1.5em,noitemsep]
  \item[\textbf{1.}] \textbf{Empirical evidence of modality-specific purposes and trade-offs:} Systematic evidence of functions and usage patterns for speech, visual annotations, and remote screen control in software instruction, revealing that modalities with spatial and temporal contiguity afford higher instruction precision and a trade-off between precision and preserving agency.

  \item[\textbf{2.}] \textbf{Agency-aware modality selection:} Demonstration of how human teachers adaptively determine modality choice to maintain \textit{calibrated agency} in students, balancing the need for instruction progression with students' cognitive engagement and their sense of ownership of their digital territories.

\end{enumerate}

Our findings lend further support to the contiguity principle~\cite{mayer2002multimedia,mayer2023past} in multimedia learning; and the value of learner agency~\cite{gee2003videogames} from the perspectives of Cognitive Load Theory; and joint attention~\cite{kang2024jointattention,oura2024supporting,schneider2017mutualgaze} in the context of one-to-one tutoring of complex software. We further highlight precision-agency trade-off and digital territory ownership as two additional design constraints for on-screen software guidance that has direct access to the student's interface.

Together, these findings advance our understanding of multimodal instruction and inform the design of adaptive AI software assistance systems. The identified patterns also suggest novel design concepts (as illustrated in Fig.\ \ref{fig:design-ideas}), including ghost cursor interactions, adaptive annotation fading, timeline scrubbing, and interactive references. These address cognitive and usability challenges in software guidance and offer design possibilities for future AI tutoring systems that mirror and transcend human teaching strategies.

\section{Related Work}
This work draws from foundational learning theories (Cognitive Load Theory, help-seeking research, and Computer-Supported Collaborative Learning) and intersects three research areas: multimodal learning theory, software guidance systems, and adaptive tutoring.

\subsection{Foundational Learning Theories}

Our work builds on foundational learning theories that span from basic cognitive mechanisms to more social and collaborative perspectives on learning. At the cognitive level, Cognitive Load Theory (CLT) explains how the structure and complexity of information presentation affect learners' ability to form durable knowledge~\cite{sweller1982effects, sweller2011cognitive}. CLT distinguishes between intrinsic load (inherent to the material), extraneous load (imposed by poor instructional design), and germane load (effort devoted to schema construction). This framework is particularly relevant for software learning. Learners must manage both procedural knowledge (how to operate the interface) and conceptual understanding (why certain actions achieve desired outcomes) while navigating complex visual interfaces.

Building on these cognitive foundations, work on help regulation examines how learners and instructional systems manage the amount, timing, and form of assistance during learning~\cite{pea2018social, aleven2006metacognitive, aleven2007assistance, roll2007helpseeking}. Rather than focusing only on when learners ask for help, this perspective emphasizes how learners monitor their understanding, decide how much to rely on available support, and how tutors or systems adjust the level of scaffolding and feedback over time. Work in intelligent tutoring systems has explored the ``assistance dilemma''---the challenge of determining when and how much help to provide to optimize learning outcomes~\cite{aleven2007assistance}. In software learning contexts, help regulation is further complicated by the need to manage not only how much guidance is provided, but also which modalities of assistance (text, video, interactive guidance) are emphasized at different moments. Recent studies have examined how computing students' help-related behaviors and preferences are affected by generative AI tools~\cite{hou2024effects}, and how LLM-based assistants can participate in this ongoing regulation of help~\cite{khurana2024and}.

At a more social and distributed level, Computer-Supported Collaborative Learning (CSCL) research investigates how technology mediates learning through interaction with others~\cite{schneider2017mutualgaze, kang2024jointattention}. CSCL work highlights principles such as scaffolding, distributed cognition, and joint attention, showing how tools can structure collaborative activity so that learners co-construct knowledge. Although our study focuses on one-on-one human teaching rather than peer collaboration, these insights inform how we think about AI tutoring systems that model effective teaching practices observed in social settings and coordinate attention across shared digital workspaces.

\subsection{Software Task Assistance and Learning}

The problem of learning new features in complex software systems has captured the attention of researchers since the early days of computing~\cite{stewart1976displays}. Some of the pioneering work in HCI studied how people used software manuals and found that many tended to engage in minimum or no learning from the manuals and relied on self-exploration to learn the features. While these users could learn enough to perform the tasks of their interest, they often failed to reach expert proficiency as their self-exploration did not necessarily reveal the advanced features that boost productivity~\cite{carroll1987paradox}. This phenomenon was known as the Paradox of the Active User. The observation inspired the minimalist approach of instruction design, which emphasized practical and concise content~\cite{carroll1990nurnberg,lazonder1993minimal}, and a series of new interaction technique designs aimed at supporting novice-to-expert transition~\cite{cockburn2014supporting}. Many of these built on the idea of scaffolding to help learners gradually build proficiency, such as Marking Menu~\cite{kurtenbach1994user}, OctoPocus~\cite{bau2008octopocus}, and CommandMap~\cite{scarr2012improving}.

While the work above discussed what to present in software tutorials, another line of research mainly focused on how to organize and present them. Some early work along this line studied the effect of animated demonstrations for explaining software operations and found mixed outcomes~\cite{harrison1995comparison,palmiter1991evaluation}. A later approach closely related to this work used in-context tutorials that present guidance directly on the user interface element to operate, including stencil-based tutorials~\cite{kelleher2005stencils}, ToolClips~\cite{grossman2010toolclips}, and more comprehensive systems such as Sketch-Sketch Revolution~\cite{fernquist2011sketch}. However, prior work has primarily focused on static, single-modality guidance. Our work addresses this limitation by exploring dynamic, multimodal guidance that adapts both content and communication channels based on instructional context and learner needs.

\subsection{Multimodality in Education}

Multimodal communication plays a central role in learning, with research showing that coordinated use of complementary channels can improve comprehension.
Mayer's cognitive theory of multimedia learning explains how learners process information through partially separate visual and auditory channels~\cite{mayer1998cognitive, mayer2001cognitive, mayer2002multimedia}. Key principles include the modality principle, which favors narration over on-screen text when visuals are present, and the contiguity principle, which stresses aligning words and pictures in space and time~\cite{mayer2001cognitive, mayer2002multimedia}. These principles show how distributing information across channels can increase working memory capacity, while poorly coordinated redundancy can be counterproductive~\cite{mayer2001cognitive, mayer2023past}. In our work, we treat Mayer's framework as a set of cognitive baseline and use empirical analyses of human teaching to examine richer patterns of modality sequencing and coordination in software instruction.

Empirical studies of in-person teaching demonstrate how these principles appear in practice. Instructors often synchronize speech with visual demonstrations or gestures to direct attention and clarify abstract concepts~\cite{cook2013consolidation, singer2005children, alibali2013teachers}. Tutoring works mainly by eliciting students’ constructive responses rather than tutor explanations, as students continued to learn effectively even when tutors were prevented from explaining \cite{chi2001tutoring}. Work on software learning specifically highlights how teachers commonly combine pointing, screen annotations, and verbal explanations to guide attention and support procedural understanding~\cite{chi2012mixt, liu2023instrumentar, chilana2018supporting}. Studies of help-seeking in remote or informal software-learning contexts further underscore the importance of diagnosing when help is needed and how learners coordinate assistance~\cite{joshi2020micromentor, tanprasert2024helpcall,Kiani2020IWJ}.

Recent work in AI tutoring systems has explored how to model these multimodal teaching patterns. Systems like ~\cite{graesser2004autotutor} and interactive sketchpad~\cite{lee2025interactive} interfaces have shown benefits from combining visual annotations with verbal communication. However, most existing AI tutoring systems focus on single modalities or simple combinations, with limited coordination patterns compared to human teaching~\cite{shridhar2022automatic, kazemitabaar2023novices, wang2024tutor}. The challenge of modality selection and coordination in AI systems has been addressed in various domains. In robotics, research has explored when to use visual, auditory, or haptic feedback for different types of guidance~\cite{wang2025explainmr}. In educational technology, studies have examined how to automatically select appropriate modalities based on task characteristics and learner needs~\cite{ipsita2022towards}. However, there remains a gap in understanding how these principles apply specifically to software learning contexts in which spatial awareness and procedural knowledge are critical.

\subsection{Adaptiveness of Guidance Tools}

Research on adaptive tutoring systems investigates how to tailor guidance to learners by adjusting what help is provided and when it is offered~\cite{huang2021adaptutar, wiethuchter2024balancing}. Classic intelligent tutoring systems demonstrated that dynamically adapting feedback and content sequencing to learner performance can improve outcomes compared to static instructional materials, motivating subsequent work on more fine-grained forms of adaptiveness~\cite{weerasinghe2022arigato}. In software use, this idea has been extended to context-aware and proactive help systems that adjust guidance based on the user's current interface state, task goals, and inferred intent. For example, in-context tooltips and stencil-based overlays surface just-in-time explanations of relevant UI elements~\cite{grossman2010toolclips, kelleher2005stencils, fourney2014intertwine, chilana2012lemonaid, fraser2019replay}. Machine learning methods further broaden the scope of adaptiveness by using reinforcement learning to optimize interface configurations over time and natural language processing to tune the style and specificity of explanations to user preferences~\cite{gaspar2024reinforcement, gaspar2023learning, liu2025designing, spain2025applying}. Within this broader line of work, some systems instantiate adaptive guidance through pedagogical agents, such as animated characters or on-screen tutors that deliver instruction and feedback~\cite{graesser2004autotutor, johnson2000animated, graesser1999autotutor, graesser2005autotutor}. Studies of these systems examine when agents should intervene, how much support they should offer, and how to coordinate speech and gesture to scaffold learning~\cite{kapur2008productive, buisine2007speechgesture}, highlighting that adaptiveness involves managing not only content but also the timing and multimodal form of interventions.

Despite these advances, most adaptive systems primarily adjust \emph{what} content is presented rather than \emph{how} it is communicated or how open-ended tasks are structured. Software learning tasks are often loosely constrained, demand fine-grained spatial and procedural knowledge, and allow multiple valid strategies, making it more difficult to determine appropriate intervention points and forms of support. While some work has explored switching between presentation formats (e.g., text versus video) based on user preferences, there is limited research on dynamically adapting the choice and coordination of communication modalities based on task characteristics, user expertise, or learning context~\cite{weerasinghe2022arigato}. This gap is particularly salient for software learning, where different tasks (spatial arrangement, procedural execution, conceptual explanation) may benefit from different combinations of visual, verbal, and demonstrative guidance. Our work addresses this challenge by analyzing how human teachers adapt their modality choices and coordination patterns across instructional situations, with the goal of informing future adaptive guidance tools.
\section{Methodology}
This observational study employed a qualitative, exploratory approach to examine modality choice patterns during software skill instruction. We utilized a real-time teaching environment where instructors could view, add sketch annotations, and control students' screens to provide direct guidance and intervention when permitted.

\subsection{Participants}
Ten teacher-student pairs participated in the study \edit{(N=10)}. Participants ranged in age from 20 to 35+ years (teachers: M=25yo, SD=2.86; students: M=25.7, SD=3.72) and included 12 females, 7 males, and 1 participant identifying as other (see Table~\ref{tab:demographics}). Teachers included five experts with 2+ years of Figma experience and five with intermediate proficiency (6 months to 2 years). Students comprised five beginners with 1-6 months of Figma experience and five with no prior experience.

Teachers were recruited based on strong Figma proficiency, with a preference for candidates with prior teaching experience. Five teachers had over two years of Figma experience, and six had prior university-level software teaching experience with class sizes of 25 or more students. In total, nine of the ten teachers had some form of teaching background: the six with university teaching experience, \edit{plus three additional teachers (P1, P6, P8) who had K-12 tutoring experience with 1-5 students. One teacher (P5) had no prior teaching experience but possessed advanced Figma skills.} Recruitment occurred through targeted outreach on social media and university networks. Participants received \$30 in compensation. The study was approved by the institution’s ethics review board.

\begin{table*}[t]
    \centering
    \newcommand{\roundedcolorbox}[2]{\tcbox[colback=#1,colframe=#1,boxrule=0pt,arc=2pt,left=0.15em,right=0.15em,top=0.05em,bottom=0.05em,on line]{#2}}
    \caption{Participant demographics (N=10: 10 teachers, 10 students; ages 20-35+; 12F, 7M, 1O). Figma experience: Advanced (>2 years), Intermediate (6 months - 2 years), Beginner (<6 months), None. Teaching experience color-coded: software (darker), non-software (lighter). Students taught: 25+ (light grey). Course types: (1) Univ. software/technical; (2) Univ. non-software; (3) K-12 STEM; (4) K-12 tutoring. \textit{Note:} Gen. = Gender; Exp. = Experience; Adv. = Advanced; Int. = Intermediate; Beg. = Beginner.}
    \label{tab:demographics}
    \Description{Demographics for 20 participants (10 teachers, 10 students). Teachers: ID, Age, Gen., Figma Exp. (green color-coded: white=None, light=Beginner, medium=Intermediate, dark=Advanced), Software Teaching Exp. and Other Teaching Exp. (red color-coded: darker=software, lighter=non-software), number of students taught (25+ shaded grey). Students: ID, Age, Gen., Figma Exp. (green color-coded).}
    \small
    \renewcommand{\arraystretch}{1.2}
    \setlength{\aboverulesep}{0pt}
    \setlength{\belowrulesep}{0pt}
    \setlength{\extrarowheight}{0pt}
    \begin{tabular}{@{}c c c >{\raggedright\arraybackslash}p{1.8cm} >{\raggedright\arraybackslash}p{3cm} >{\raggedright\arraybackslash}p{2.6cm} >{\raggedright\arraybackslash}p{1.6cm} | c c c >{\raggedright\arraybackslash}p{1.8cm}@{}}
    \toprule
    \multicolumn{7}{c|}{\textbf{Teachers}} & \multicolumn{4}{c}{\textbf{Students}} \\
    \cmidrule{1-11}
    \textbf{ID} & \textbf{Age} & \textbf{Gen.} & \textbf{Figma Exp.} & \textbf{Software Teaching Exp.} & \textbf{Other Teaching Exp.} & \textbf{\# Students} & \textbf{ID} & \textbf{Age} & \textbf{Gen.} & \textbf{Figma Exp.} \\
    \midrule
    \arrayrulecolor{white}
P1-T & 20 & F & \cellcolor{green!40}Advanced & -- & \cellcolor{red!15}K-12 tutoring & 1-5 & P1-S & 29 & F & \cellcolor{green!10}Beginner \\\hline
P2-T & 27 & F & \cellcolor{green!40}Advanced & \cellcolor{red!30}Univ.-level software instruction (Figma) & \cellcolor{red!15}Univ.-level non-software instruction & \cellcolor{gray!10}200+ & P2-S & 22 & M & \cellcolor{white}-- \\\hline
P3-T & 24 & F & \cellcolor{green!20}Intermediate & \cellcolor{red!30}Univ.-level software instruction (Python/Pandas/Tableau) & \cellcolor{red!15}K--12 tutoring & \cellcolor{gray!10}50+ & P3-S & 25 & F & \cellcolor{white}-- \\\hline
P4-T & 28 & M & \cellcolor{green!20}Intermediate & \cellcolor{red!30}Univ.-level software instruction (SQL, Python) & \cellcolor{red!15}K-12 tutoring & \cellcolor{gray!10}75+ & P4-S & 24 & M & \cellcolor{green!10}Beginner \\\hline
P5-T & 30 & F & \cellcolor{green!40}Advanced & -- & -- & -- & P5-S & 27 & F & \cellcolor{white}-- \\\hline
P6-T & 23 & F & \cellcolor{green!20}Intermediate & -- & \cellcolor{red!15}K-12 tutoring & 1-5 & P6-S & 23 & F & \cellcolor{green!10}Beginner \\\hline
P7-T & 27 & M & \cellcolor{green!40}Advanced & \cellcolor{red!30}Univ.-level software instruction (Figma) & -- & \cellcolor{gray!10}25+ & P7-S & 35+ & M & \cellcolor{white}-- \\\hline
P8-T & 22 & M & \cellcolor{green!40}Advanced & -- & \cellcolor{red!15}K-12 tutoring & 1-5 & P8-S & 23 & M & \cellcolor{green!10}Beginner \\\hline
P9-T & 24 & F & \cellcolor{green!20}Intermediate & \cellcolor{red!30}Univ.-level software instruction (Unity, Figma) & \cellcolor{red!15}K--12 STEM workshops (Robotics, Drones, Chatbots) & \cellcolor{gray!10}100+ & P9-S & 23 & F & \cellcolor{green!10}Beginner \\\hline
P10-T & 25 & O & \cellcolor{green!20}Intermediate & \cellcolor{red!30}Univ.-level software instruction (Figma) & \cellcolor{red!15}K--12 STEM workshops (Robotics) & \cellcolor{gray!10}75+ & P10-S & 26 & F & \cellcolor{white}-- \\
    \arrayrulecolor{black}
    \bottomrule
    \end{tabular}
    \vspace{-\baselineskip}
    \end{table*}

\subsection{Experiment Design and Environment}

The study used a structured teaching session in which teachers received preparatory materials two days in advance but retained full autonomy over how they taught. Each session included two sequential tutorials followed by independent test tasks. Sessions were conducted in a controlled lab setting where teacher and student used separate laptops connected to the same Zoom call, enabling screen sharing, annotation, and remote control. All interactions were screen- and audio-recorded; detailed hardware specifications are provided in Appendix~\ref{sec:appendix-technical}.

The instructional platform supported three communication modalities: speech, visual annotations, and remote screen control. We selected these not as an exhaustive set for tutoring systems, but as distinct and representative modalities for examining how instructors deployed them, coordinated them, and navigated trade-offs (e.g., precision vs.\ agency) during live instruction. This bounded interaction space allowed natural teaching interactions to emerge while keeping technical conditions consistent.

\subsection{Procedure}

Each teaching session followed a standardized procedure lasting approximately 1.5 hours per teacher-student pair. Two days before each session, teachers received preparatory materials including complete task instructions, technical setup information, and study expectations (See Appendix \ref{sec:appendix-study-tasks}). While teachers were encouraged to teach naturally with no restrictions on their approach, they were instructed to ensure students could produce the same outputs as specified in the tutorials. Teachers retained full autonomy over their instructional approach and modality choices throughout both tasks.

The session began with a 15-minute pre-session setup including informed consent, technical verification, and participant introductions. Teachers were then briefed on Zoom's screen sharing, annotation, and remote-control features and given time to practice to ensure baseline technical familiarity. No formal assessment of teachers' prior experience with Zoom features was conducted, as the study aimed to capture natural teaching behaviors rather than controlled feature usage.

The teaching portion consisted of \textbf{Task 1 (Weather Icons)} (15--30 minutes of teaching, 10 minutes of post-task testing, and a 10-minute discussion), followed by a 5-minute break and \textbf{Task 2 (Weather Cards)} (15--30 minutes of teaching, 10 minutes of post-task testing, and a final 10-minute discussion). Semi-structured interviews were conducted after each task and at the end of each session to capture both teacher and student perspectives on teaching effectiveness, modality choices, learning experiences, and communication challenges.

\subsection{Task Design}

\begin{figure}[t!]
    \centering
    \captionsetup{width=\linewidth}
    \includegraphics[width=\linewidth]{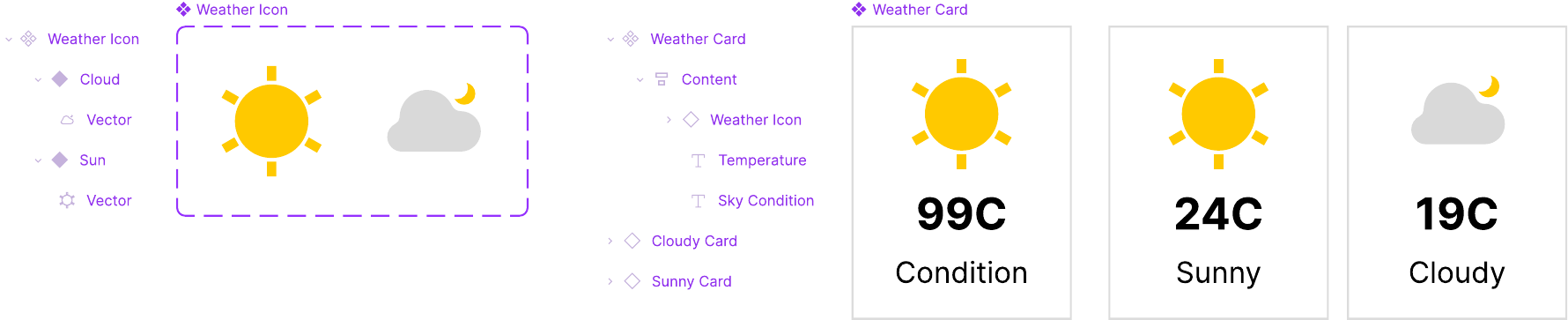}
    \caption{Two Figma design tasks: Task 1: Weather Icons Creation (left) and Task 2: Weather Cards Design (right), showing progression from fundamental skills to advanced component usage}
    \label{fig:task-comparison}
    \Description{Two-part Figma design workflow. Task 1 (left): Weather Icon component hierarchy with Sun and Cloud vector sub-components; displays yellow sun with rays and gray cloud with moon. Task 2 (right): Weather Card component hierarchy with nested Weather Icon, Temperature, and Sky Condition elements; shows three card instances (sun icon with "99C/Condition", sun icon with "24C/Sunny", cloud icon with "19C/Cloudy") demonstrating progression from basic icons to composite components.}
\end{figure}

The study employed a progressive approach centered on designing weather-themed components in Figma. We selected Figma because it represents modern applications with rich functionality and minimalist interfaces that can pose an initial learning hurdle. To calibrate the task difficulty and ensure appropriate skill coverage, we conducted pilot studies with participants who had no prior Figma experience. These pilots helped us establish a foundational set of basic features that any beginner should master. They also identified more advanced features that would challenge learners without being inaccessible. This deliberate calibration ensured the tasks were sufficiently challenging to elicit diverse teaching strategies while remaining approachable for novice learners. The two-task sequence follows a natural learning progression from concrete manipulation to abstract system thinking, with Lesson 1 emphasizing direct manipulation and Lesson 2 requiring more conceptual understanding. Both tasks provided rich opportunities for various communication modalities. Vector manipulation required precise pointing and annotation, while component concepts benefited from conceptual explanations and demonstrations (see Figure~\ref{fig:task-comparison}). Lessons 1 and 2 consisted of seven and eight steps respectively. More details about the two lessons can be found in the Appendix~\ref{sec:appendix-study-tasks}.

Following each lesson, students completed independent tests that required applying learned skills to new but similar tasks (e.g., creating additional weather icons or designing new component layouts). These post-lesson tests were video-recorded to assess skill transfer and retention and to provide immediate feedback to teachers.

\subsection{Data Collection and Analysis}

We collected synchronized screen and audio recordings from all sessions, capturing the student’s Figma interface, teacher annotations, remote-control actions, and all verbal interactions. These recordings constituted the primary dataset for examining how instructors coordinated speech, visual annotations, and screen control during live software instruction.

\textbf{Multimodal Data Extraction.}
Sessions were segmented into lesson steps to provide temporal structure for subsequent analyses. For each session, we extracted (1) visual annotations drawn through Zoom’s annotation tools, (2) remote-control segments where teachers directly manipulated the student’s interface, and (3) speech transcripts generated via automated speech recognition and manually corrected for accuracy.
Automated detection of annotations and technical details of speech and video processing are provided in Appendix~\ref{sec:technical-implementation}.

\textbf{Coding Framework:}
For the modality usage patterns, we performed content analysis over the data collected. 
To enable cross-modal analysis, speech segments were temporally aligned with corresponding visual annotations and remote control demonstrations through shared step markers and timestamps. This alignment allowed us to examine how different modalities complemented each other during instruction, such as identifying instances where verbal commands were accompanied by visual annotations or where explanations coincided with remote control demonstrations.
Two researchers first conducted open coding on two sessions to generate fine-grained descriptions of instructional functions for the three modalities. 
Upon discussion, they refined and grouped the codes into major function categories.
Using this coding scheme, they independently coded the rest of the data and achieved an inter-rater reliability (Cohen's Kappa) of 0.82. 
The team discussed again to resolve discrepancies and generate the final modality usage frequency counts.

For higher-level teacher-student dynamics, we followed a similar workflow but took the inductive thematic analysis approach, where two researchers reviewed two sessions to produce the potential codes and used affinity diagramming to group them into initial themes. 
They then independently analyzed the rest of the data with a discussion half-way to share new findings and refine the codes. 
Our final behavior analysis reached an inter-rater reliability of 0.77.

\section{Results}
Our analysis reveals three key findings about multimodal teaching in software instruction. First, teachers relied on three communication modalities---speech, annotation, and screen control---each with distinct strengths and limitations. Second, teachers adapted their communication in response to student feedback and task demands. Third, modality effectiveness varied by context, requiring strategic combinations: speech provided the baseline scaffolding, annotations added spatial precision, and screen control supported spatially and temporally complex demonstrations.

\subsection{Study Overview and Participant Performance}

\definecolor{speechgreen}{HTML}{95C78C}
\definecolor{annoblue}{HTML}{4DABF5}
\definecolor{rcorange}{HTML}{E4B36E}
\definecolor{rectifyred}{HTML}{D97A7A}

\begin{table*}[t!]
    \centering
    \caption{Teaching Modality Categories and Examples from Study}
    \label{tab:mm-comparison}
    \Description{Table of teaching modality categories and examples. Three modality types with 11 categories total: Speech only (5: Direct Attention, Suggesting Value, Command Action, Explain Concept, Micro-guidance); Visual Annotations with speech (4: Highlight Target + Command, Write Value + Command, Draw + Explain Arrangement, Draw + Explain Concept); Remote Screen Control (Demonstration with speech, Rectification without speech). Columns are Modality Type and Combination, Category, Description, and Examples with participant quotes (P1--P9).}

    \small
    \renewcommand{\arraystretch}{1.5}
    \setlength{\aboverulesep}{0pt}
    \setlength{\belowrulesep}{0pt}
    \setlength{\extrarowheight}{0pt}
    \begin{tabular}{|>{\centering\arraybackslash}p{0.15\textwidth}
                    |>{\centering\arraybackslash}p{0.08\textwidth}
                    |>{\centering\arraybackslash}p{0.15\textwidth}
                    |m{0.25\textwidth}
                    |m{0.25\textwidth}|}
        \toprule
        
        \multicolumn{2}{|c|}{\textbf{Modality Type and Combination}} & \textbf{Category} & \textbf{Description} & \textbf{Examples} \\
        
        \midrule
        
        \multicolumn{2}{|c|}{\multirow{5}{*}{\textbf{Speech Only}}} 
            & \cellcolor{speechgreen!25}\textbf{Direct Attention} 
            & Verbal cues to focus student attention on specific interface elements 
            & \textit{"at the top of the circle"} (P1), \textit{"look at this button here"} (P5) \\\cmidrule{3-5}
            
        \multicolumn{2}{|c|}{} & \cellcolor{speechgreen!25}\textbf{Suggesting Value} 
            & Providing specific numerical or categorical values for students to use 
            & \textit{"change the height to 64"} (P4), \textit{"set the width to 200"} (P6) \\\cmidrule{3-5}
            
        \multicolumn{2}{|c|}{} & \cellcolor{speechgreen!25}\textbf{Command Action} 
            & Direct verbal instructions for students to perform specific actions 
            & \textit{"click on this icon"} (P2), \textit{"drag this element ..."} (P7) \\\cmidrule{3-5}
            
        \multicolumn{2}{|c|}{} & \cellcolor{speechgreen!25}\textbf{Explain Concept} 
            & Verbal explanations of concepts, relationships, or reasoning 
            & \textit{"components help standardize your design"} (P3), \textit{"this creates a reusable template"} (P8) \\\cmidrule{3-5}
            
        \multicolumn{2}{|c|}{} & \cellcolor{speechgreen!25}\textbf{Micro-guidance} 
            & Brief confirmations, encouragements, or directional cues 
            & \textit{"Yes, correct, yeah, yeah"} (P2), \textit{"good, keep going"} (P9) \\
        
        \midrule
        
        \multirow{4}{*}{\textbf{\shortstack{Visual\\Annotations}}} & \multirow{4}{*}{+ Speech} 
            & \cellcolor{annoblue!25}\raisebox{-0.6em}{\textbf{\shortstack{Highlight Target\\+ Command}}} 
            & Combining verbal instructions with visual annotations to guide actions 
            & \textit{"So make a circle with that tool. Okay."} \textit{(Circle shape tool)} (P2) \\\cmidrule{3-5}
        
        & & \cellcolor{annoblue!25}\raisebox{-0.6em}{\textbf{\shortstack{Write Value\\+ Command}}} 
            & Providing specific values while highlighting relevant interface elements 
            & \textit{"This is the position it is on like this page"} \textit{(Circle coordinate field)} (P3) \\\cmidrule{3-5}
            
        & & \cellcolor{annoblue!25}\raisebox{-0.6em}{\textbf{\shortstack{Draw + Explain\\Arrangement}}} 
            & Using visual annotations to show spatial relationships and layout while explaining verbally 
            & \textit{"Put this here, and then this goes below it"} \textit{(Arrows showing positioning)} (P4) \\\cmidrule{3-5}
            
        & & \cellcolor{annoblue!25}\raisebox{-0.6em}{\textbf{\shortstack{Draw + Explain\\Concept}}} 
            & Using visual annotations to clarify concepts while explaining verbally 
            & \textit{"This is like if you just put this sun, they won't know what it is"} \textit{(Arrow pointing toward text)} (P4) \\
        
        \midrule
        
        \multirow{2}{*}{\textbf{\shortstack{Remote Screen\\Control}}} & + Speech 
            & \cellcolor{rcorange!25}\textbf{Demonstration} 
            & Explaining actions while performing them remotely on student's screen 
            & \textit{"So I'm taking over the screen. So you can..."} (P6) \\\cmidrule{2-5}
        
         & - & \cellcolor{rectifyred!25}\textbf{Rectification} 
            & Taking control without explanation to fix or adjust student work 
            & NA\\
        
        \bottomrule
        
    \end{tabular}
\end{table*}

Ten teacher-student pairs completed two sequential Figma design tasks, generating 12.4 hours of recorded footage. We refer to participants by pair ID and role (e.g., P1-T for the teacher and P1-S for the student). All participants completed both lessons, with session durations ranging from 61--92 minutes (M=71.5, SD=9.2). Lesson 1 required 20--31 minutes (M=24.8, SD=3.4), while Lesson 2 took 12--30 minutes (M=19.2, SD=4.1). The most time-intensive step was sun ray creation (Step 3) in Lesson 1 (M=8.2 minutes). Demonstrating cascading changes for auto-layout in Lesson 2 (Step 8) generated the highest speech output (5,748 words) and screen control usage; the ray creation step also required the most annotations (85 instances).

\subsection{Learning Outcomes and Skill Transfer}

Students performed well on the independent transfer tasks, indicating that the lessons enabled skill transfer to novel problems. All students (10/10) completed Test~1 successfully, and nine students (9/10) completed Test~2. Although a few students initially hesitated on unfamiliar steps, they generally resolved these challenges and completed the tasks.

\subsection{Modality Usage Pattern}

All participants employed multiple communication modalities, with 7 out of 10 using all three (speech, annotation, and screen control). Speech was universal, while 8 participants used annotations and 8 used screen control. Usage patterns varied substantially: 3 teachers were heavy annotation users, while 5 used annotations lightly; similarly, 3 relied heavily on screen control, with 5 using it sparingly. Notably, some teachers expressed strong preferences against certain modalities: P2-T avoided screen control, stating \textit{"if I control too much, the student don't get the experience,"} while P9-T was hesitant about both annotation and control, preferring not to \textit{"touch their stuff."} Overall, modality use was adaptive rather than uniform, shaped by instructional goals, student responses, and individual teaching philosophies. Below, we detail the functions and limitations of each modality.

\definecolor{C0}{HTML}{FFFFFF}
\definecolor{C1}{HTML}{FBFCFF}
\definecolor{C2}{HTML}{E7F0FB}
\definecolor{C3}{HTML}{D1E1F7}
\definecolor{C4}{HTML}{B5CFF3}
\definecolor{C5}{HTML}{9ABDEF}

\begin{table*}[h]
    \centering
    \small
    \raggedright
    \hyphenpenalty=10000
    \exhyphenpenalty=10000
    \caption{Frequency distribution of visual annotations by type and periods of remote control across design task steps for Weather Icons and Weather Cards tasks. Color intensity indicates relative frequency within each column, with darker shades representing higher values.}
    \label{fig:direct-attention-command-annotations}
    \Description{Table of frequency of visual annotations by type and remote control periods by task step. Columns: Step, Description, Total annotations, Highlight Target & Command, Suggest Value & Command, Draw & Explain Arrangement, Draw & Explain Concept, Demonstration (seconds), Rectification (seconds). Two sections: Weather Icons (Pre-Lesson plus 7 steps) and Weather Cards (Pre-Lesson plus 8 steps). Cell color intensity (white to dark blue) indicates relative frequency within column. Totals row: 426 annotations (312, 23, 49, 37 by type), 1784 s demonstration, 874 s rectification. Orange shading in Demo and Rectify columns.}
    \setlength{\aboverulesep}{0pt}
    \setlength{\belowrulesep}{0pt}
    \setlength{\extrarowheight}{0pt}
        \begin{tabular}{c|c|p{3.8cm}|p{0.8cm}|p{1.4cm}|p{1.4cm}|p{1.8cm}|p{1.4cm}|p{1.3cm}|p{1.3cm}}
    \toprule
    & \multirow{2}{*}{\textbf{Step}} & \multirow{2}{*}{\textbf{Description}} & \multicolumn{5}{c|}{\textbf{Number of Visual Annotations}} & \multicolumn{2}{c}{\textbf{Periods of Remote Control}} \\
    \cmidrule{4-8} \cmidrule{9-10}
    & & & \textbf{Total} & \textbf{Highlight Target \& Command} & \textbf{Suggest Value \& Command} & \textbf{Draw \& Explain Arrangement} & \textbf{Draw \& Explain Concept} & \textbf{Demo. (s)} & \textbf{Rectify. (s)} \\
    \midrule
    \multirow{8}{*}{\rotatebox{90}{\textbf{Weather Icons}}} & \multicolumn{2}{c|}{Pre-Lesson: Introduction}
                                            & \cellcolor{C1}3  & \cellcolor{C0}0  & \cellcolor{C0}0  & \cellcolor{C3}3  & \cellcolor{C0}0 & \cellcolor{orange!10}202 & 0 \\
    \cmidrule{2-10}
    & 1 & Prepare Your Canvas               & \cellcolor{C1}4  & \cellcolor{C1}4  & \cellcolor{C0}0  & \cellcolor{C0}0  & \cellcolor{C0}0  & \cellcolor{orange!10}198 & 0 \\
    & 2 & Create the Circle                 & \cellcolor{C3}14 & \cellcolor{C2}9  & \cellcolor{C5}3  & \cellcolor{C3}2  & \cellcolor{C0}0  & 0 & 0 \\
    & 3 & Create the Rays                   & \cellcolor{C5}85 & \cellcolor{C5}65 & \cellcolor{C5}7  & \cellcolor{C5}12 & \cellcolor{C2}1  & \cellcolor{orange!40}354 & \cellcolor{orange!25}196 \\
    & 4 & Create the Cloud Shape            & \cellcolor{C4}34 & \cellcolor{C3}12 & \cellcolor{C5}3  & \cellcolor{C5}18 & \cellcolor{C2}1  & \cellcolor{orange!20}136 & 0 \\
    & 5 & Create the Crescent Shape         & \cellcolor{C3}23 & \cellcolor{C3}11 & \cellcolor{C3}2  & \cellcolor{C4}6  & \cellcolor{C3}4  & 0 & 0 \\
    & 6 & Flatten the Icon                  & \cellcolor{C1}5  & \cellcolor{C2}5  & \cellcolor{C0}0  & \cellcolor{C0}0  & \cellcolor{C0}0  & \cellcolor{orange!10}39 & \cellcolor{orange!10}39 \\
    & 7 & Create a Component Set            & \cellcolor{C5}39 & \cellcolor{C5}33 & \cellcolor{C0}0  & \cellcolor{C0}0  & \cellcolor{C5}6  & \cellcolor{orange!25}206 & \cellcolor{orange!45}412 \\
    \midrule
    \multirow{9}{*}{\rotatebox{90}{\textbf{Weather Cards}}} & \multicolumn{2}{c|}{Pre-Lesson: Introduction}
                                            & \cellcolor{C2}11 & \cellcolor{C1}1  & \cellcolor{C0}0  & \cellcolor{C0}0  & \cellcolor{C4}5  & 0 & \cellcolor{orange!10}2 \\
    \cmidrule{2-10}
    & 1 & Card Frame                        & \cellcolor{C1}3  & \cellcolor{C1}2  & \cellcolor{C0}0  & \cellcolor{C2}1  & \cellcolor{C0}0  & \cellcolor{orange!10}87 & 0 \\
    & 2 & Add Weather Icon                  & \cellcolor{C2}12 & \cellcolor{C3}11 & \cellcolor{C2}1  & \cellcolor{C0}0  & \cellcolor{C0}0  & \cellcolor{orange!10}85 & \cellcolor{orange!10}85 \\
    & 3 & Add Temperature Text              & \cellcolor{C2}12 & \cellcolor{C2}10 & \cellcolor{C2}1  & \cellcolor{C2}1  & \cellcolor{C0}0  & 0 & 0 \\
    & 4 & Add Condition Label               & \cellcolor{C5}43 & \cellcolor{C5}37 & \cellcolor{C3}2  & \cellcolor{C4}4  & \cellcolor{C0}0  & 0 & 0 \\
    & 5 & Create Component                  & \cellcolor{C4}36 & \cellcolor{C5}35 & \cellcolor{C0}0  & \cellcolor{C0}0  & \cellcolor{C2}1  & 0 & 0 \\
    & 6 & Create Two Card Instances         & \cellcolor{C3}27 & \cellcolor{C4}21 & \cellcolor{C2}1  & \cellcolor{C0}0  & \cellcolor{C4}5  & 0 & 0 \\
    & 7 & Cascading Changes - Font Weight   & \cellcolor{C4}28 & \cellcolor{C4}25 & \cellcolor{C0}0  & \cellcolor{C2}1  & \cellcolor{C2}2  & 0 & 0 \\
    & 8 & Cascading Changes - Auto Layout   & \cellcolor{C5}47 & \cellcolor{C4}31 & \cellcolor{C5}3  & \cellcolor{C2}1  & \cellcolor{C5}12 & \cellcolor{orange!45}477 & \cellcolor{orange!25}140 \\
    \midrule
    \multicolumn{3}{r|}{\textbf{Total}} & \textbf{426} & \textbf{312} & \textbf{23} & \textbf{49} & \textbf{37} & \textbf{1784} & \textbf{874} \\
    \bottomrule
    \end{tabular}

    \end{table*}
    
    \vspace{-\baselineskip}

\subsection{Speech Only}
\noindent
\begin{tabular}{@{}p{5em}@{\hspace{0.6em}}p{\dimexpr\linewidth-5em-0.6em\relax}@{}}%
  \begin{minipage}[t]{\linewidth}
    \vspace{0pt}\includegraphics[height=6em,width=5em,keepaspectratio,alt={Illustration showing speech bubbles to represent instruction delivered solely through verbal communication}]{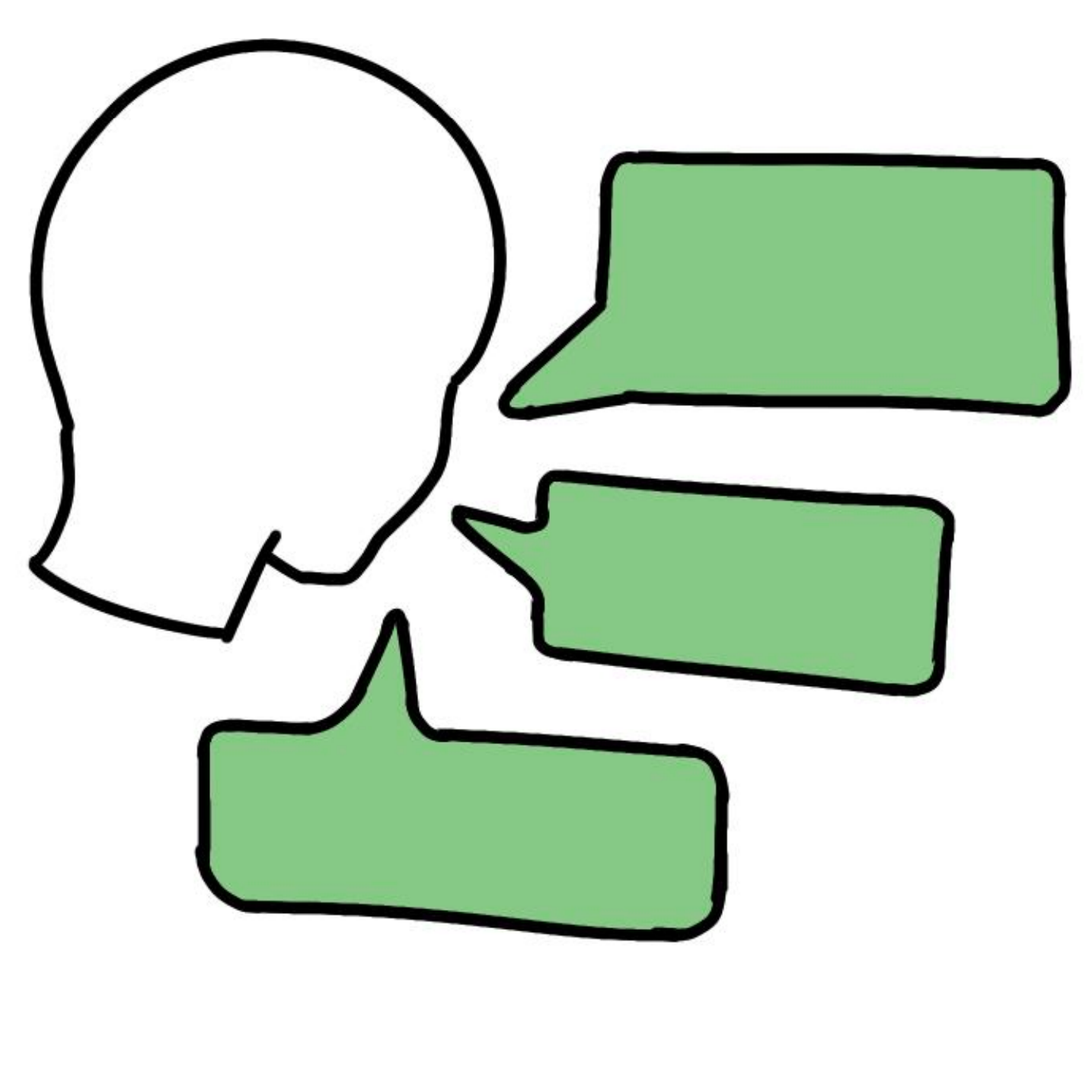}
  \end{minipage}
  &
  \begin{minipage}[t]{\linewidth}
    \vspace{0pt}Speech served as the foundational modality for all instruction, providing the essential scaffolding upon which other modalities built. This foundational role was evident in several ways: speech was universally employed across all participants and sessions, while other modalities were used selectively.
  \end{minipage}
\end{tabular}\par\medskip

Students consistently relied on verbal guidance for procedural support; as one student put it, when teachers \textit{"tell me where to press, I would remember this."} Teachers used speech to establish context, explain relationships between actions, and provide real-time feedback that students could process alongside their interface interactions. Speech also coordinated modality switches: teachers often announced transitions (e.g., ``let me show you'' before demonstrations or ``I'll circle this'' before annotations), making other modalities feel deliberate rather than abrupt. Our analysis identified five key interaction patterns: directing attention to specific interface elements, suggesting precise values for adjustments, commanding specific actions, explaining conceptual relationships between interface components, and providing micro-guidance.

\subsubsection{Direct Attention}

\raisebox{-0.2em}{\includegraphics[height=1em,alt={Small label-style icon indicating a verbal cue used to redirect the student’s visual attention to a specific interface location}]{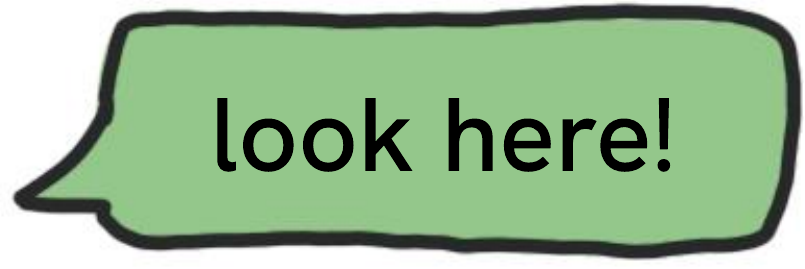}}
Teachers used direct language to guide users' visual focus toward specific interface elements. Teachers provided both detailed explanations and concise commands to direct user attention. For instance, one teacher (P1-T) directed attention to interface elements: \textit{``you can change the size at the side here''} while another teacher (P2-T) used a more focused directive: \textit{``focus on the corner handles''}.

\subsubsection{Suggesting Value}

\raisebox{-0.2em}{\includegraphics[height=1em,alt={Tag icon showing a numeric value, representing teacher speech that specifies exact parameters such as dimensions or numerical settings}]{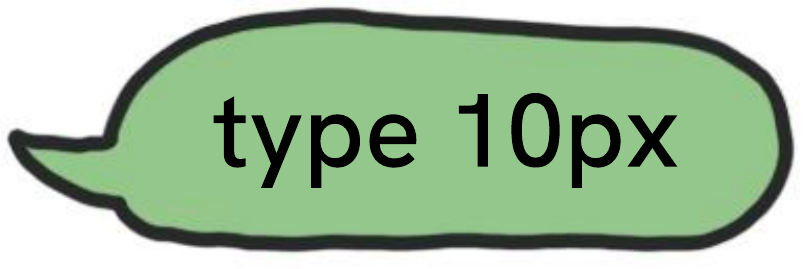}}
Teachers provided specific numerical or categorical values to guide interface adjustments. Teachers used direct value specifications such as ``change it to <value>'' or ``set the width to 200'' to guide users toward precise configurations. For example, one teacher (P4-T) provided instruction: \textit{``Can you change a bit and height to 64?''} while another teacher (P1-T) simply stated: \textit{``130 by 160''} with the value implied. This demonstrates how teachers used value-specification language to provide efficient and actionable guidance.

\subsubsection{Command Action}

\raisebox{-0.2em}{\includegraphics[height=1em,alt={Clickable-style badge representing direct action commands, such as instructing the student to press or activate a specific interface control}]{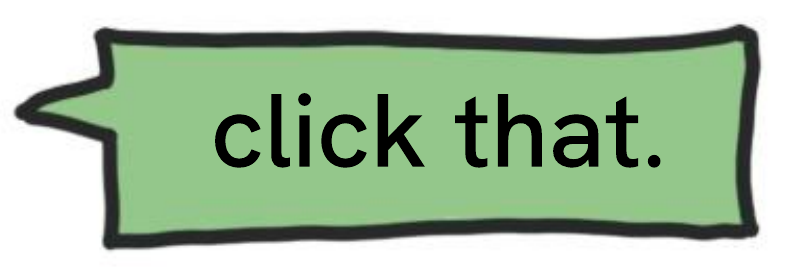}}
Teachers provided step-by-step verbal commands for interface manipulation, using prompts like ``Circle coordinate field'' or ``Circle stroke panel'' sometimes followed by explanatory context. For instance, P1-T offered a detailed prompt such as \textit{``Circle coordinate field''} while another teacher used a more focused directive: \textit{``You can click ... you can set all of them.''} These instructions helped guide students through specific interface operations.

\subsubsection{Explain Concept}

\raisebox{-0.2em}{\includegraphics[height=1em,alt={Icon representing conceptual explanation, indicating speech used to describe underlying principles, relationships, or system logic}]{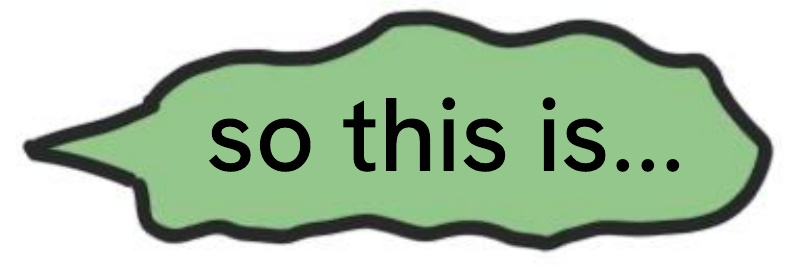}}
Teachers described abstract interface concepts and component interactions to clarify how system elements relate. Teachers provided explanations for complex relationships such as ``explaining parent child relationship'' or ``describing how components interact'' to help users understand the underlying system architecture. For example, one teacher (P1-T) explained alignment feedback by noting, \textit{``(Figma) will tell you when you’re at the center''} while another teacher (P2-T) used a more accessible conceptual summary: \textit{``It's like a parent-child relationship.''} This demonstrates how teachers used conceptual explanations to help users understand complex interface relationships.

\begin{figure*}[t!]
    \centering
    \small
    \caption{Annotation Styles and Their Characteristics}
    \vspace{5pt}
    \label{fig:annostyles}
    \Description{Table of seven annotation styles used in software instruction: Circles/Rectangles (highlighting elements), Arrows (pointing to UI elements), Lines (connecting elements), Brackets (grouping elements), Text (providing values or labels), Ticks/Crosses (confirming completion), and Drawings (freehand sketches). Each column shows the style name, its function, and a visual example.}
    \resizebox{\textwidth}{!}{%
        \begin{tabular}{@{}*{7}{>{\raggedright\arraybackslash}p{2.2cm}}@{}}
            \includegraphics[width=1.8cm]{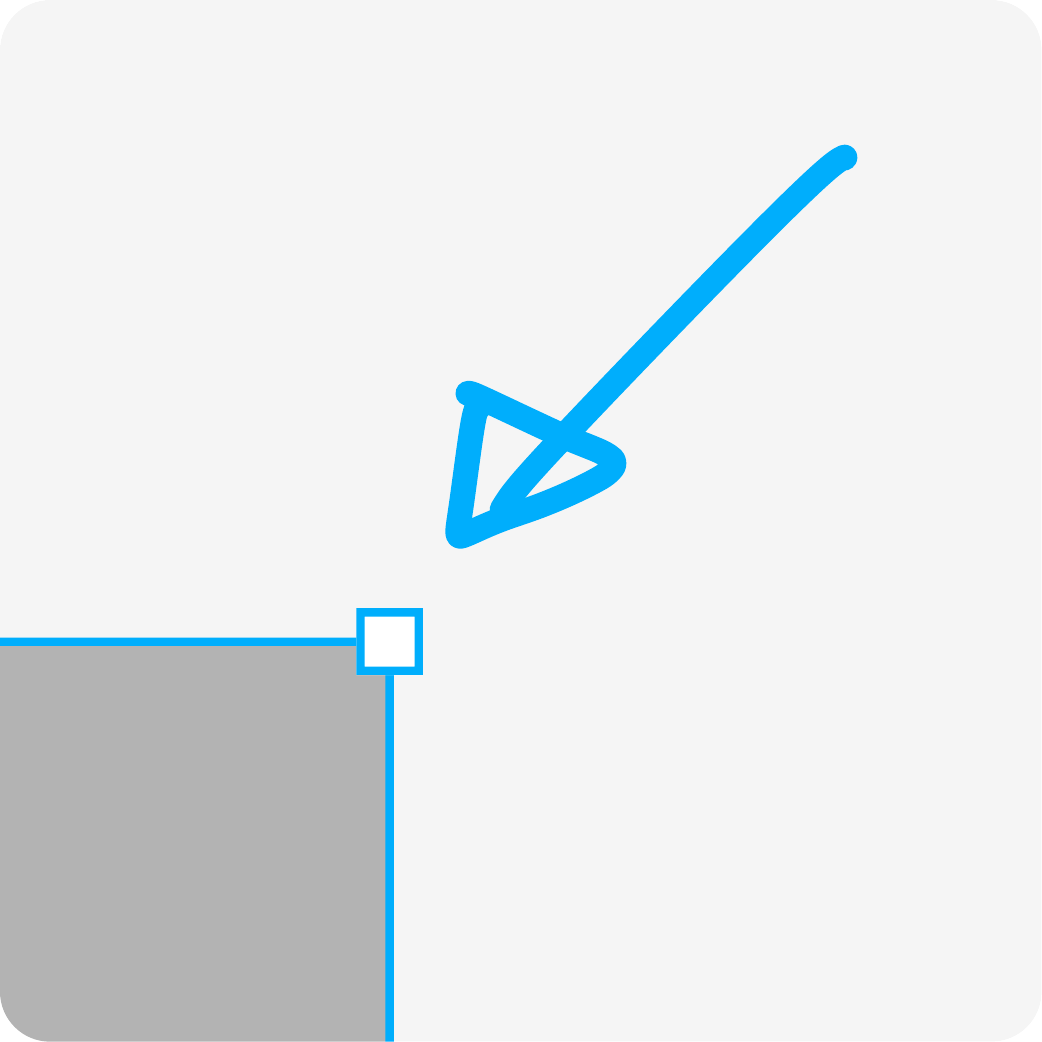} & 
            \includegraphics[width=1.8cm]{images/anno-styles/arrow.pdf} & 
            \includegraphics[width=1.8cm]{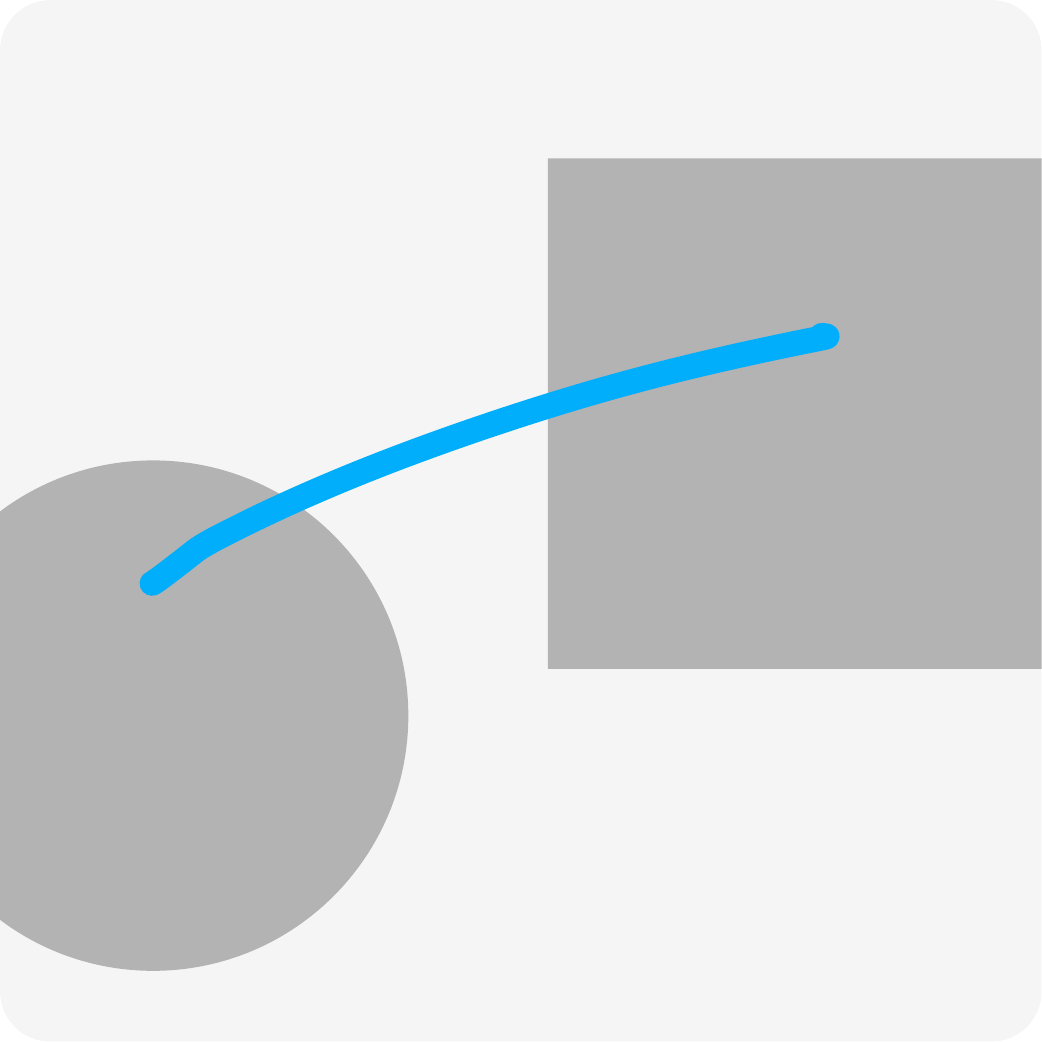} & 
            \includegraphics[width=1.8cm]{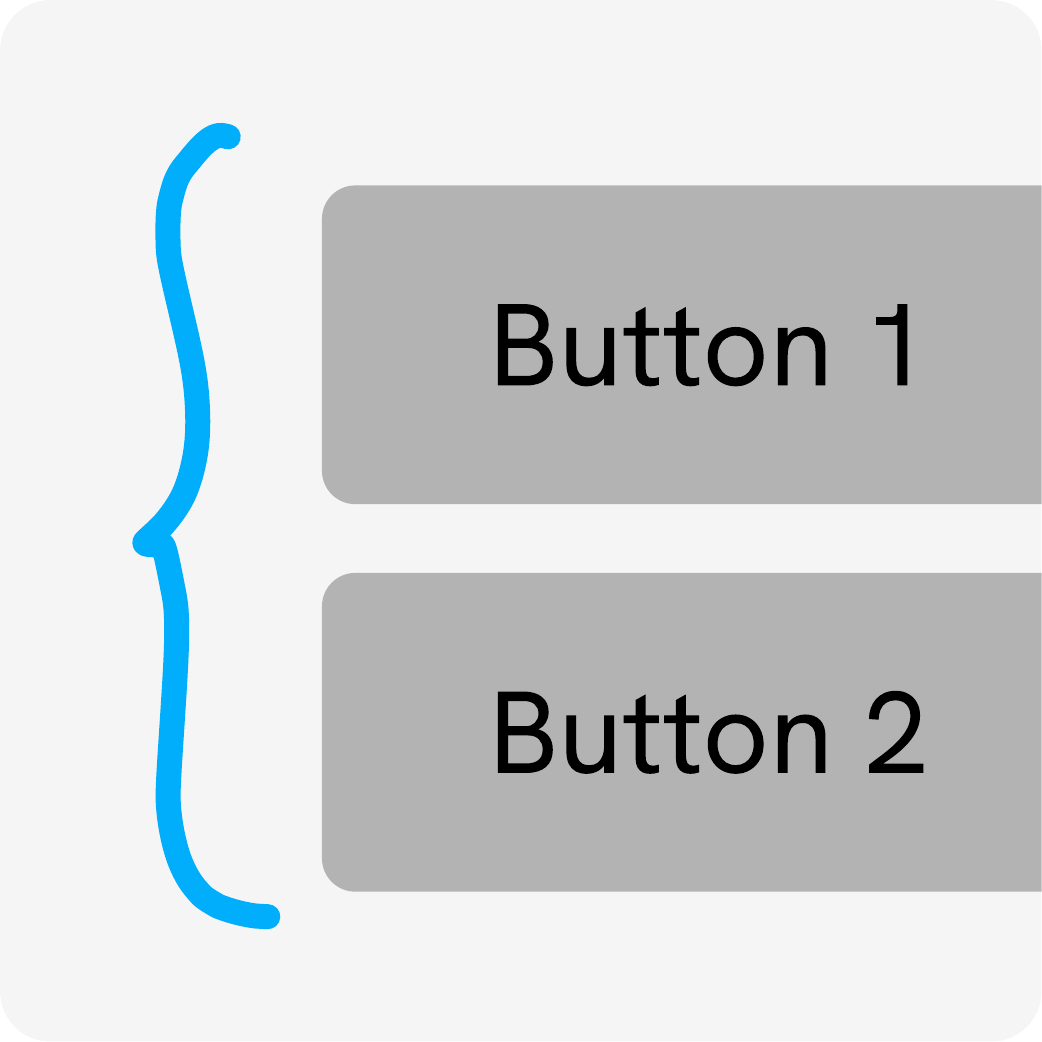} & 
            \includegraphics[width=1.8cm]{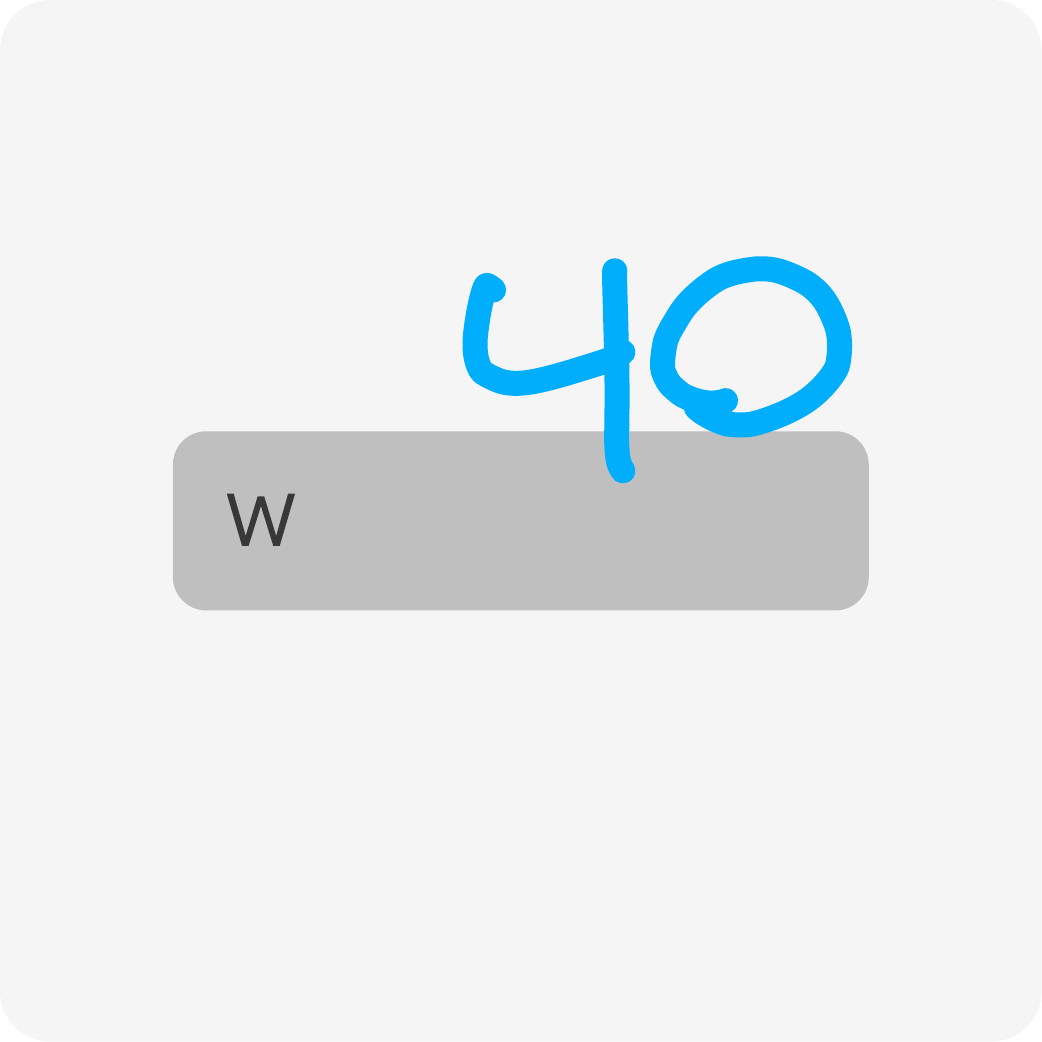} & 
            \includegraphics[width=1.8cm]{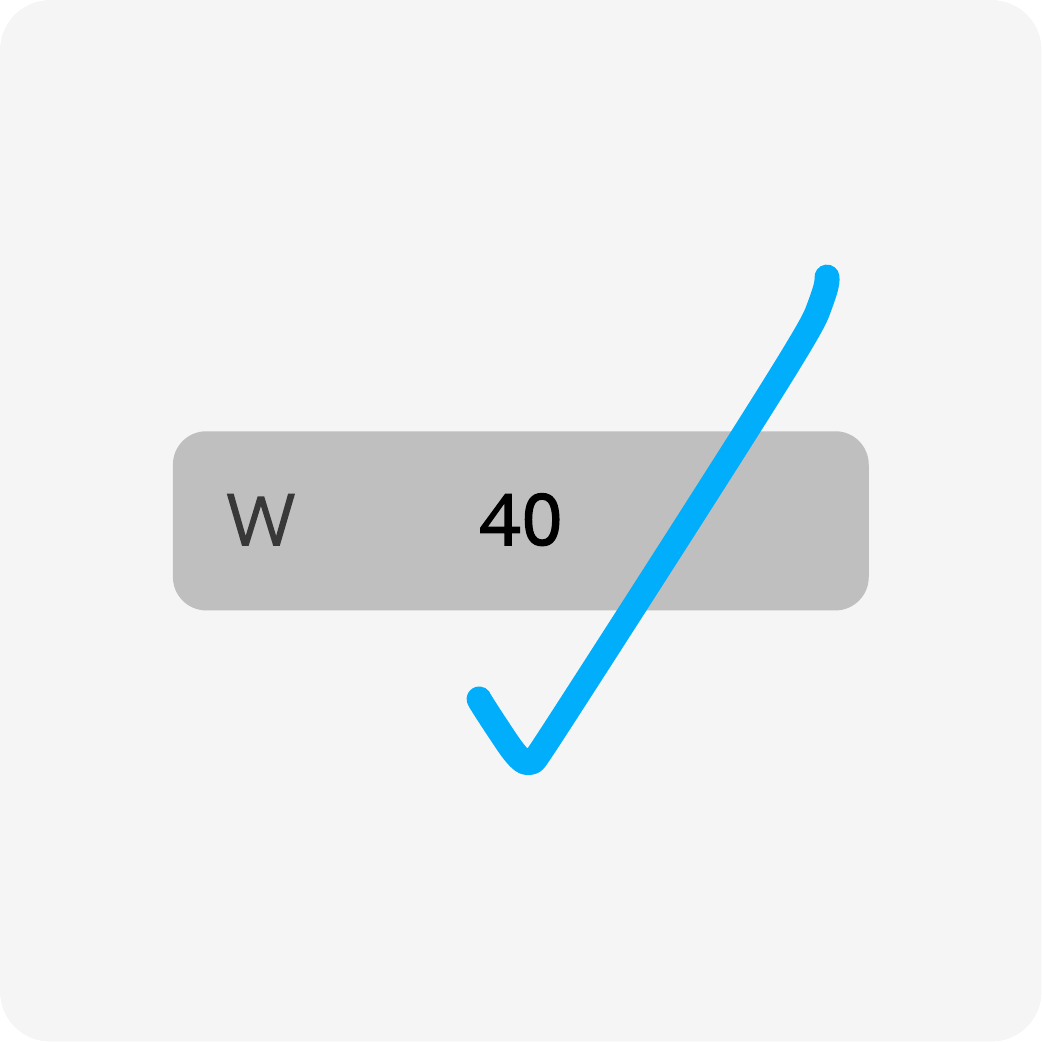} & 
            \includegraphics[width=1.8cm]{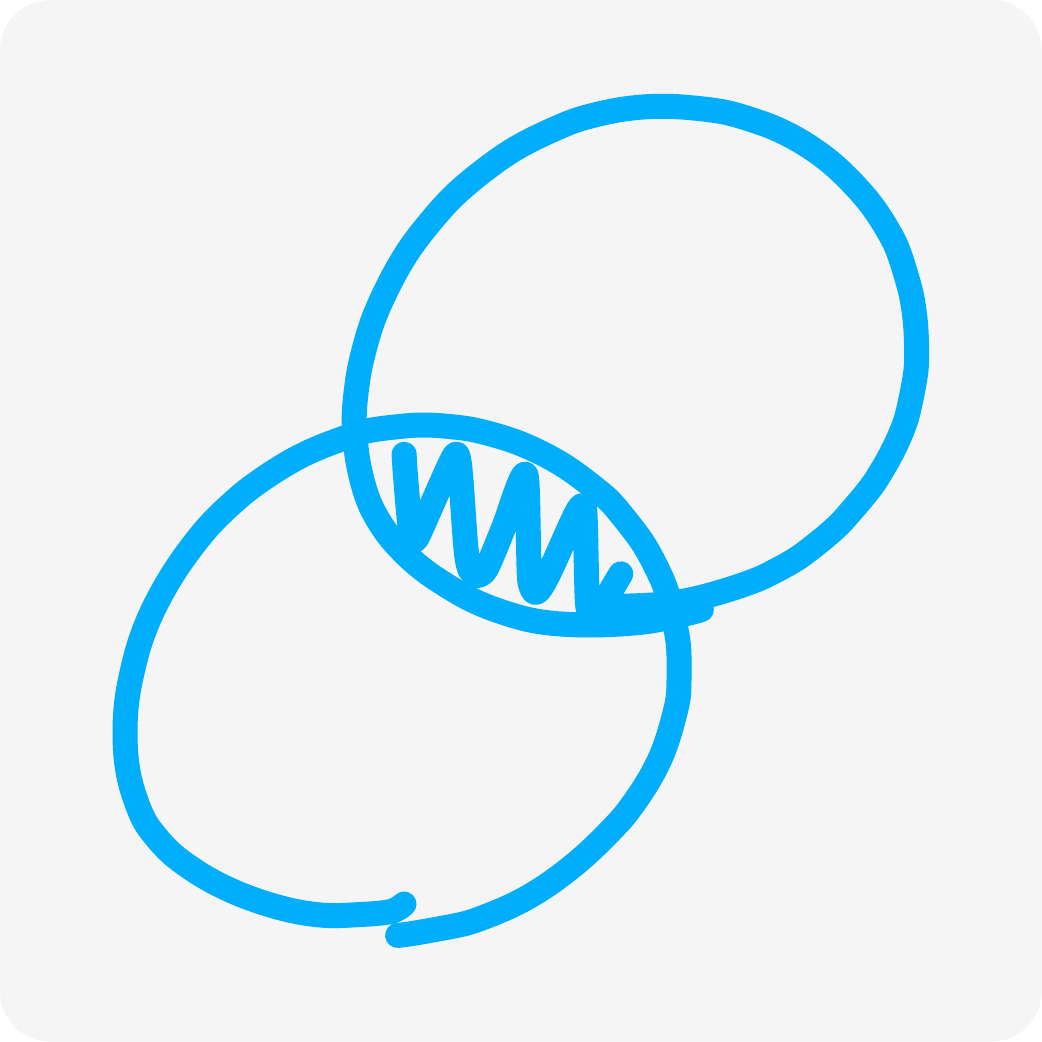} \\[0.3em]
            
            \textbf{Circles/Rectangles} & 
            \textbf{Arrow} & 
            \textbf{Line} & 
            \textbf{Bracket} & 
            \textbf{Text} & 
            \textbf{Ticks/Crosses} & 
            \textbf{Drawings} \\
            
            \footnotesize Draw attention to elements & 
            \footnotesize Direct focus to locations & 
            \footnotesize Connect or point to elements & 
            \footnotesize Group related elements & 
            \footnotesize Provide values or labels & 
            \footnotesize Confirm completion & 
            \footnotesize Freehand sketches \\
        \end{tabular}
    }
\end{figure*}

\subsubsection{Micro-guidance}

\raisebox{-0.2em}{\includegraphics[height=1em,alt={Small conversational badge representing short verbal acknowledgements or directional micro-cues like "left," "okay," or "yes, that’s right"}]{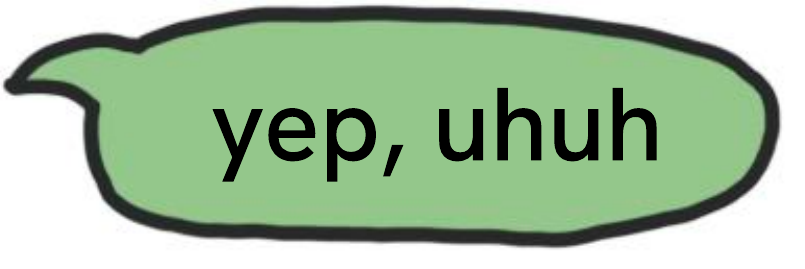}}
Teachers provided real-time feedback and micro-instructions through brief verbal cues and confirmations during user interactions. Teachers used short, directive phrases such as ``yep, uhuh, nope, left, ok'' or encouraging statements like ``that's right, keep going'' to guide users through immediate actions and provide validation. For example, one teacher (P1-T) provided encouraging micro-guidance: \textit{``That's right, keep going''} while another teacher (P2-T) used a focused directive: \textit{``yep, uhuh, nope, left, ok.''} This demonstrates how teachers used micro-guidance to provide real-time feedback and focused instructions during user interactions.

\subsubsection{Limitations}

Speech-only communication revealed several significant limitations that impacted instructional effectiveness. First, verbal instruction alone often proved insufficient for complex tasks, with students struggling to translate abstract descriptions into concrete actions. Teachers frequently found themselves unable to effectively communicate precise interface locations, leading to confusion about "where to press" or which specific elements to manipulate. One student (P3-S) expressed frustration: \textit{"I don't know where to click, I'm lost in the interface."} Second, speech suffered from information overload and memory limitations. The sequential nature of verbal instruction made it difficult for students to retain multiple steps or refer back to previous instructions. Third, speech lacked the precision needed for exact positioning and measurement tasks, requiring teachers to provide extensive verbal descriptions that often led to errors or misunderstandings. One teacher (P4-T) struggled with this limitation: \textit{"It's very hard to translate the text to application... to explain what it actually means in design."} Finally, because speech provides limited persistent visual context, students had to continuously coordinate listening with searching the interface, increasing cognitive load.

\subsection{Annotation + Speech}

\noindent
\begin{tabular}{@{}p{5em}@{\hspace{0.6em}}p{\dimexpr\linewidth-5em-0.6em\relax}@{}}%
  \begin{minipage}[t]{\linewidth}
    \vspace{0pt}\includegraphics[height=6em,width=5em,keepaspectratio,alt={Illustration of a hand drawing on a digital canvas while speech bubbles appear, representing the combined use of visual annotations with verbal instruction}]{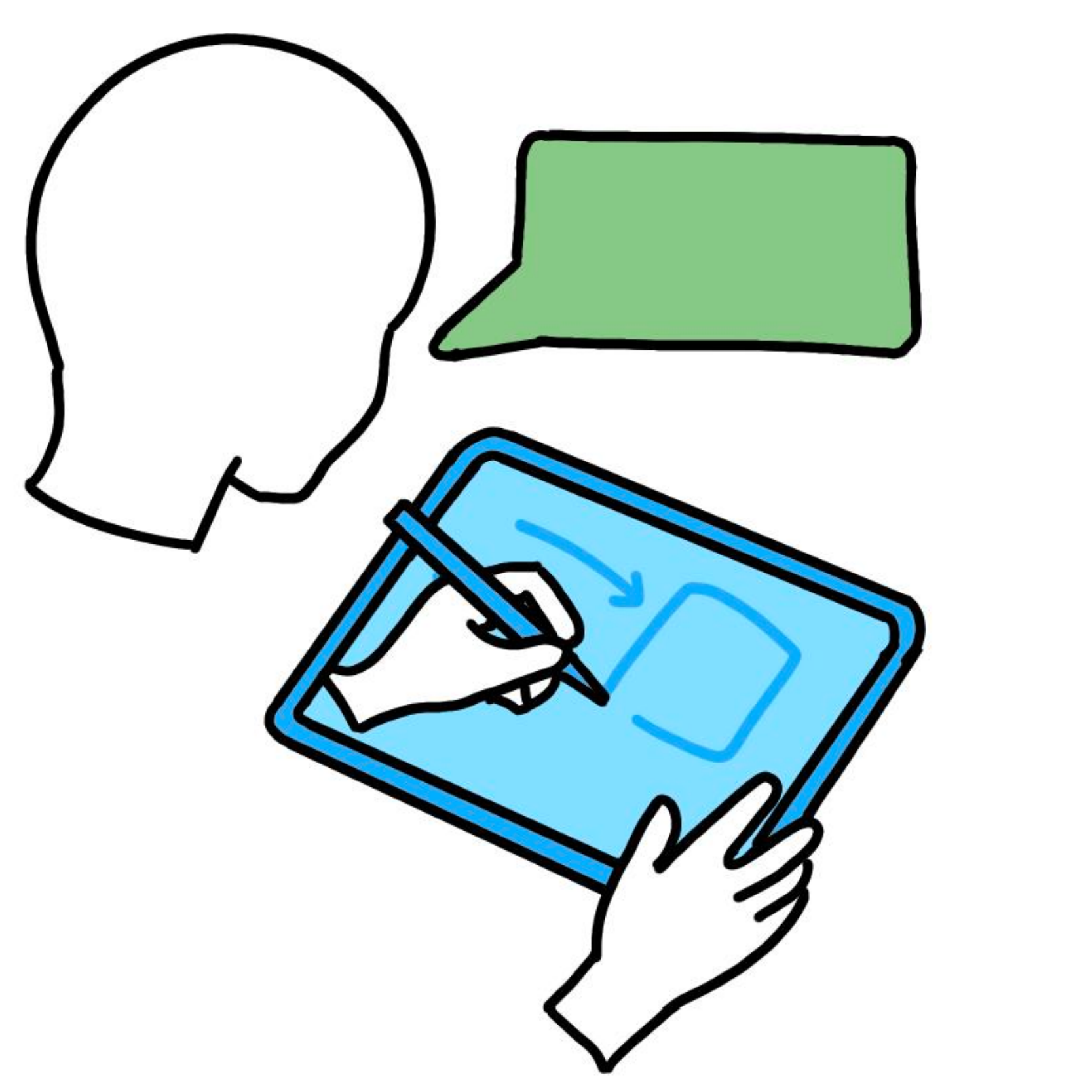}
  \end{minipage}
  &
  \begin{minipage}[t]{\linewidth}
    \vspace{0pt}Annotations emerged as a visual communication tool that effectively complemented speech for spatially sensitive instructions. Teachers frequently used annotations to address spatial precision challenges that speech alone could not adequately convey. Annotations proved particularly effective for element finding and spatial guidance, and for facilitating workflow visualization.
  \end{minipage}
\end{tabular}\par\medskip

Teachers employed various visual annotation styles (see Table~\ref{fig:annostyles}) including circles, arrows, lines, and freehand drawings to achieve different instructional purposes. The coordination between annotation and speech enabled additional information delivery, with teachers able to provide both visual and verbal context simultaneously. The instances of annotations identified can be categorized into the following categories (see Table~\ref{tab:speech-annotations}): (1) Highlight Target + Command Action, (2) Suggest Value + Command Action, (3) Draw + Explain Arrangements, and (4) Draw + Explain Concept.

\subsubsection{Highlight Target + Command}

\raisebox{-0.2em}{\includegraphics[height=1em,alt={Icon depicting an annotated highlight mark, representing when teachers visually mark an interface element while verbally instructing the student to interact with it}]{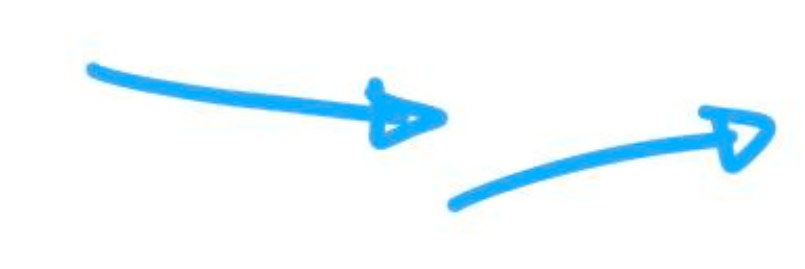}}
This multimodal interaction pattern (see Table~\ref{tab:speech-annotations}) combines visual highlighting with verbal instruction to guide user actions. This approach involves circling or highlighting specific interface elements while providing verbal commands such as ``look here and click this'' to focus user attention and direct their actions.

For example, one teacher (P1-T) circled the mask button while saying \textit{"This one is like if you take a picture it will..."} to guide students to the correct Boolean operation. Another teacher (P2-T) circled the Boolean operations button while instructing \textit{"yeah yeah okay so can you go over like this and this one yeah okay then you put union okay"} to help students select the right tool for combining shapes. Teachers frequently used this technique to highlight specific interface elements that students needed to interact with, such as circling the frame tool while saying \textit{"So you try to go down to the bar here and click on this"} (P10-T).

The data shows heavy use of this pattern across both lessons, with 85 annotations in Lesson 1 and 43 in Lesson 2, reflecting greater reliance on visual–verbal guidance during early learning. In Lesson 1, usage peaked in Step 3 (85 annotations) and Step 4 (34 annotations), which involved initial setup and complex manipulation. In Lesson 2, annotations were more distributed, with Step 4 (43 annotations) and Step 5 (37 annotations) being the most intensive. This shift suggests that teachers moved from detailed element-by-element highlighting to more streamlined multi-element cues as students became more familiar with the interface.

\begin{table*}[t!]
  \centering
  \small
  \caption{Annotation Types and Their Corresponding Speech Patterns}
  \label{tab:speech-annotations}
  \Description{Table of four annotation types and corresponding speech patterns: Highlight Target \& Command (directing focus with action instructions), Write Value \& Command (providing parameter values with instructions), Draw \& Explain Arrangement (spatial relationships and layout), Draw \& Explain Concept (visual demonstration with conceptual explanation). Each row gives the type, description, and example images from the study.}

  \setlength{\aboverulesep}{0pt}
  \setlength{\belowrulesep}{0pt}
  \setlength{\extrarowheight}{0pt}
  \begin{tabular*}{\textwidth}{|p{0.479\textwidth}|p{0.479\textwidth}|}
  \toprule

  \begin{minipage}[t]{\linewidth}
    \raggedright
    \textbf{Highlight Target \& Command}\\
    Directing student focus to interface elements with clear action instructions

    {\raggedright\noindent
      \setlength{\tabcolsep}{4pt}%
      \begin{tabular}{@{}l l l@{}}
        \begin{minipage}[t]{\widthof{\includegraphics[height=1.6cm,keepaspectratio]{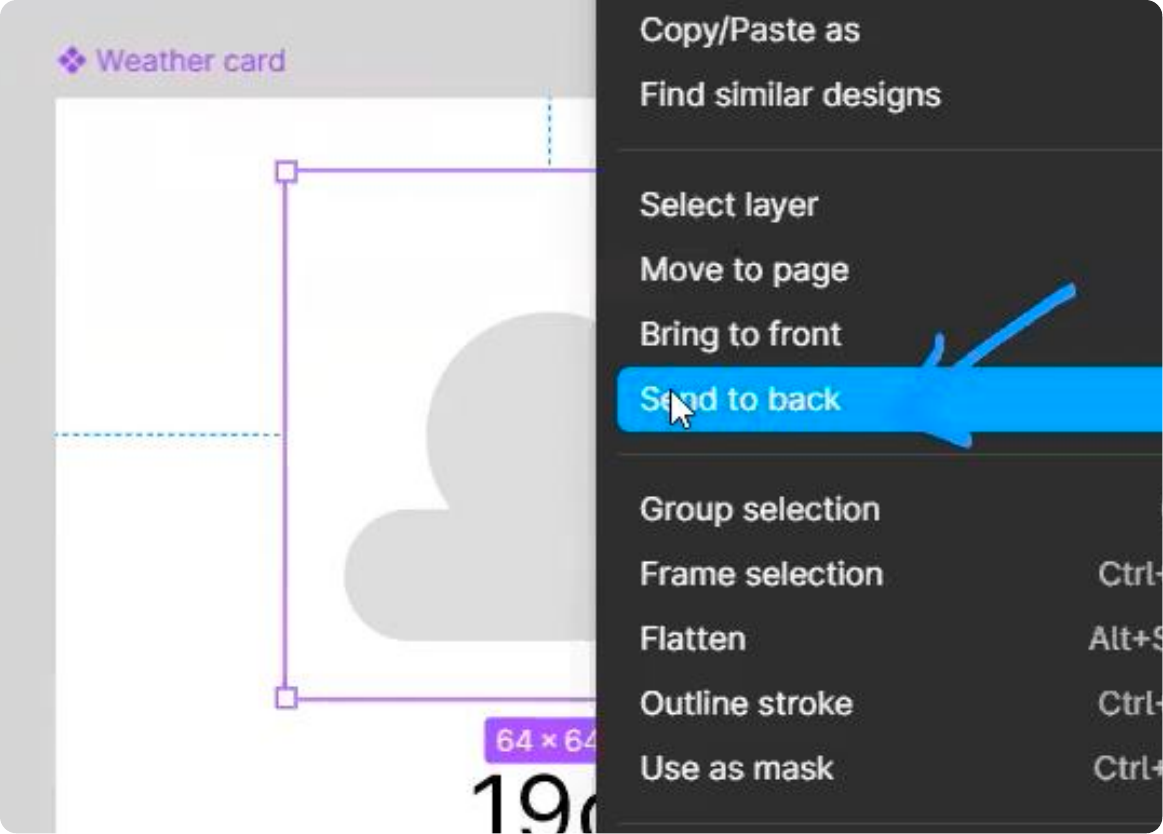}}}%
          \centering
          \includegraphics[height=1.6cm,keepaspectratio]{images/anno-examples/eg-highlight-1.pdf}\par
          \raggedright\footnotesize\itshape `try to double click it and send it back' [P6]
        \end{minipage} &
        \begin{minipage}[t]{\widthof{\includegraphics[height=1.6cm,keepaspectratio]{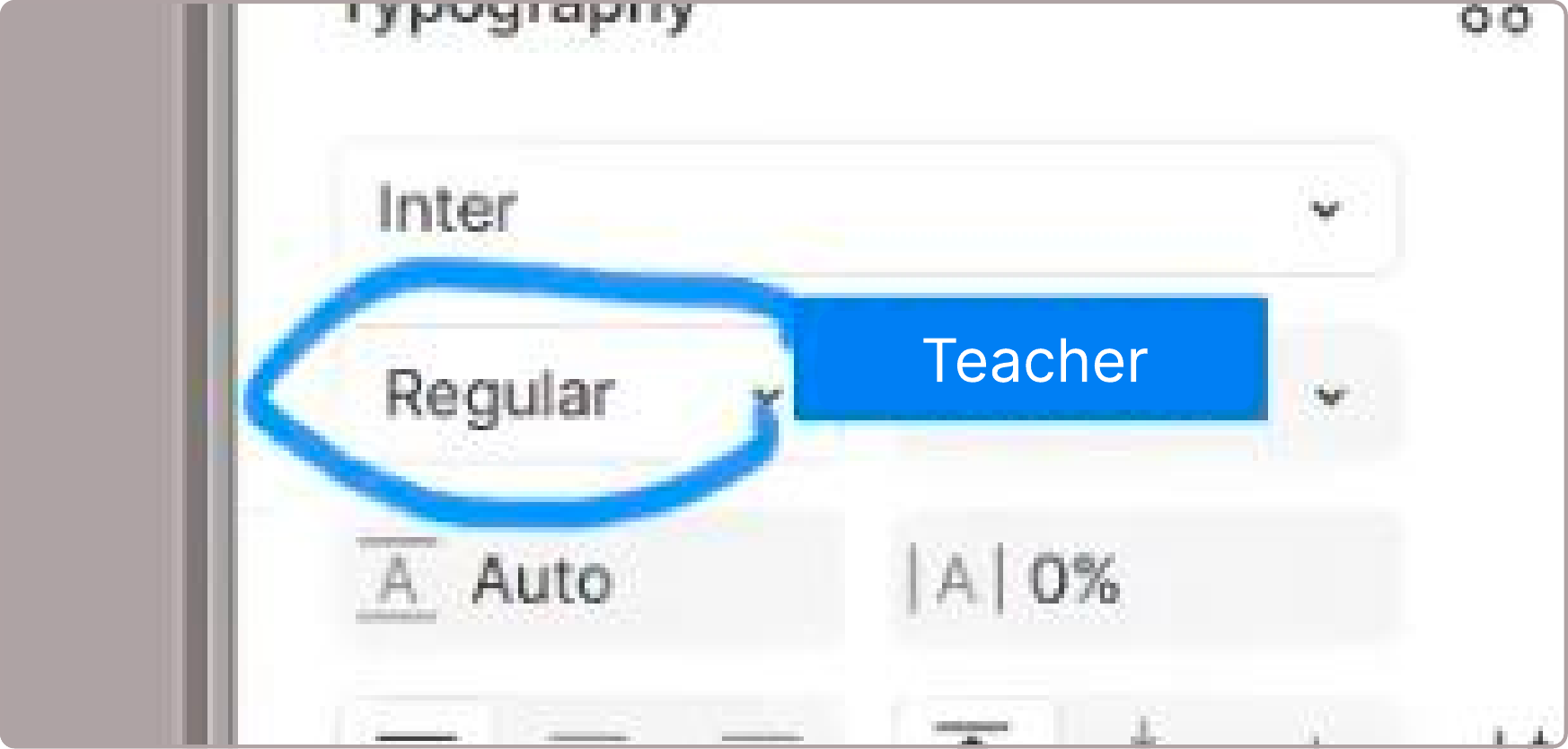}}}%
          \centering
          \includegraphics[height=1.6cm,keepaspectratio]{images/anno-examples/eg-highlight-2.pdf}\par
          \raggedright\footnotesize\itshape `And you can... this is... so all the text' [P1]
        \end{minipage} &
        \begin{minipage}[t]{\widthof{\includegraphics[height=1.6cm,keepaspectratio]{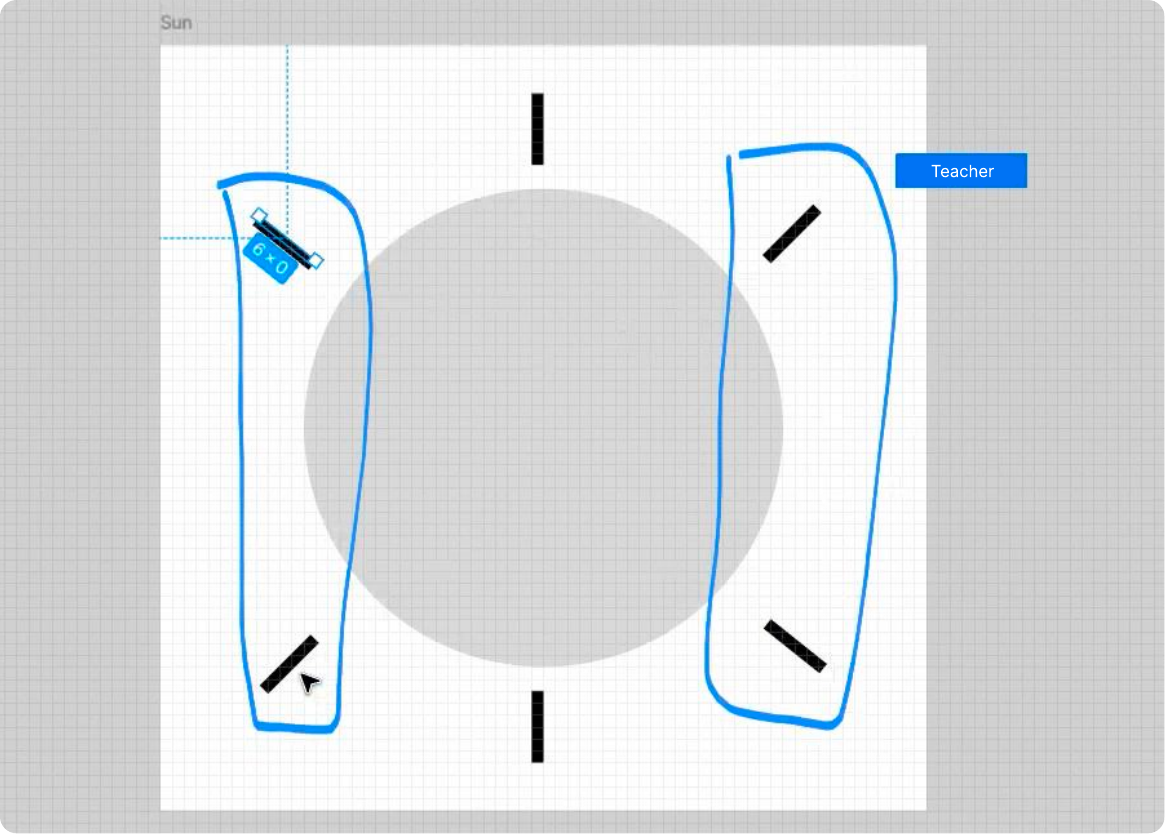}}}%
          \centering
          \includegraphics[height=1.6cm,keepaspectratio]{images/anno-examples/eg-highlight-3.pdf}\par
          \raggedright\footnotesize\itshape `can you remove these' [P6]
        \end{minipage}
      \end{tabular}%
    }
  \end{minipage}
  &
  \begin{minipage}[t]{\linewidth}
    \raggedright
    \textbf{Suggest Value \& Command}\\
    Providing specific parameter values combined with action instructions

    {\raggedright\noindent
      \setlength{\tabcolsep}{4pt}%
      \begin{tabular}{@{}l l l@{}}
        \begin{minipage}[t]{\widthof{\includegraphics[height=1.6cm,keepaspectratio]{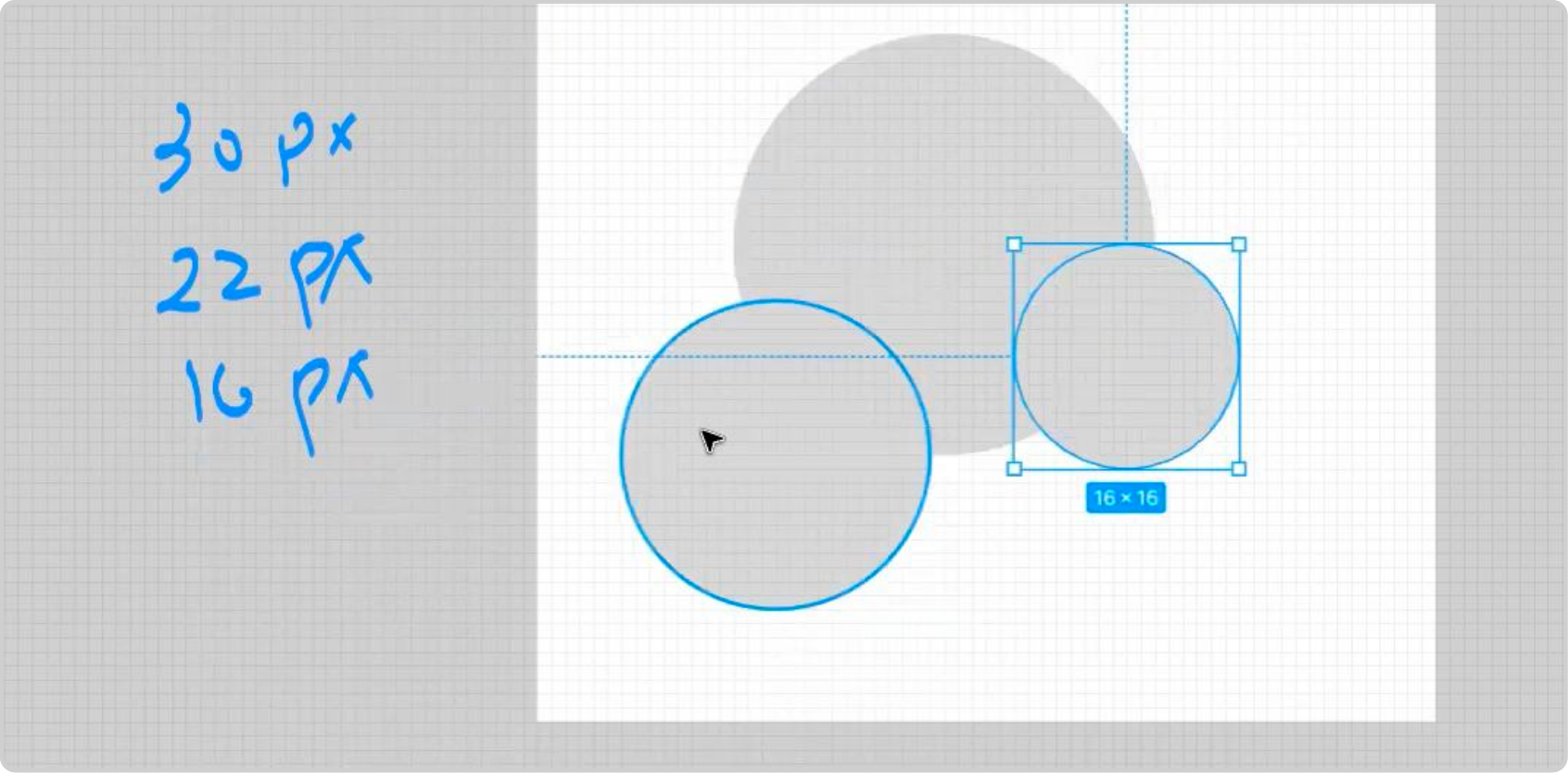}}}%
          \centering
          \includegraphics[height=1.6cm,keepaspectratio]{images/anno-examples/eg-suggest-val-1.pdf}\par
          \raggedright\footnotesize\itshape `...one that is 16 pixels. Yes! So you can align it' [P10]
        \end{minipage} &
        \begin{minipage}[t]{\widthof{\includegraphics[height=1.6cm,keepaspectratio]{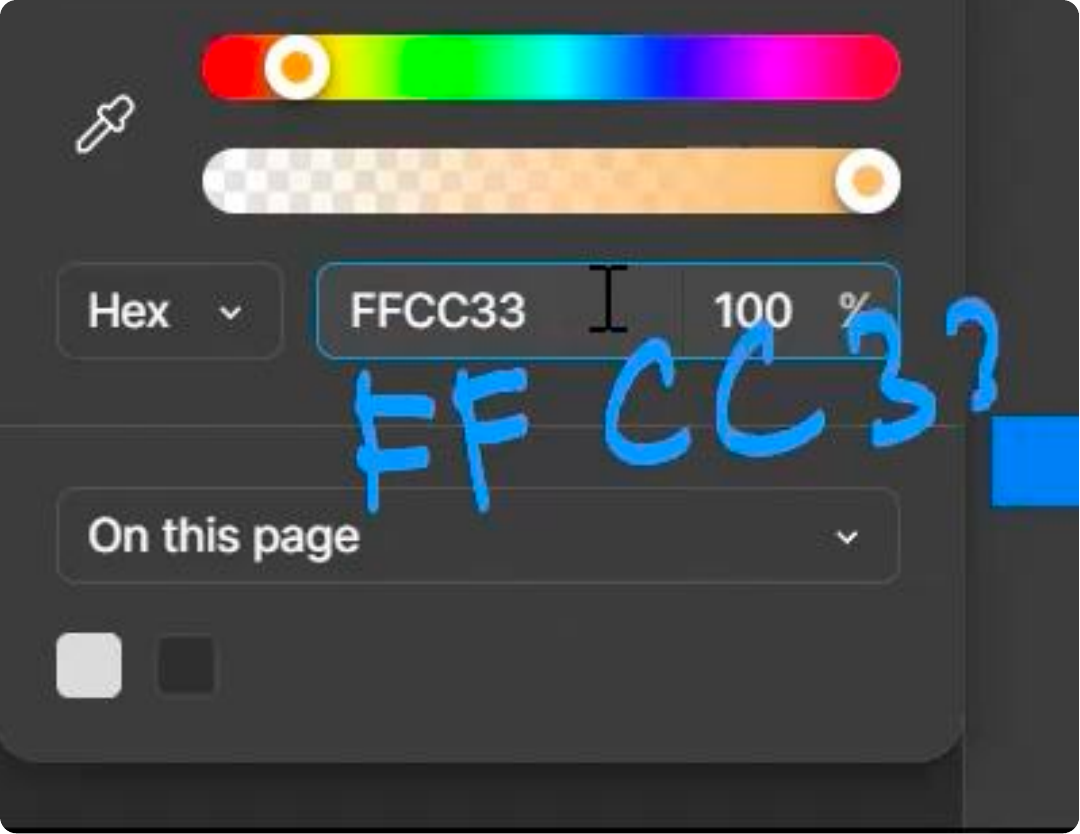}}}%
          \centering
          \includegraphics[height=1.6cm,keepaspectratio]{images/anno-examples/eg-suggest-val-2.pdf}\par
          \raggedright\footnotesize\itshape `... put in FFCC33. Oh, okay, great' [P2]
        \end{minipage} &
        \begin{minipage}[t]{\widthof{\includegraphics[height=1.6cm,keepaspectratio]{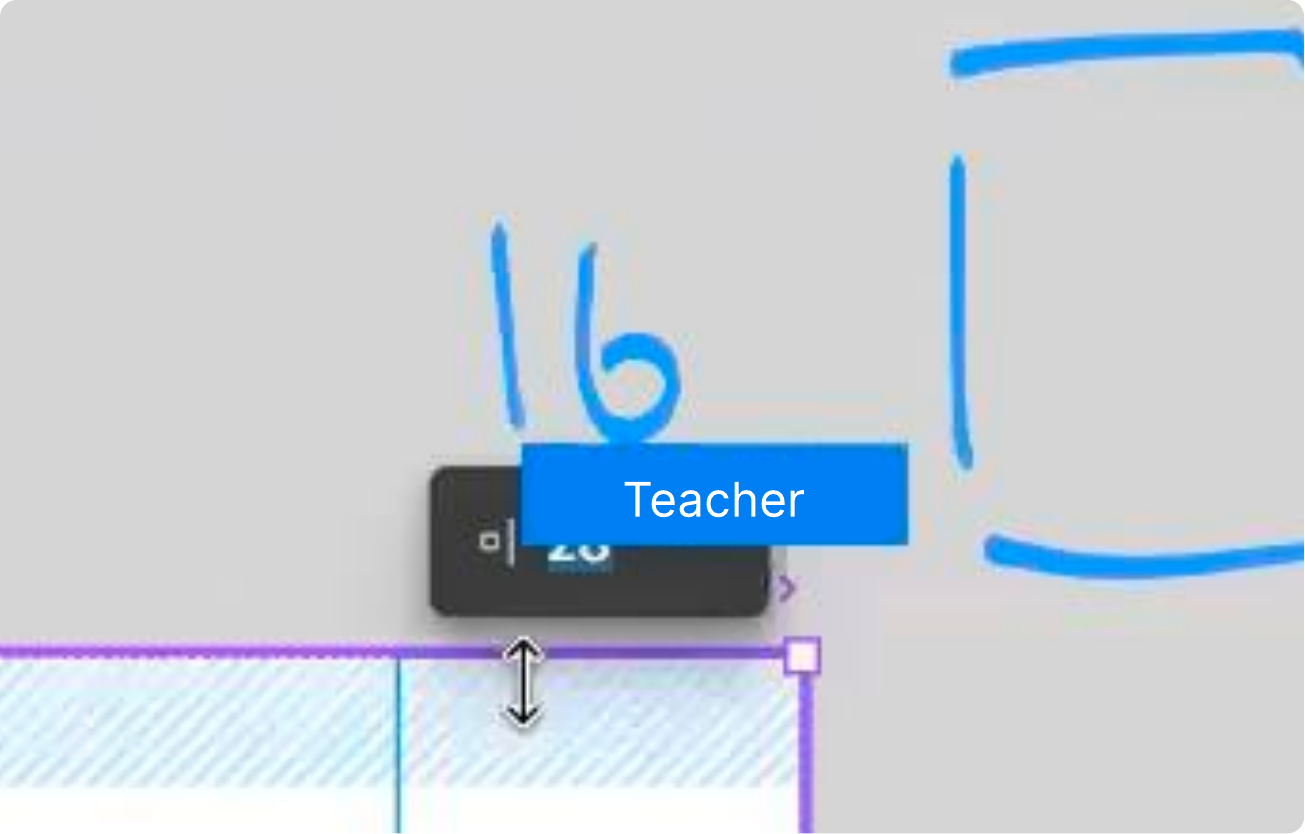}}}%
          \centering
          \includegraphics[height=1.6cm,keepaspectratio]{images/anno-examples/eg-suggest-val-3.pdf}\par
          \raggedright\footnotesize\itshape `... update this to 16 pixels and we can do this for' [P6]
        \end{minipage}
      \end{tabular}%
    }
  \end{minipage}
  \vspace{0pt}
  \\\midrule

  \begin{minipage}[t]{\linewidth}
    \raggedright
    \textbf{Draw \& Explain Arrangement}\\
    Visual demonstrations of spatial relationships and organizational structures

    {\raggedright\noindent
      \setlength{\tabcolsep}{4pt}%
      \begin{tabular}{@{}l l l@{}}
        \begin{minipage}[t]{\widthof{\includegraphics[height=1.6cm,keepaspectratio]{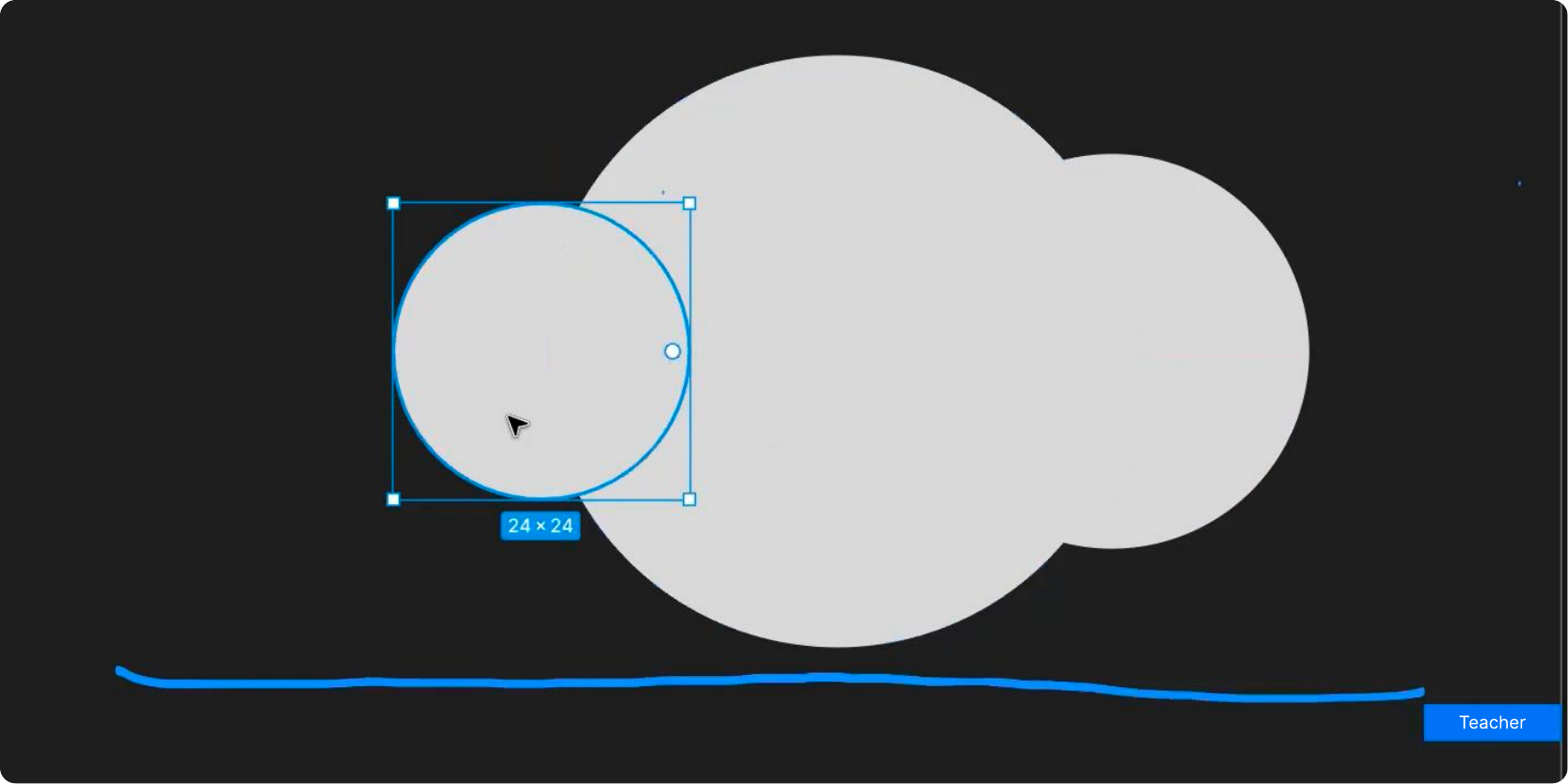}}}%
          \centering
          \includegraphics[height=1.6cm,keepaspectratio]{images/anno-examples/eg-arrange-1.pdf}\par
          \raggedright\footnotesize\itshape `as a master file and any changes you make, you see will be applied to it.' [P10]
        \end{minipage} &
        \begin{minipage}[t]{\widthof{\includegraphics[height=1.6cm,keepaspectratio]{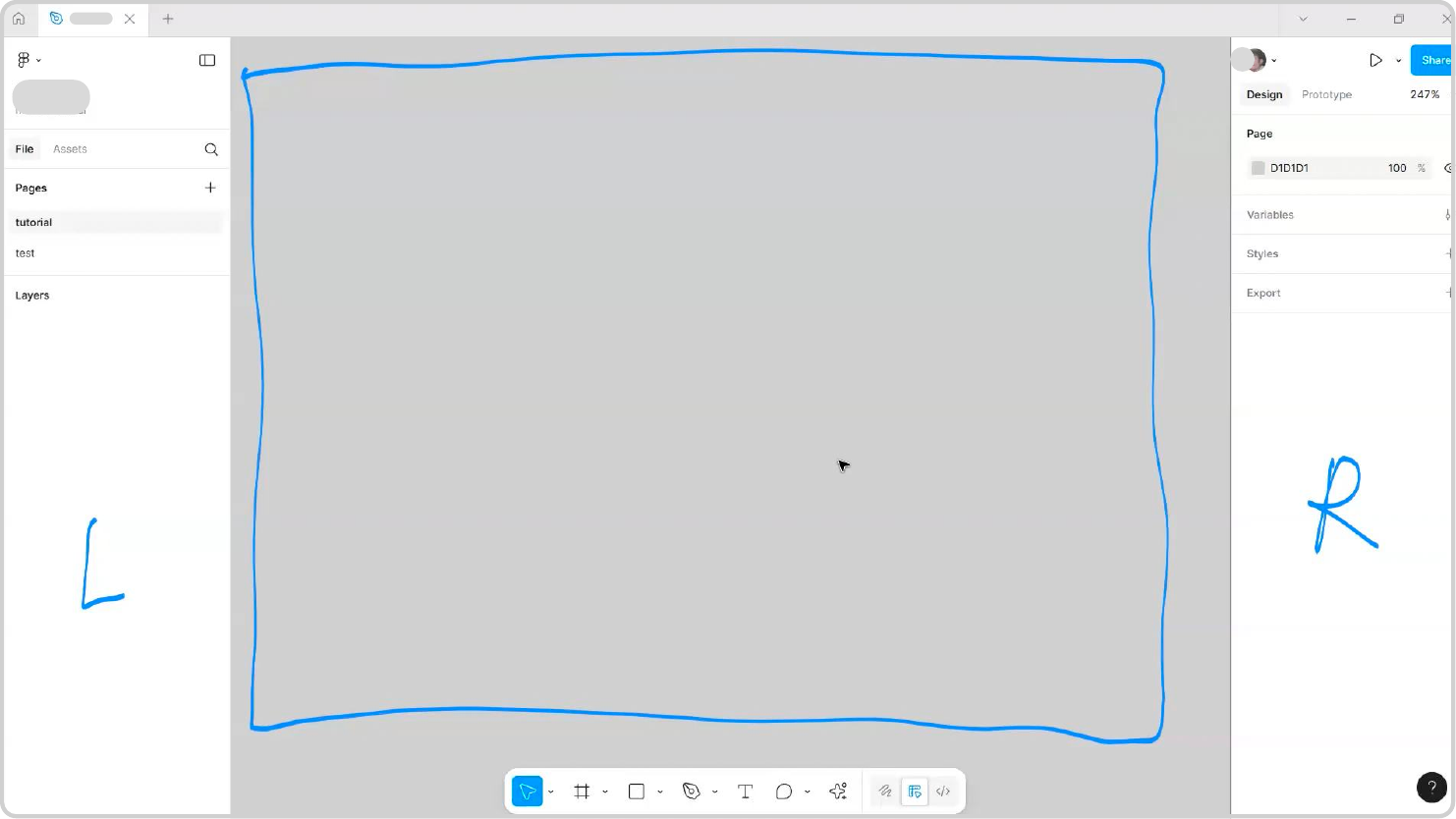}}}%
          \centering
          \includegraphics[height=1.6cm,keepaspectratio]{images/anno-examples/eg-arrange-2.pdf}\par
          \raggedright\footnotesize\itshape `... inheritance but once the instance right currently this is blue for example' [P8]
        \end{minipage} &
        \begin{minipage}[t]{\widthof{\includegraphics[height=1.6cm,keepaspectratio]{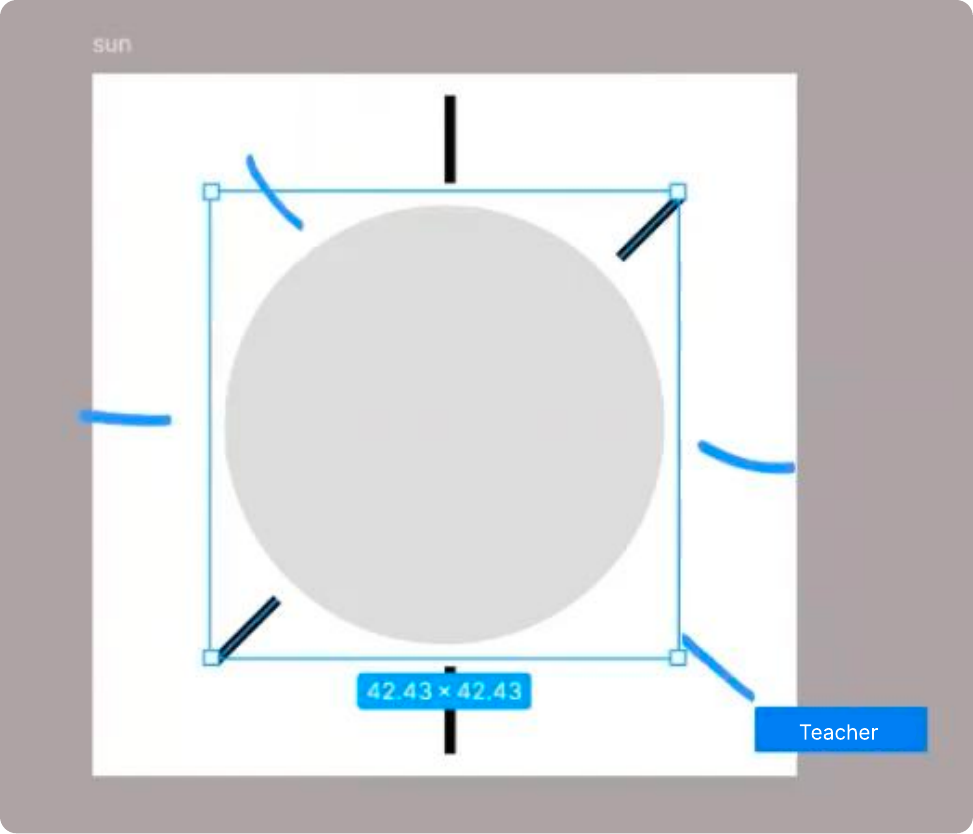}}}%
          \centering
          \includegraphics[height=1.6cm,keepaspectratio]{images/anno-examples/eg-arrange-3.pdf}\par
          \raggedright\footnotesize\itshape `the two shapes that are together' [P1]
        \end{minipage}
      \end{tabular}%
    }
  \end{minipage}
  &
  \begin{minipage}[t]{\linewidth}
    \raggedright
    \textbf{Draw \& Explain Concept}\\
    Visual demonstrations paired with conceptual explanations

    {\raggedright\noindent
      \setlength{\tabcolsep}{4pt}%
      \begin{tabular}{@{}l l l@{}}
        \begin{minipage}[t]{\widthof{\includegraphics[height=1.6cm,keepaspectratio]{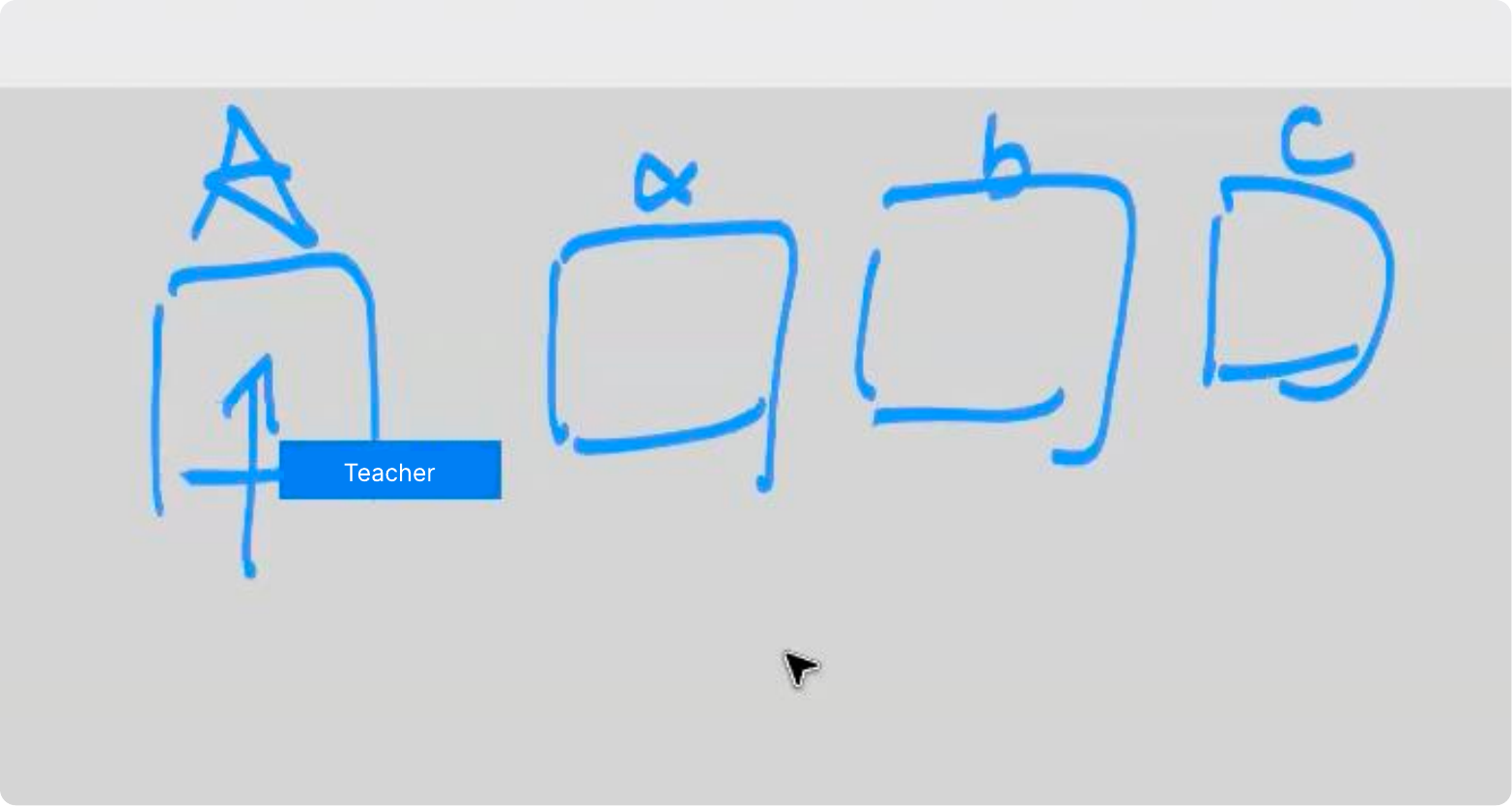}}}%
          \centering
          \includegraphics[height=1.6cm,keepaspectratio]{images/anno-examples/eg-concept-1.pdf}\par
          \raggedright\footnotesize\itshape `They are ... in the same line so ... The bottom should be in the same.' [P2]
        \end{minipage} &
        \begin{minipage}[t]{\widthof{\includegraphics[height=1.6cm,keepaspectratio]{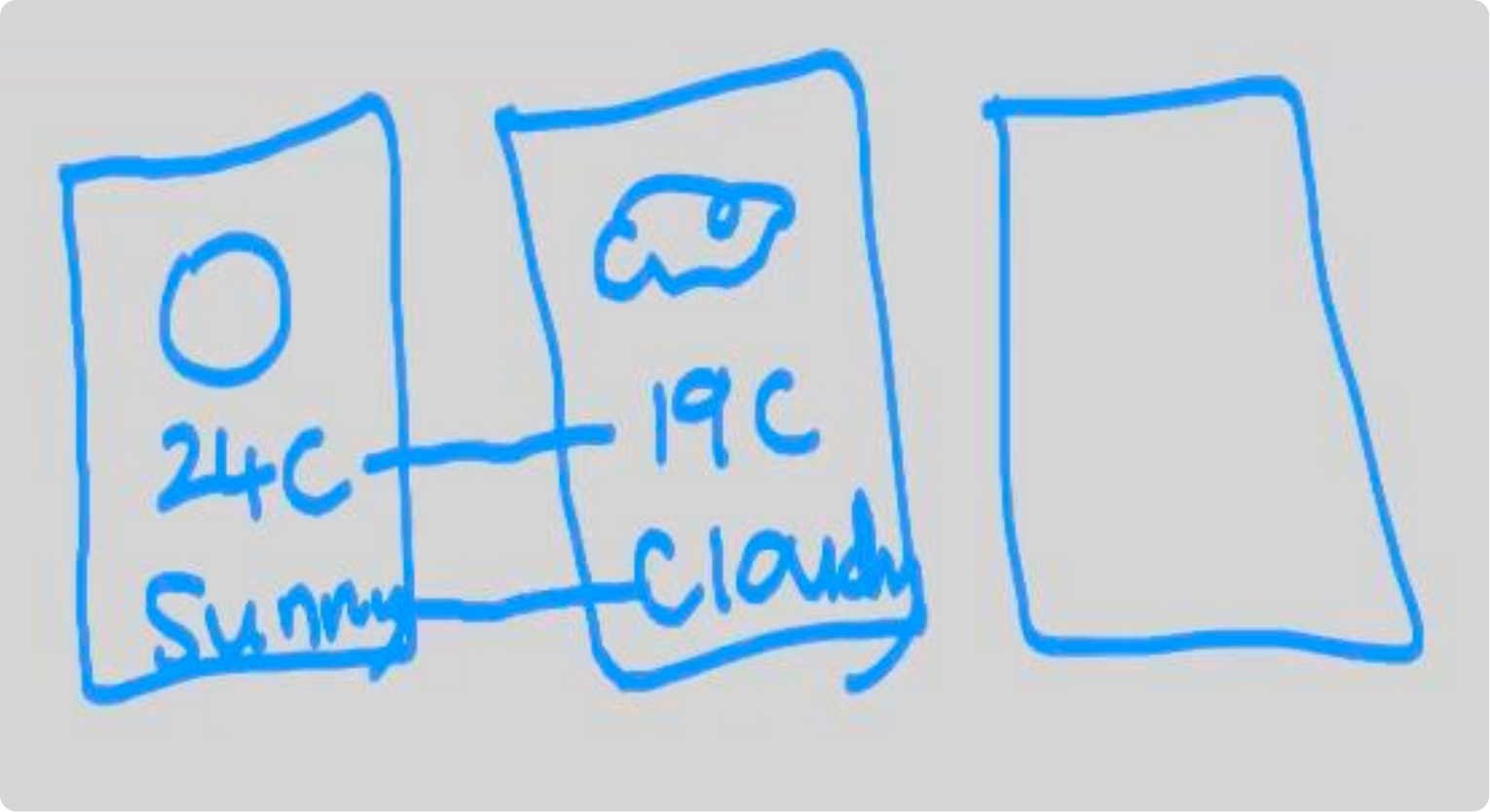}}}%
          \centering
          \includegraphics[height=1.6cm,keepaspectratio]{images/anno-examples/eg-concept-2.pdf}\par
          \raggedright\footnotesize\itshape `you probably kind of juggle between these three parts' [P6]
        \end{minipage} &
        \begin{minipage}[t]{\widthof{\includegraphics[height=1.6cm,keepaspectratio]{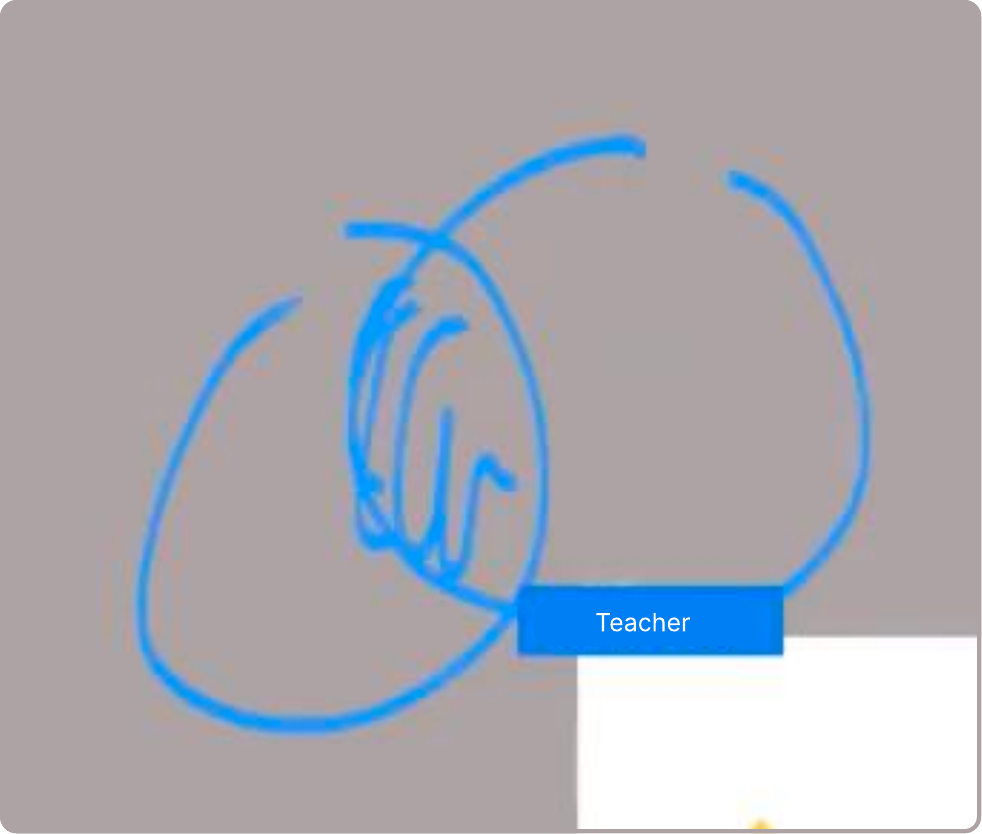}}}%
          \centering
          \includegraphics[height=1.6cm,keepaspectratio]{images/anno-examples/eg-concept-3.pdf}\par
          \raggedright\footnotesize\itshape `So duplicate it so there's one at like this' [P1]
        \end{minipage}
      \end{tabular}%
    }
  \end{minipage}
  \vspace{2pt}
  \\ \bottomrule

  \end{tabular*}
\end{table*}

\subsubsection{Suggest Value + Command}

\raisebox{-0.2em}{\includegraphics[height=1em,alt={Value-tag drawing illustrating annotation of specific numerical parameters combined with spoken commands directing how those values should be applied}]{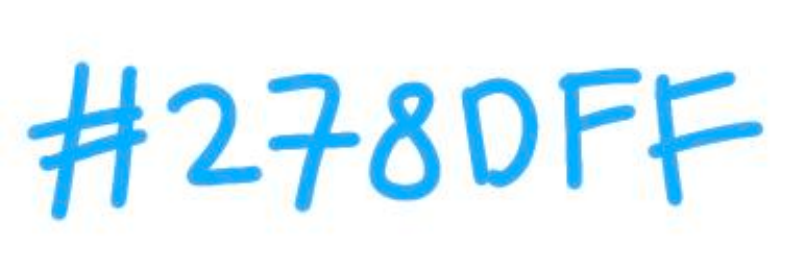}}
Teachers employed this annotation-speech approach (see Table~\ref{tab:speech-annotations}) to provide visual indication of specific values combined with verbal guidance. This technique involves writing exact input values near the corresponding input fields as a form of highlighting, such as ``look here and change to 40'' or ``set the width to 200'' with a circle around the coordinate field.

For example, one teacher (P2-T) drew two circles with "12" in each while explaining, \textit{"you draw two circles of 12 each. 12 each, okay. Yes."} to specify the exact pixel measurements needed for the crescent shape. Another teacher (P10-T) wrote "FFCC33" next to the crescent while saying \textit{"Yes, now you're filling the colorless yellow and you can perhaps reposition the color"} to provide the exact hex color code. Teachers also used this technique for positioning, such as when one teacher (P8-T) wrote "4 pixels" while explaining the spacing requirements for the cloud shape.

The data shows a total of 16 annotations across both lessons, with 13 occurring in Lesson 1 and only 3 in Lesson 2, indicating a significant decrease in usage as teachers refined their communication methods. The context of usage reveals that this approach is primarily employed for dimension specifications, coordinate settings, and numerical parameter adjustments where precision is crucial. Peak usage occurs in Lesson 1, Step 3 (5 annotations) and Step 4 (3 annotations), typically during tasks requiring exact measurements or positioning. The benefits of this approach include improved precision through the elimination of ambiguity, reduced effort in searching for input fields, faster execution through direct visual reference, and error reduction by preventing the misplacement of values.

\subsubsection{Draw + Explain Arrangements}

\raisebox{-0.2em}{\includegraphics[height=1em,alt={Icon showing sketched layout elements, representing teacher-drawn diagrams used to explain spatial relationships or structural arrangements in the interface}]{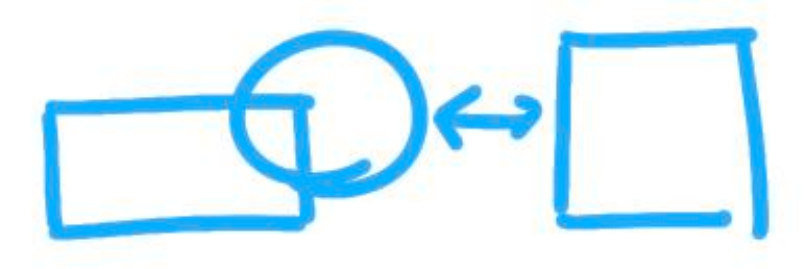}}
This multimodal interaction pattern (see Table~\ref{tab:speech-annotations}) combines visual sketching with verbal explanation to demonstrate spatial relationships and visual organization. This approach involves drawing examples, layouts, and arrangements on the canvas while providing verbal context about spatial positioning and structural relationships.

For example, one teacher (P10-T) drew a complete sun icon on the canvas while explaining \textit{"So that would be our reference. So to do this, you draw this line and this line"} to show students the final visual structure they were creating. Another teacher (P1-T) demonstrated ray duplication by sketching 4 additional rays around a circle while saying \textit{"So duplicate it so there's one at like this"} to illustrate the spatial arrangement needed for the sun icon. Teachers also used this technique to show complex layouts, such as one teacher (P2-T) who sketched a circle with 6 rays while explaining \textit{"Also the item that we are trying to create is like a sign. Okay. Yeah, so they will have six ways, right?"} to help students visualize the cloud icon structure.

The data reveals usage patterns that show this technique was most frequently employed during initial setup phases and complex visual tasks. In Lesson 1, peak usage occurred during Step 3 (ray creation) and Step 4 (cloud icon assembly), where teachers needed to demonstrate spatial relationships between multiple elements. The technique decreased in Lesson 2, suggesting that teachers adapted their communication methods as students became more familiar with spatial concepts.

\subsubsection{Draw + Explain Concepts}

\raisebox{-0.2em}{\includegraphics[height=1em,alt={Sketch-style conceptual diagram icon indicating visual explanations used to teach abstract or structural concepts, such as parent–child component relationships}]{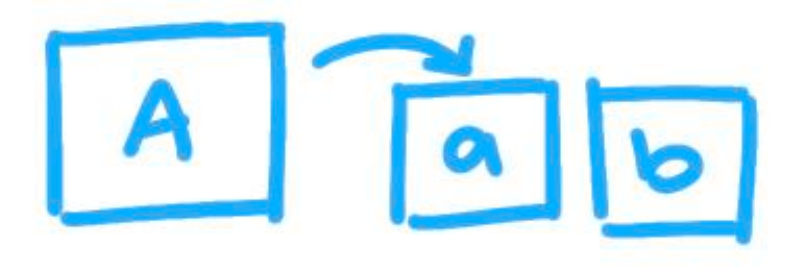}}
Teachers used this annotation-speech approach (see Table~\ref{tab:speech-annotations}) to combine visual demonstration with conceptual explanation when addressing complex software concepts. Teachers used this technique to sketch out examples and processes that explain abstract concepts like component relationships, parent-child hierarchies, and auto-layout principles.

For instance, one teacher (P1-T) drew a Venn diagram to explain the intersect operation, saying \textit{"the two shapes that are together,"} while sketching overlapping circles to demonstrate how Boolean operations work. Another teacher (P2-T) drew two circles with "12" in each to explain the concept of pixel measurements, saying \textit{"you draw two circles of 12 each. 12 each, okay. Yes."} This visual representation helped students understand both the spatial relationship and the numerical precision required.

The parent-child component relationship was particularly challenging to explain conceptually. One teacher (P10-T) drew a diagram showing the master-instance relationship while explaining, \textit{"as a master file, any modification that you make to the master file that you see will be applied to it."} (as in \ref{fig:annostyles}) Another teacher (P6-T) used annotations to distinguish between master and instance components, writing "main" above the weather card instance and "secondary" above the main component, then explaining, \textit{"one might be the master; one might be the second instance."}

This approach bridges abstract concepts with concrete examples through visual representation. Visual annotations provide tangible representations of complex relationships while eliminating the need for users to mentally map verbal descriptions to interface elements. Usage patterns show that this technique peaks during Lesson 1 Step 7 (6 annotations) and Lesson 2 Step 8 (12 annotations), typically during phases where users need to understand complex relationships of the component set and the auto-layout feature. Notably, teachers used 5 annotations during Pre-Lesson Introductions for Lesson 2, drawing out the Weather Cards at the start of the lesson to show students the end product; this pattern emerged after teachers received feedback during semi-structured interviews following Lesson 1.

\subsubsection{Limitations}

Despite the effectiveness of annotation-speech combinations, several limitations emerged that impacted their utility. First, annotation-only approaches were often insufficient for complex tasks, requiring teachers to \textit{"take control"} to provide adequate guidance when visual highlighting alone could not convey the necessary procedural steps (P5). Second, annotations could sometimes create visual clutter, especially when multiple elements needed highlighting or when previous annotations remained visible and confused current instructions. Finally, the effectiveness of annotations was highly dependent on the teacher's drawing skills and ability to create clear, recognizable visual cues, with poorly executed annotations potentially adding confusion rather than clarity.

\subsection{Screen Control + Speech}

\noindent
\begin{tabular}{@{}p{5em}@{\hspace{0.6em}}p{\dimexpr\linewidth-5em-0.6em\relax}@{}}%
  \begin{minipage}[t]{\linewidth}
    \vspace{0pt}\includegraphics[height=6em,width=5em,keepaspectratio,alt={Illustration of a hand controlling a digital interface combined with speech bubbles, representing remote-control demonstrations paired with verbal explanation}]{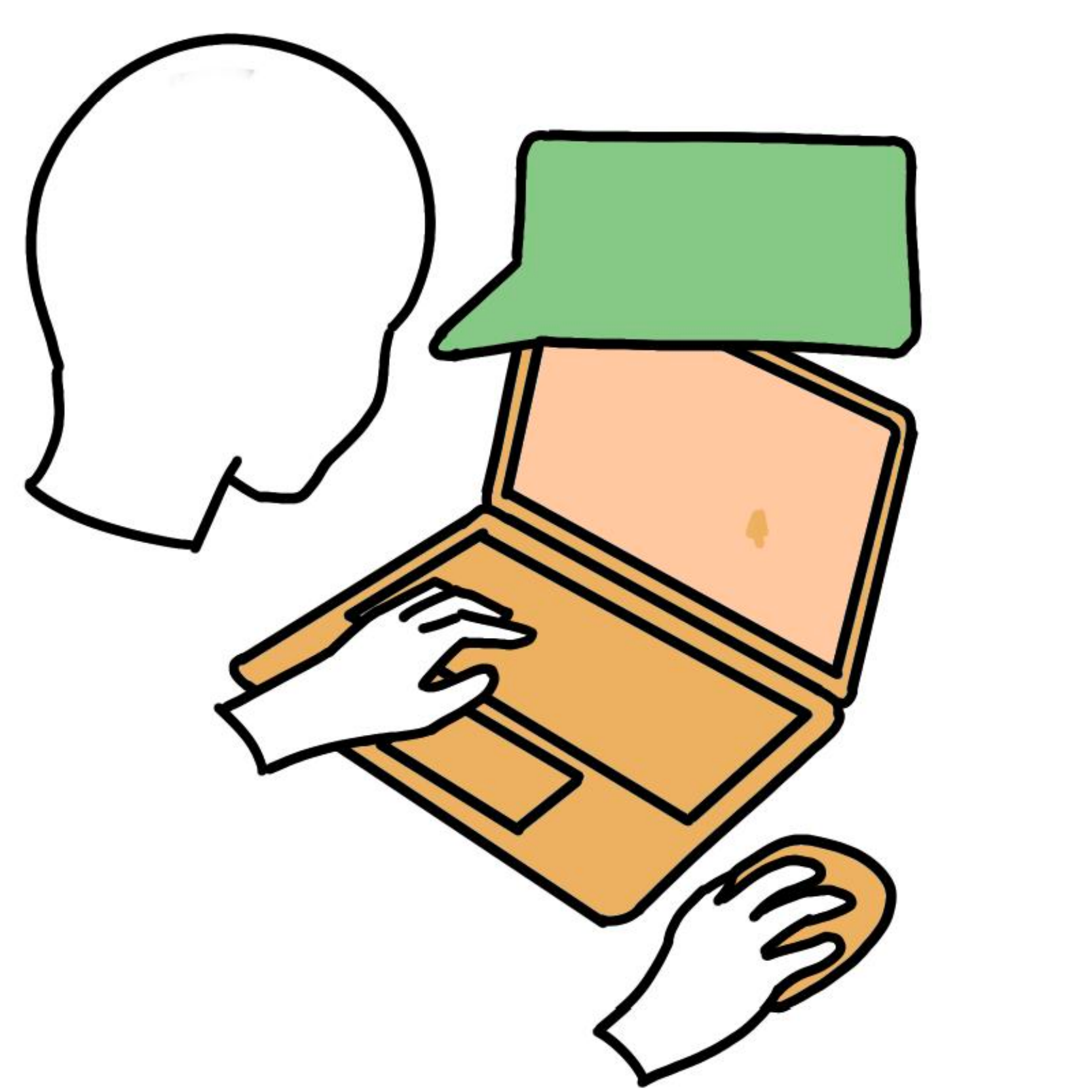}
  \end{minipage}
  &
  \begin{minipage}[t]{\linewidth}
    \vspace{0pt}Screen control functionality proved most effective when used strategically to complement speech for complex tasks. It was particularly valuable for demonstrating precise sequences of interface interactions that speech alone could not adequately convey, helping students \textit{"figure out where to press"} when they were lost.
  \end{minipage}
\end{tabular}\par\medskip

Screen control excelled at showing with both the spatial precision of where to click and the temporal precision of when to perform actions, addressing the dual challenges that teachers struggled with in verbal instruction. Our analysis reveals two primary usage patterns: demonstration of procedures and tool use, and rectification of student work. Screen control was most effective when used sparingly and purposefully, with teachers noting it was \textit{"unnecessary when student can self-correct."}

\subsubsection{Demonstration of Procedures and Tool Use}

\raisebox{-0.2em}{\includegraphics[height=1em,alt={Icon showing a pointing gesture, representing teacher-controlled demonstrations that show step-by-step tool use within the interface}]{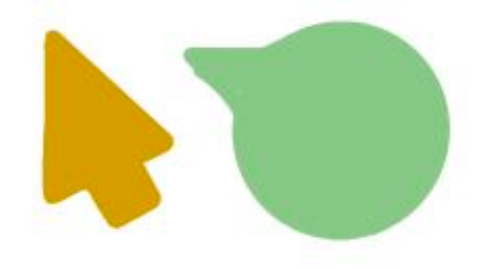}}
Procedure and tool use demonstration occurred when teachers physically manipulated interface elements while explaining their actions, such as \textit{``I will be adjusting the <shape> because...''} This method combines direct manipulation with contextual verbal guidance. We observed 19 demonstrations in Lesson 1 and 7 in Lesson 2, indicating a heavier reliance on hands-on guidance early in learning. Demonstrations appeared mainly in complex steps: Steps 1, 3, and 4 of Lesson 1, and Steps 1, 2, and 8 of Lesson 2. The highest frequencies occurred in Step 3 of Lesson 1 (11 demonstrations) and Step 8 of Lesson 2 (5 demonstrations), reflecting moments where verbal instruction alone would be insufficient for students to proceed.

\begin{figure*}[t!]
  \centering
  \caption{Teacher demonstrating radial pattern creation through hands-on interface manipulation: Shows step-by-step duplication (Ctrl+C/Ctrl+V) and constrained rotation (Shift+15°) of rectangular objects around a central circle, with accompanying verbal guidance explaining each action.}
  \label{fig:side-by-side-comparison}
  \Description{A three-panel sequence illustrating a software instruction process for duplicating and rotating an object around a central point. Left: initial setup with one rectangular object above a circle. Middle: a duplicated and rotated copy with Shift-based constrained rotation. Right: multiple evenly spaced rotated duplicates around the circle.}

  \begin{subfigure}[t]{0.32\textwidth}
    \centering
    \includegraphics[width=\linewidth]{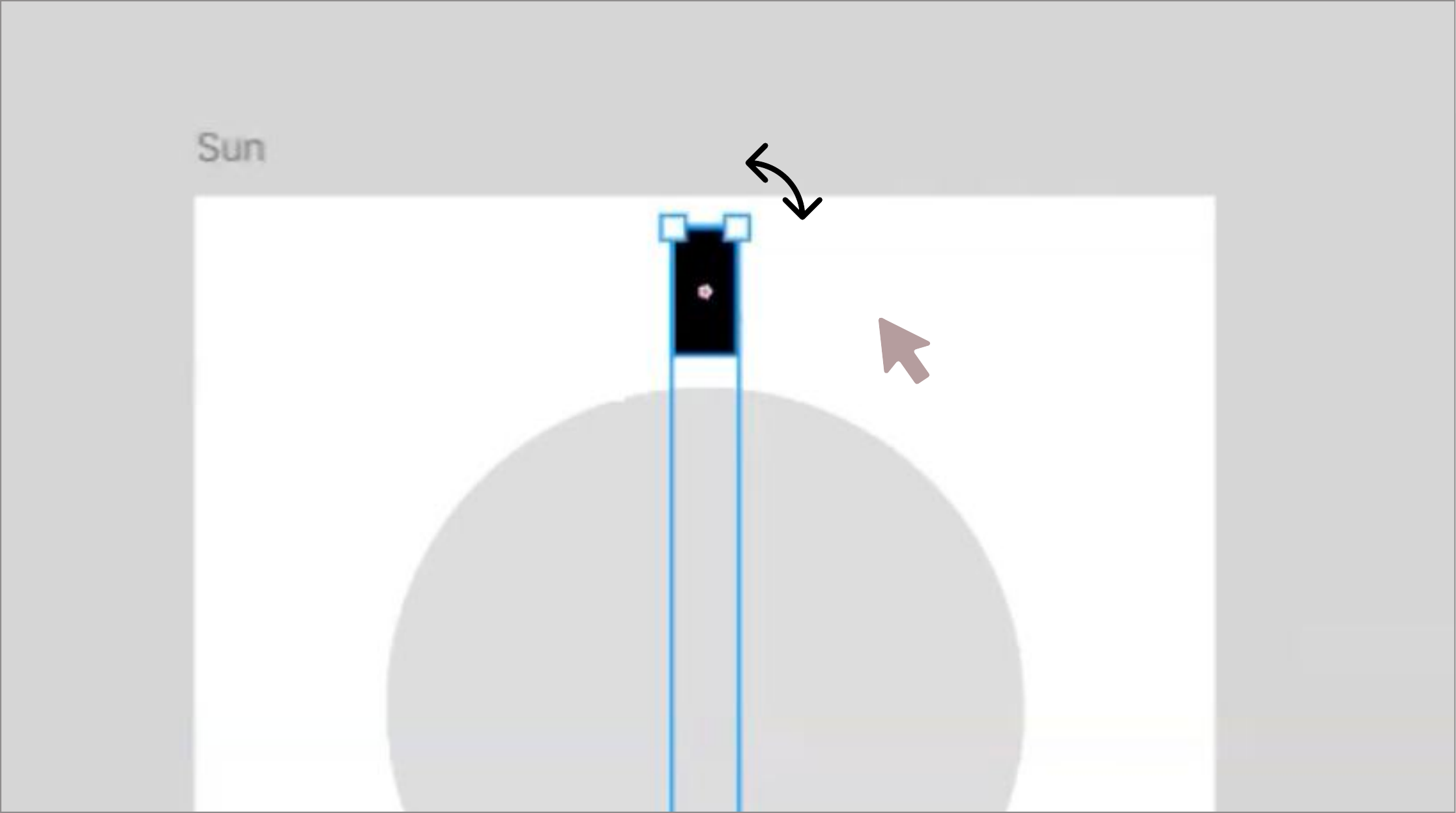}
    \subcaption{Teacher: ``So, once you are over this, you will see something like that. So, before that, we are going to create a new one.''}
    \label{fig:side-by-side-comparison-a}
  \end{subfigure}
  \hfill
  \begin{subfigure}[t]{0.32\textwidth}
    \centering
    \includegraphics[width=\linewidth]{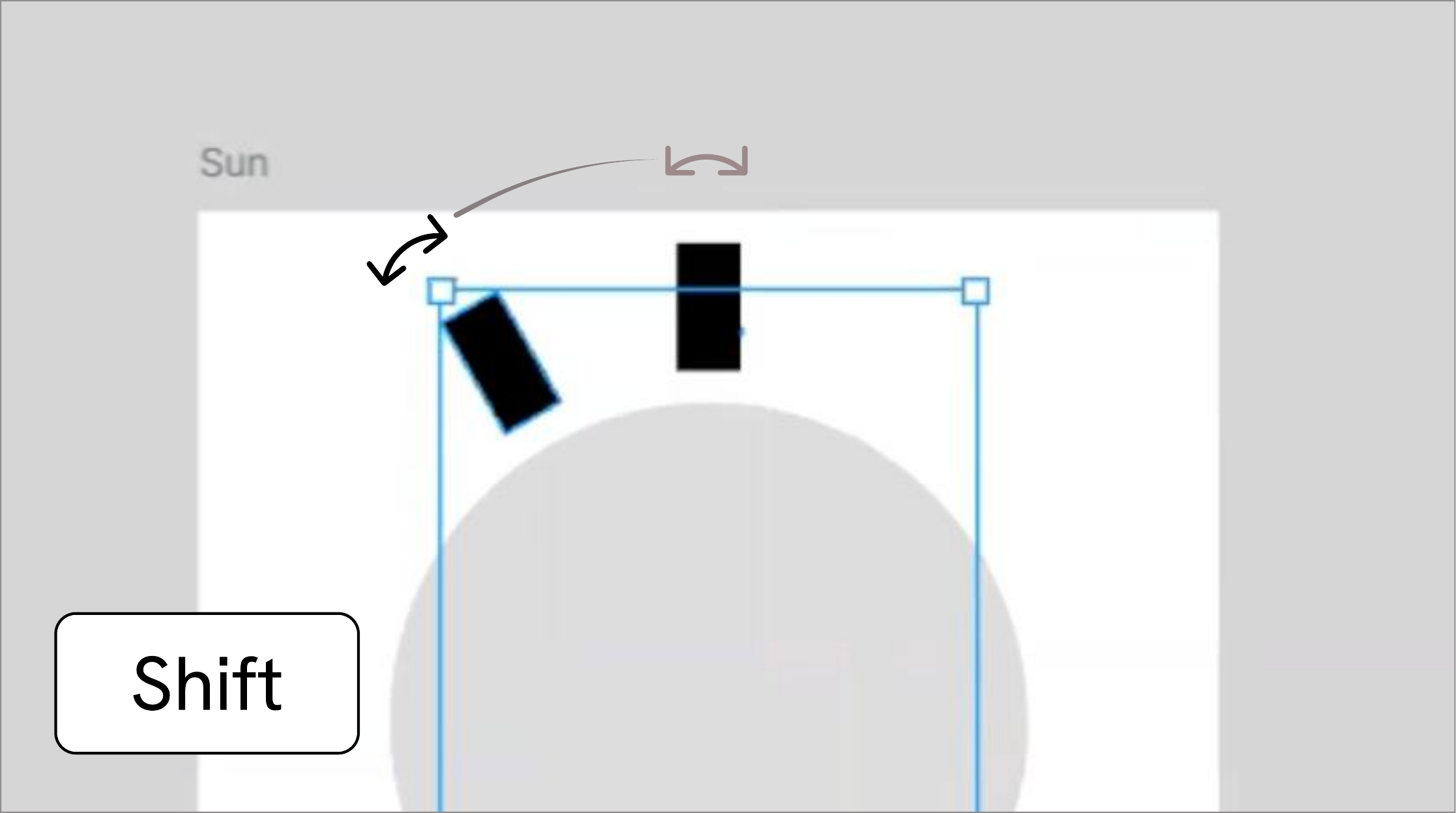}
    \subcaption{Teacher: ``Ctrl+C, Ctrl+V... And then we are going to rotate it. So, once... While you are clicking on the `Shift', it will rotate by 15 degrees.''}
    \label{fig:side-by-side-comparison-b}
  \end{subfigure}
  \hfill
  \begin{subfigure}[t]{0.32\textwidth}
    \centering
    \includegraphics[width=\linewidth]{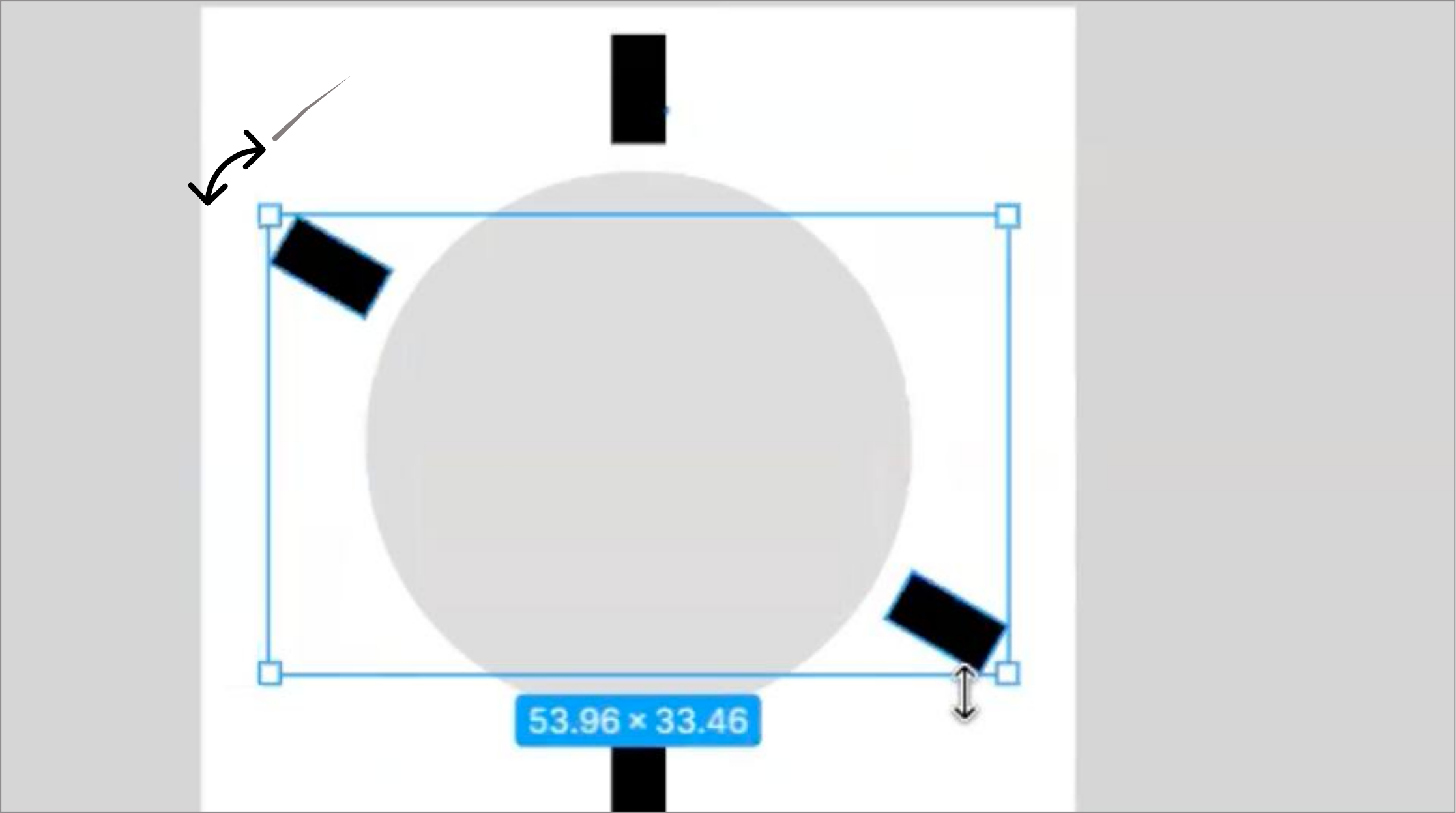}
    \subcaption{Teacher: ``So, one move is just like 15 degrees. So, can you create another duplicate one and...''}
    \label{fig:side-by-side-comparison-c}
  \end{subfigure}
\end{figure*}

\subsubsection{Rectification}

\raisebox{-0.2em}{\includegraphics[height=1em,alt={Icon with correction symbols indicating teacher intervention to fix errors or restore the interface after incorrect student actions}]{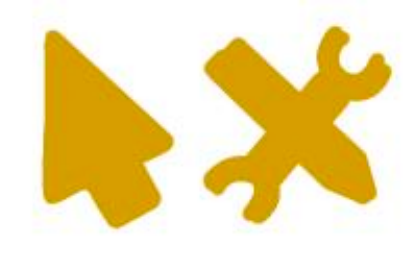}}
Problem resolution through screen control occurred when teachers needed to correct interface states or when speech-based guidance was no longer effective. These interventions typically arose when verbal instructions could not communicate the required actions or when immediate correction was necessary. Frequency data shows clear shifts: Lesson 1 had 19 demonstrations and 0 rectifications, while Lesson 2 had 1 demonstration and 6 rectifications. Screen control totaled 608 seconds in Lesson 1 and 213 seconds in Lesson 2, indicating heavier reliance on direct control during early learning when students were less familiar with the interface.

Specific examples of rectification included teachers taking control to fix incorrect layer arrangements, such as when one student P3-S accidentally created overlapping circles instead of using Boolean operations for the cloud icon. P1-T intervened: \textit{"Let me fix this for you"} and quickly corrected the layer structure while explaining \textit{"we need to subtract these circles to create the cloud shape."} Another common rectification scenario occurred when students struggled with precise positioning, such as when a student (P7-S) was unable to align sun rays properly around the central circle. The teacher (P4-T) took control and demonstrated: \textit{"I'll show you how to use the rotation tool to get these rays evenly spaced."}

Teachers held mixed views about rectification. Some saw it as necessary, with P6-T noting: \textit{"Sometimes you just have to step in and fix it, especially when they're getting frustrated."} Others preferred restraint, as P9-T noted: \textit{"I try to let them figure it out first, but if they're really stuck, I'll take control briefly to show them the right way."} Students generally appreciated the help when they were struggling, with P2-S reflecting: \textit{“It helped when they fixed it, but I wish they explained why it was wrong.”}

\subsubsection{Limitations}

Screen control, while effective for specific intervention scenarios, carried significant limitations that impacted its overall utility. First, students experienced \textit{"shock/anxiety"} when teachers took unexpected control, creating negative emotional responses that could hinder learning (P4). This sudden loss of agency often disrupted the natural flow of instruction and made students feel disempowered. One student (P6-S) described the experience: \textit{"It was jarring when the teacher suddenly took control without warning."} Second, teachers expressed strong concerns about "over-control" that could undermine student autonomy and learning opportunities. One teacher (P2-T) explicitly avoided screen control, stating \textit{"I try not to use this unless... because I feel like if I control too much, the student don't get the experience themselves."}

\subsection{Adaptiveness of Teachers}

\noindent
\begin{tabular}{@{}p{4em}@{\hspace{0.6em}}p{\dimexpr\linewidth-4em-0.6em\relax}@{}}%
  \begin{minipage}[t]{\linewidth}
    \vspace{0pt}\includegraphics[height=4.2em,width=4em,keepaspectratio,alt={Illustration of a head with arrows indicating adjustment, representing teachers dynamically adapting modalities and strategies based on student needs}]{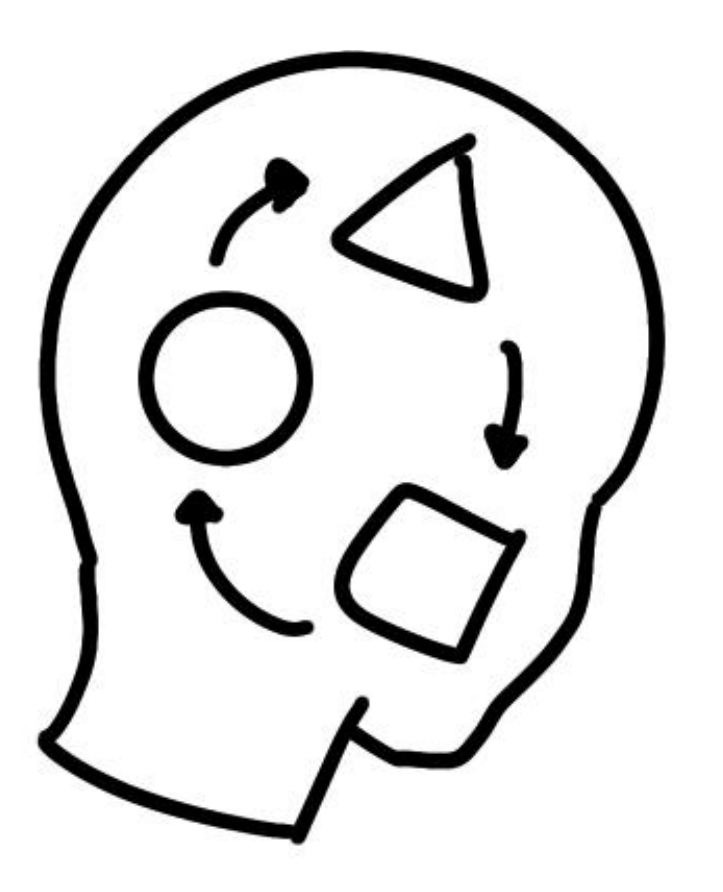}
  \end{minipage}
  &
  \begin{minipage}[t]{\linewidth}
    \vspace{0pt}All students performed well on the post-lesson tests (all 10/10 completed Test 1, and 9/10 completed Test 2), indicating that all teachers in the study were effective. The following analysis examines the adaptive teaching strategies that contributed to these successful outcomes.
  \end{minipage}
\end{tabular}\par\medskip

Beyond combining modalities, teachers adapted their instructional approach based on student responses and progression. We identified four dimensions of this adaptiveness: (1) instruction refinement, (2) error response, (3) modality switching, and (4) student adaptation. Effective teachers monitored comprehension through verbal and visual cues and adjusted their strategies in real time. This was especially clear from Lesson 1 to Lesson 2, as teachers refined their communication based on what they had learned about each student in the first lesson.

\subsubsection{Instruction Refinement Over Time}

Teachers refined their communication style as they became more familiar with both the student and the interface. Explanations shifted from verbose to concise. For example, one teacher (P1-T) began with a long, step-by-step description of how to resize a frame but later reduced this to a simple directive: \textit{``Can you make a frame 120 by 160?''} Similar patterns appeared in value specifications. Early instructions such as \textit{``So we can change it here… okay, now can you change the width and height to 64?''} later became brief statements like \textit{``130 by 160.''} Action commands moved from multi-step walkthroughs to short, targeted prompts. Conceptual explanations also shifted from lengthy descriptions to concise metaphors or summaries.

\subsubsection{Error Response and Recovery}

Teachers adapted their responses when students struggled or made mistakes. Rather than taking control immediately, effective teachers began with verbal correction. When this was not enough, they used brief demonstrations through screen control, then returned control to the student. For example, when one student (P5-S) could not recover from an error, the teacher demonstrated how to return to the last correct state, then undid the fix so the student could perform it themselves. This pattern shows how teachers calibrated intervention based on the severity and persistence of the difficulty.

\subsubsection{Modality Switching Based on Student Response}

The flattening versus union operation illustrates this adaptive teaching. Although the two operations look the same visually, they produce different layer structures. When students questioned why flattening was necessary after union, teachers recognized that demonstration alone was insufficient and added verbal explanations about the underlying system architecture. Modality choices also shifted as students gained familiarity. Teachers used fewer detailed annotations and demonstrations over time and moved toward concise verbal guidance that assumed greater student knowledge.

\subsubsection{Individual Student Adaptation}

Teachers showed awareness of individual student differences and adapted their approach accordingly. Some students required multiple repetitions and additional examples, while others demonstrated rapid comprehension. Teachers adjusted their pacing and detail level based on these individual characteristics. One teacher (P2-T) reflected on the challenge of adapting to individual needs: \textit{``I think it's very hard to translate the text to application... it's very hard for me to explain, like, to the student, like, what it actually means in design.''} This awareness of their own limitations in explanation led teachers to experiment with different communication approaches until they found methods that worked for specific students.

\section{Discussion}
Our analysis of instructor behavior reveals how teaching modalities function in software learning contexts.
We first identify a fundamental constraint: software learning requires instruction to simultaneously convey both spatial and temporal precision; yet, no single modality excels at both without compromising student agency. This precision--agency trade-off creates a central tension that teachers navigate through strategic modality selection and coordination.
We then examine how teachers regulate student agency, pushing for progress while preserving student autonomy.
These findings confirm established learning principles, including the spatial and temporal contiguity principles~\cite{mayer2002multimedia,mayer2023past} and the benefits of student agency from the angles of cognitive load theory~\cite{sweller1988cognitive,sweller2011cognitive} and joint attention~\cite{kang2024jointattention,oura2024supporting}.
Additionally, we identify precision-agency trade-off and digital territory ownership as two new design constraints unique to on-screen, in-situ software learning.
Finally, we explore design opportunities for AI tutoring systems that can navigate these constraints and go beyond the limits of the three modalities examined to provide stronger support for communication clarity, student memory, and intrusion minimization.

\subsection{Spatio-Temporal Precision and Its Tradeoffs}

Across lessons, teachers faced a recurring challenge: conveying instructions that were precise in both space and time. Software learning required students to execute exact interface sequences that specified where to act and in what order; yet, speech alone often failed to convey both dimensions (e.g., \textit{"Try to make the bottom of the 16 on the same level as the bottom of the 30 and 22"} (P3-T)). Students had to map words to spatial locations, time their actions, and sequence steps correctly, which increased cognitive load and created opportunities for misinterpretation.

We observed that teachers addressed these precision demands by distributing work across modalities, consistent with Cognitive Load Theory and multimedia learning principles of spatial and temporal contiguity~\cite{sweller2011cognitive, mayer2001cognitive, mayer2002multimedia}. Visual annotations supported spatial precision by linking speech to specific on-screen targets. This reduces extraneous cognitive load~\cite{sweller1988cognitive, sweller2011cognitive} and helps students identify elements that were difficult to infer verbally. Demonstrations provided both spatial and temporal precision by showing where to act and in what order. This created temporal contiguity between narration and action. However, their spatial cues were brief and required teachers to take control, which some viewed as \textit{"too intrusive"} (P8-T). Speech remained flexible but was limited for conveying fine-grained spatial and temporal detail. In practice, teachers used annotations for spatial guidance and demonstrations for temporal sequences, suggesting their choices reflected implicit use of the contiguity principles.

Software learning requires instruction that conveys both spatial and temporal specificity, yet no single modality supports both well. This creates a core design constraint for AI tutoring systems. Effective support will require combining or switching modalities as precision demands change. Modalities that provide both forms of precision often reduce student agency. Here, \textbf{student agency} refers to students’ ability to initiate and control their own actions and shared visual focus. From a Cognitive Load Theory view, agency concerns who drives cognitive processing. From a joint attention view, it concerns who directs gaze and action. Personal computing interfaces mostly follow a single-locus-of-control paradigm. As a result, increasing precision through annotations or demonstrations can clarify instruction but also shift attention and control away from students. This \textbf{precision–agency trade-off} frames our analysis of how teachers regulate help to maintain student ownership while ensuring progress.

\subsection{"\textit{Am I Helping Them Too Much?}": How Teachers Regulate Student Agency}

Teachers were not driven only by procedural efficiency. They also sought to maintain students’ agency and a sense of ownership, consistent with prior learning science theories of agency, locus-of-control, and just-in-time information~\cite{gee2003videogames}. Teachers calibrated their interventions so that students retained control wherever possible while receiving sufficient guidance to continue making progress.

\subsubsection{The Value of Preserving Student Agency}
Despite the precision demands of software learning, teachers commonly switched between modalities to prioritize student agency. Speech allowed guidance without taking over, annotations highlighted where to act, and demonstrations were used sparingly to avoid overriding student control. Teachers often paused before intervening, treating cursor movement or menu exploration as signs of effortful processing \cite{roediger2011critical}. As one teacher (P4-T) noted, they wanted to \textit{"give them time to explore"} because students \textit{"might be on their way to figuring out (on their own)."} Another teacher (P5-T) reflected that \textit{"the student is like, curious and exploratory on their own. It's easier for me (to teach and guide)... as opposed to like a student that's just literally waiting for your next command."}  This highlights the value of proactive learning behaviors and curiosity-driven exploration.

In this context, we define an ideal state of agency, and this state involves instructor participation. It does not imply independent student action, since novices often lack the procedural and spatial knowledge needed to initiate tasks. Teachers therefore establish goals, highlight relevant interface regions, and scaffold the initial problem space. Ideal agency reflects a form of supported autonomy in which students retain decision-making authority while receiving enough structure to act meaningfully.

From a Cognitive Load Theory perspective, ideal agency is co-constructed by teacher and student. Teachers provide cues and constraints that reduce extraneous load while allowing students to perform the cognitive operations required for schema formation. This balance supports germane processing by reducing unproductive search yet preserving opportunities for mental integration. As one student (P9-S) noted, \textit{"he was specifying the left and the right panel and circling which button I should press... all this was very clear,"} showing how annotations reduce extraneous load by eliminating unproductive search. Agency mismatches, whether through over-annotation or excessive autonomy, increased extraneous load and reduced germane processing. They prevented students from coordinating the cognitive steps needed for understanding.

From a joint-attention perspective, ideal agency is co-regulated through shared control of focus and action. Students signal their attentional priorities through cursor movements and verbalizations, and teachers align their guidance to these signals. Teachers shape the perceptual environment through annotations or demonstrations, while students determine the pace and ordering of task-relevant actions. As P6-T noted, \textit{“If there was no movement in the mouse… I’ll circle the button to bring focus to it,”} demonstrating how teachers respond to student signals to maintain coordinated attention. Excessive autonomy and teacher takeover can disrupt the shared attentional frame, destabilizing the co-regulation needed for a coherent task trajectory.

\subsubsection{Digital Territorial Boundaries}

To navigate this tension, teachers managed help by tuning agency in situ, aligning with assistance-dilemma research~\cite{aleven2007assistance}. These choices were shaped not only by task demands but by social norms around permission and digital territoriality~\cite{scott2004territoriality}. Some teachers hesitated to intrude, as P9-T reflected \textit{``I shouldn't have stepped in... I would rather annotate.''}. In practice, annotations acted as a ``soft'' boundary crossing that marked the student's workspace without taking it over. By contrast, remote control constituted a ``hard'' boundary crossing, temporarily reassigning ownership to the teacher. These differences help explain why teachers reached for annotations first and reserved remote control for severe breakdown. Help-regulation was therefore not merely a technical choice but one shaped by interpersonal norms and shared territorial boundaries.

  \subsubsection{Calibrated Agency as a Design Goal}
  Calibrated agency means keeping control appropriately balanced rather than tilted towards either teacher or student. Control should align with task precision, pacing demands and students' current competence. Human teachers seek to preserve student agency and repair breakdowns through help-regulating practices. AI teachers should treat calibrated agency as a central design goal rather than an afterthought~\cite{graesser2004autotutor, huang2021adaptutar}. Systems should default to preserving student control, increasing guidance or intrusiveness only when precision demands or breakdowns exceed student's ability to proceed. Effective systems will need to support adaptive modality choice, modulate intrusiveness in real time, and encode help-regulation policies that mirror expert teachers' timing norms.

  Working on the same digital workspace creates unique challenges for AI tutoring systems. Unlike human teachers who share a workspace with implicit social contracts that signal when intervention is appropriate, AI systems can take over the screen instantly, which makes territorial boundaries more fragile. This raises key design questions: when is screen takeover pedagogically justified, how should boundaries be signaled, and how can student control be returned? Because AI lacks the social presence that constrains human intrusion, territoriality norms must be explicitly encoded to preserve student agency.

  \subsection{Design Opportunities for AI Tutoring Systems}

  \begin{figure}[t!]
    \centering
    \begin{subfigure}[t]{\linewidth}
      \centering
      \includegraphics[width=\linewidth]{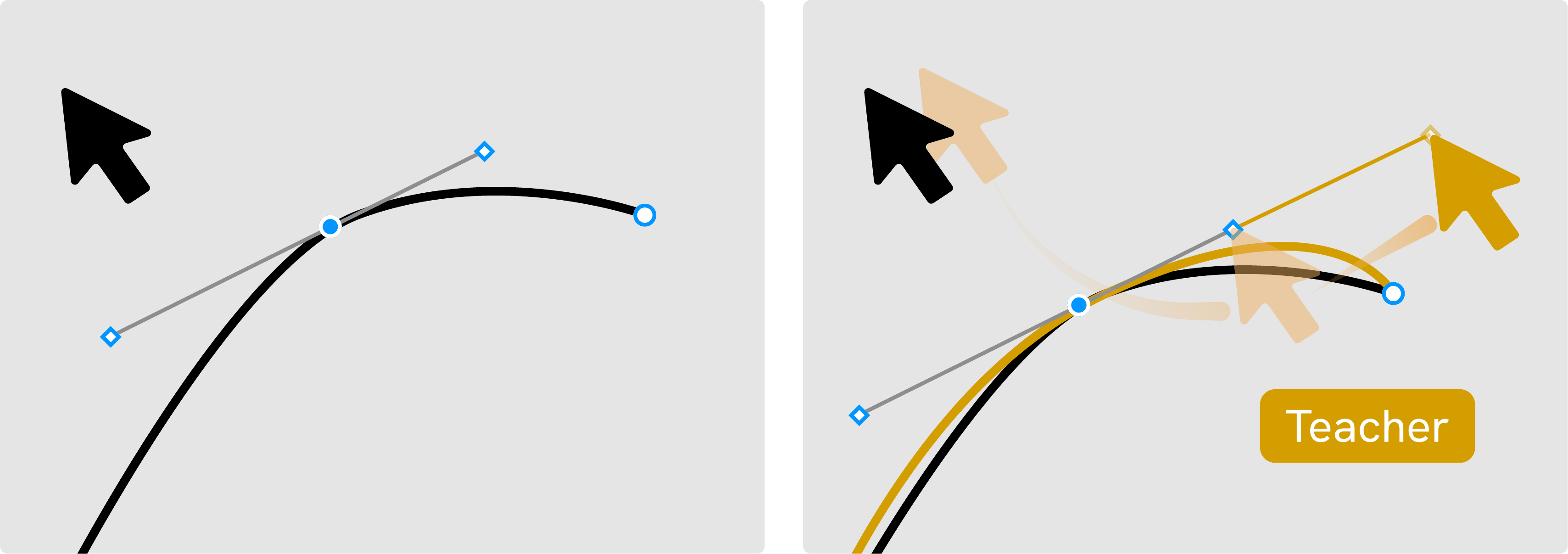}
      \subcaption{Ghost Cursor}\label{fig:design-ideas-a}
    \end{subfigure}\\[0.5em]
    \begin{subfigure}[t]{\linewidth}
      \centering
      \includegraphics[width=\linewidth]{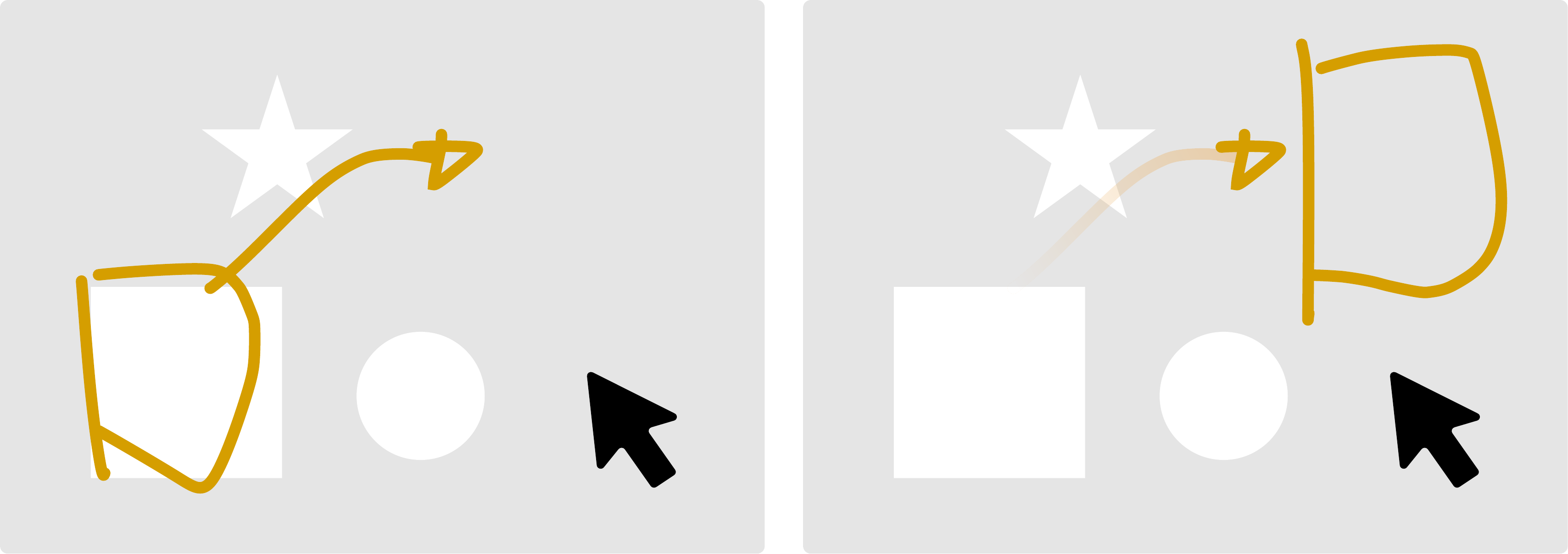}
      \subcaption{Fading Annotations}\label{fig:design-ideas-b}
    \end{subfigure}\\[0.5em]
    \begin{subfigure}[t]{\linewidth}
      \centering
      \includegraphics[width=\linewidth]{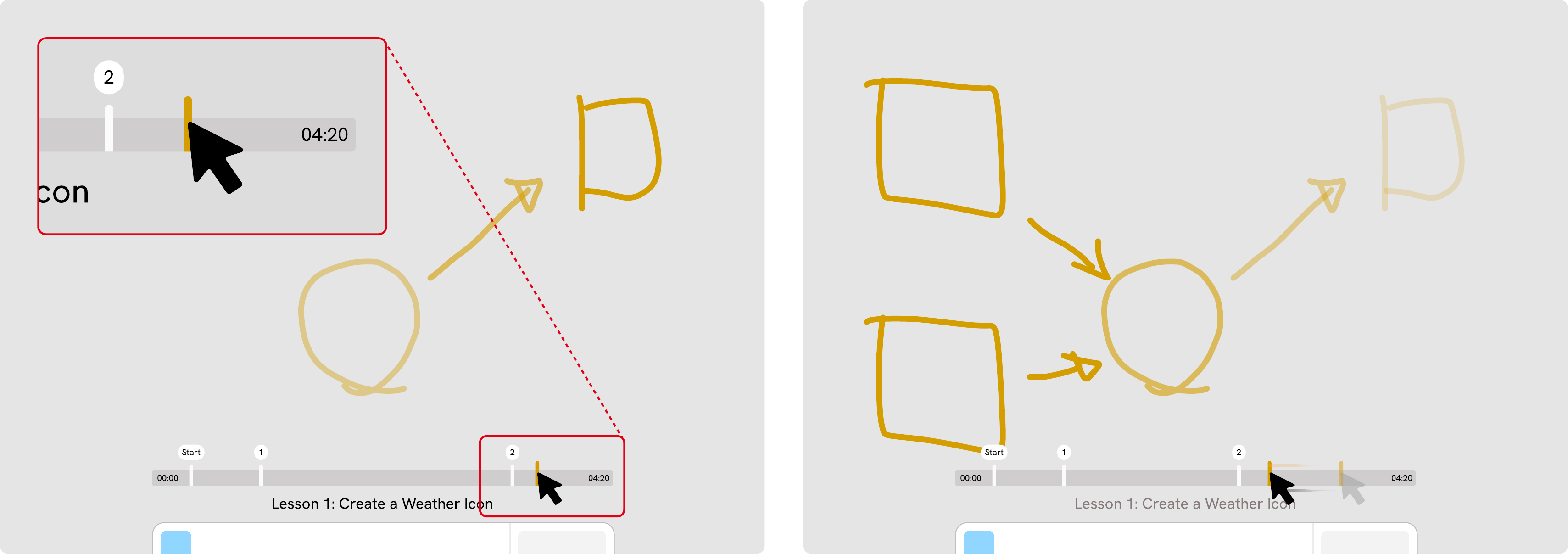}
      \subcaption{Timeline Scrubbing}\label{fig:design-ideas-c}
    \end{subfigure}\\[0.5em]
    \begin{subfigure}[t]{\linewidth}
      \centering
      \includegraphics[width=\linewidth]{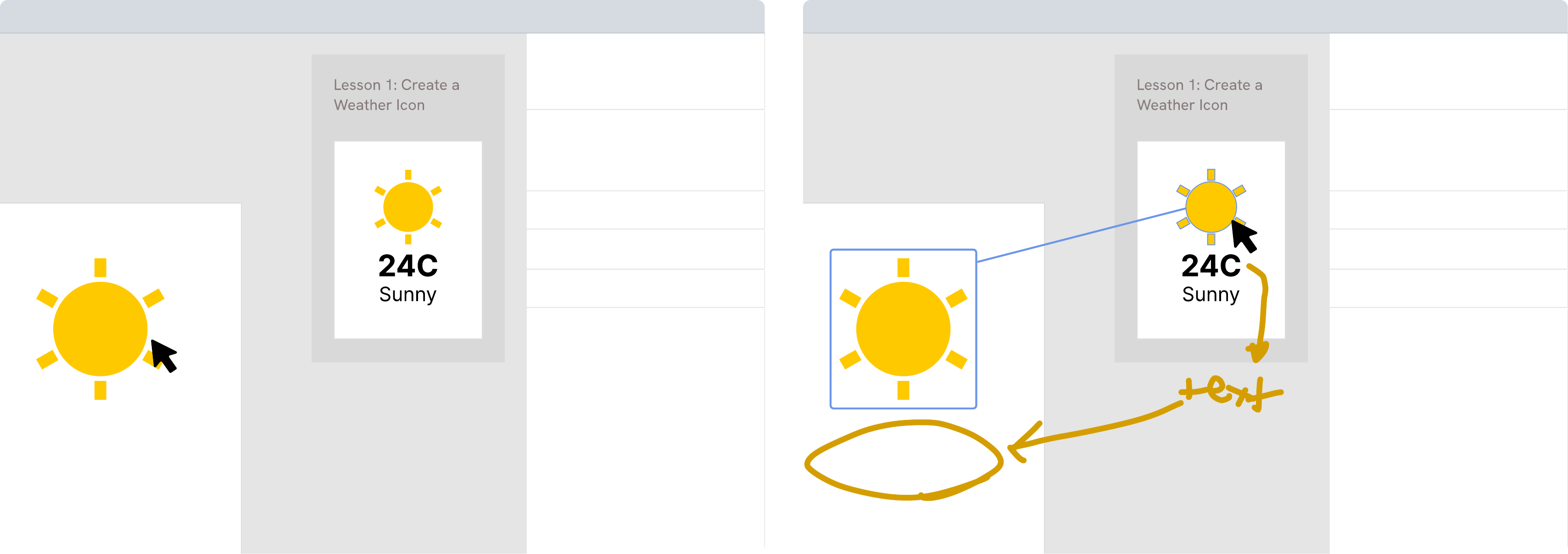}
      \subcaption{Interactive References}\label{fig:design-ideas-d}
    \end{subfigure}
    \caption{Design implications for AI tutoring systems: (a) \textbf{Ghost Cursor} - showing AI tutor's cursor position to students, (b) \textbf{Fading Annotations} - annotations that gradually disappear over time, (c) \textbf{Timeline Scrubbing} - reviewing past annotations through temporal navigation, and (d) \textbf{Interactive References} - extractable and manipulable content elements from reference images.}
    \label{fig:design-ideas}
    \Description{A 2x2 grid presenting four interaction design concepts for educational interfaces. Top-left (a) Ghost Cursor: Shows a black cursor with a semi-transparent orange ghost cursor behind it, demonstrating how a teacher's previous input can be visualized alongside current actions. Top-right (b) Fading Annotations: Displays orange hand-drawn annotations around interface elements that fade over time, showing temporal decay of visual guidance. Bottom-left (c) Timeline Scrubbing: Illustrates a timeline interface with a scrubber that allows navigation through different states of the canvas, where orange annotations fade as the timeline moves forward, with a magnified view showing timestamps. Bottom-right (d) Interactive References: Shows a sun icon that, when hovered over, reveals an information card with weather details, demonstrating how on-screen elements can be interactively referenced with contextual information through hand-drawn annotations and arrows.}
  \end{figure}

  Participant observations revealed several limitations in current tutoring interfaces. Teachers noted that direct cursor control felt \textit{``too intrusive''} (P8-T), highlighting the agency-precision tension. Students reported that heavy annotation created visual clutter, saying there were \textit{``a lot of arrows in the same space''} (P7-S). Teachers managed this by manually deleting annotations. Students also expressed difficulty remembering steps and wanted persistent notes or references. Teachers struggled to convey abstract concepts, such as parent-child component relationships, with existing tools. Other issues included low visual contrast, lack of persistent reference material, and limited contextual framing.

  Building on these limitations, our findings point to several design opportunities. A \textit{ghost cursor} feature (Fig.~\ref{fig:design-ideas-a}) could show teacher cursor movements without taking control, providing temporal and spatial contiguity while preserving student control. \textit{Fading annotations} (Fig.~\ref{fig:design-ideas-b}) would reduce visual clutter while keeping guidance long enough to support comprehension. \textit{Timeline scrubbing} (Fig.~\ref{fig:design-ideas-c}) would let students revisit annotations on demand, giving them control over the pacing of temporal information. A \textit{persistent notes sidebar} would support memory and reduce reliance on working memory during complex tasks. Finally, \textit{interactive references} (Fig.~\ref{fig:design-ideas-d}) would help students understand abstract relationships by making reference materials manipulable rather than static. Together, these ideas show how AI tutoring systems can move beyond the limitations of speech, annotation, and screen control to improve clarity, support memory, and minimize intrusion.

\section{Limitations and Future Work}
This study offers an initial view of modality coordination in on-screen software instruction, but several limitations remain. Our focus on Figma may limit generalizability to other domains (e.g., programming). Future work should test whether the same modality patterns hold across a broader range of tools. \edit{In programming environments, for example, precision is syntactic and territory concerns code ownership. Replicating this work in such domains would clarify whether spatial density, temporal coupling, and conceptual complexity predict modality emphasis and whether different coordination strategies are required.}

The controlled laboratory setup differs from real-world learning contexts and may have influenced teaching. \edit{The precision-agency trade-offs identified here likely extend beyond speech, annotation, and control: gaze highlighting (high precision, low intrusion), for example, may dominate over annotation where available; testing such modalities would reveal whether the trade-off structure is modality-invariant or reshaped by affordance cost. Several dynamics remain open: how modality balance should shift as learners progress, and when AI should proactively reduce scaffolding. A related question is whether trade-offs persist when the interaction cost of high-precision modalities is reduced (e.g., hover-triggered annotations).}

Our sample of ten teacher--student pairs limits generalization; the variation we observed suggests individual differences shape modality choices. Future work should examine how learner characteristics (e.g., prior experience) interact with modality combinations, informing personalized AI tutoring.

Finally, we analyzed teaching behavior rather than learning outcomes, and our one-to-one, synchronous setting may not reflect other assistance practices or typical classrooms. \edit{An open question is whether annotation and demonstration yield different mental models of the same procedures.} Future studies should connect modality coordination to measurable gains in learning, retention, and transfer, \edit{examine asynchronous and more lightweight forms of on-screen guidance, and explore how modalities can be orchestrated in group configurations (e.g., one-to-many classrooms) where learners may have competing needs.} Emerging technologies such as augmented reality may further expand the design space; these directions may support modality types not covered in this study.
\section{Conclusion}
Through an observational study of ten teacher--student pairs \edit{(N=10)}, we examined how human teachers coordinate speech, visual annotations, and remote screen control to teach feature-rich software procedures.
Our findings validate and extend foundational learning theories within interactive workspaces: effective modality coordination supports the spatial and temporal contiguity principles of multimedia learning and facilitates joint attention. At the same time, we identify two domain-specific design constraints---the precision--agency trade-off and digital territoriality. While high-precision modalities like remote control can reduce extraneous cognitive load during complex procedures, they can also infringe upon learner autonomy and territorial boundaries.

We argue that AI tutoring systems should be designed not merely as information-delivery engines, but as adaptive systems that balance these competing constraints. Systems should prioritize calibrated agency, dynamically escalating from low- to high-precision interventions only when necessary to maintain instructional momentum. By integrating these theoretical insights with interaction mechanisms such as ghost cursors or fading annotations, future interfaces can transcend the limitations of human teaching to provide precise and context-aware software guidance.

\begin{acks}
  We thank the HCI community at Singapore Management University for their thoughtful feedback and steady encouragement throughout this project. We are deeply grateful to Emily Aurelia, Yeo Shunyi, Justin G, and Georgia Zhang whose generosity, enthusiasm, and insightful conversations substantially strengthened our study design and analysis.

  We extend our sincere appreciation to all our study participants—both experts and novices—for dedicating their time, embracing the challenge of learning a new skill, and contributing to this work. This research is meaningfully better because of these contributions.

  This work acknowledges the support from Google (internal code: T050309).
\end{acks}

\bibliographystyle{ACM-Reference-Format}
\bibliography{references}

\appendix

\section{Appendices}
\subsection{Session Environment and Hardware Specifications}
\label{sec:appendix-technical}

All sessions were conducted in a controlled lab environment. The teacher and student were seated facing each other, separated by a lightweight divider to prevent face-to-face nonverbal cues. Both participants joined the same Zoom call using the following devices:

\begin{itemize}
    \item \textbf{Teacher:} 16-inch Legion Pro 5 16IRX9 laptop (Intel i9-14900HX, 32GB RAM)
    \item \textbf{Student:} 15-inch Legion 5 15ACH6H laptop (AMD Ryzen 7 5800H, 16GB RAM)
    \item \textbf{Annotation Tablet:} 12.9-inch iPad Pro (4th generation) with Apple Pencil Pro
\end{itemize}

All interactions were screen- and audio-recorded at full resolution, capturing Zoom annotations and remote-control manipulations.

\subsection{Full Session Procedure and Timing}

Each session followed a standardized structure lasting approximately 1.5 hours:

\begin{enumerate}
    \item \textbf{Pre-session Setup (around 15 min):} informed consent, demographic verification, technical checks, overview of study expectations, and a short orientation to Zoom’s screen sharing, annotation, and remote-control features. Teachers were also given a brief practice period to ensure basic tool familiarity.
    \item \textbf{Task 1: Weather Icons (15--30 min):} teacher-led instruction, followed by a 10-minute transfer test and a short semi-structured interview.
    \item \textbf{Break (around 5 min)}
    \item \textbf{Task 2: Weather Cards (15--30 min):} teacher-led instruction, followed by a second transfer test and a final interview.
\end{enumerate}

Teachers were instructed to teach naturally and retained full autonomy over their modality choices. No formal assessment of prior Zoom proficiency was conducted.

\subsection{Multimodal Data Processing Pipeline}
\label{sec:technical-implementation}
\begin{figure*}[!ht]
    \centering
    \includegraphics[width=\linewidth]{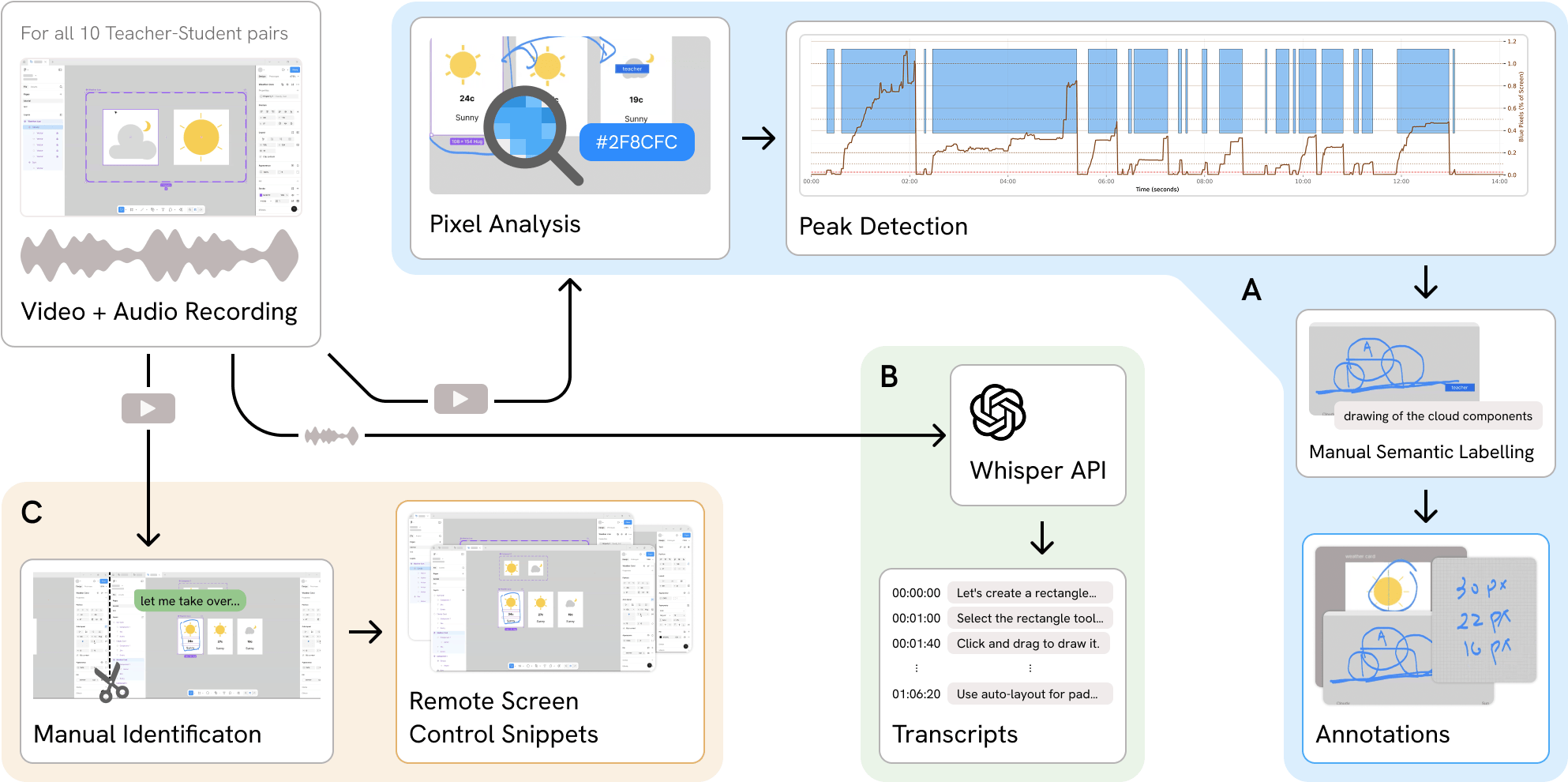}
    \caption{Multi-modal data processing pipeline showing the three parallel analysis streams: (A) visual annotation processing, (B) audio transcription, and (C) remote screen control extraction.}
    \label{fig:data-processing-pipeline}
    \Description{Workflow diagram showing multi-modal data analysis from 10 teacher-student pairs. The process splits into three parallel streams: (A) Visual Analysis using pixel detection and semantic labeling, (B) Audio Analysis using Whisper API to generate timestamped transcripts, and (C) Manual Identification of remote control snippets. Arrows show data flow between processing stages.}
\end{figure*}

See Figure~\ref{fig:data-processing-pipeline}.

\subsubsection{Step Identification and Remote-Control Extraction}

Recordings were manually reviewed to segment each session into tutorial steps defined by the task structure. Remote-control periods were identified and exported as separate clips due to their relative infrequency and pedagogical importance.

\subsubsection{Visual Annotation Detection and Processing}

Teacher annotations were detected through frame-by-frame analysis (1 fps sampling) using color-based detection for blue pixels (\#2f8cfc, Zoom's default annotation color). Frames were processed to extract overlay regions, apply color thresholding, and identify contiguous annotation periods. Activity periods were identified when pixel counts exceeded 500 pixels. Individual annotation periods were manually verified to eliminate false positives from interface elements and verify annotation boundaries.

Each completed annotation (e.g., arrow, circle, sketch) was treated as a single semantic unit. Incremental strokes (such as partial ray segments) were not counted individually unless they represented meaningful instructional units. Each annotation was timestamped and associated with its corresponding lesson step.

\subsubsection{Audio Processing and Transcription}

Audio was extracted from Zoom recordings (.m4a) and processed using OpenAI Whisper API. Remote control segments used a 5-second pre-buffer; visual annotation contexts used 7-second snippets (5s before, 2s after annotation timestamps). All transcriptions were manually corrected for accuracy and temporally aligned with lesson steps, detected annotations, and remote-control events.

\subsubsection{Data Integration and Coding Framework}

Multi-modal data streams were temporally aligned using precise timestamp matching. Critical processing steps were manually verified, and cross-validation procedures ensured consistency between automated and manual coding results. A representative teacher-student pair was selected based on modality usage diversity. Coding was conducted at utterance level, mapped to lesson step markers. Inter-coder reliability was established through independent coding with consensus resolution.

\subsection{Study Tasks}
\label{sec:appendix-study-tasks}

\begin{figure}[t]
    \centering
    \includegraphics[width=\columnwidth]{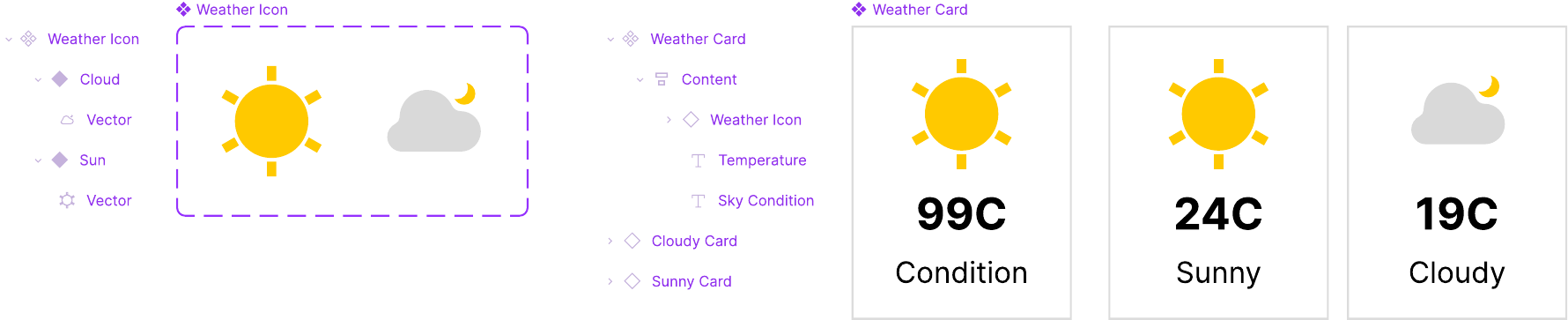}
    \caption{Overview of the two study tasks used in the teaching sessions: Task 1 (Weather Icons) and Task 2 (Weather Cards).}
    \label{fig:appendix-task-comparison}
    \Description{Two Figma task screenshots showing Task 1 (Weather Icons) and Task 2 (Weather Cards), illustrating the progression from basic vector manipulation to composing reusable components.}
\end{figure}

\textbf{Lesson 1: Weather Icons}

\textit{The first task focused on vector manipulation fundamentals, requiring participants to create two custom weather icons (sun and cloudy) from basic geometric shapes. This task was designed to provide students with a gentle introduction about the fundamental Figma concepts. Students learned to create frames, manipulate basic shapes (circles, lines), apply transformations (rotation, scaling), and use Boolean operations (union, subtract). Students created a sun icon by arranging radiating lines around a central circle, and a cloudy icon by combining multiple circles using Boolean operations to form a cloud shape with a crescent detail. Both icons required color application and final flattening into single vector objects and to produce a component set at the end.}

\textbf{Structure:} 7 steps, Estimated 15-30 minutes (Requires: Basic computer skills, No Figma experience.)
    
\textbf{Lesson 2: Weather Cards}

\textit{The second task introduced components, requiring students to create reusable weather cards that incorporated the icons from Lesson 1. Students learned more advanced concepts including component creation, instance management, auto-layout principles, typography tools, and the master-instance relationship. Students designed a weather card component that contains an icon, temperature text, and condition label. They then created two instances with different weather data, learning how component variants work and how changes to the master component propagate to all instances.}

\textbf{Structure:} 8 steps, Estimated 15-30 minutes (Requires: Completion of Lesson 1, Basic Figma familiarity.)

\begin{figure*}[!ht]
\centering
\includegraphics[width=\textwidth]{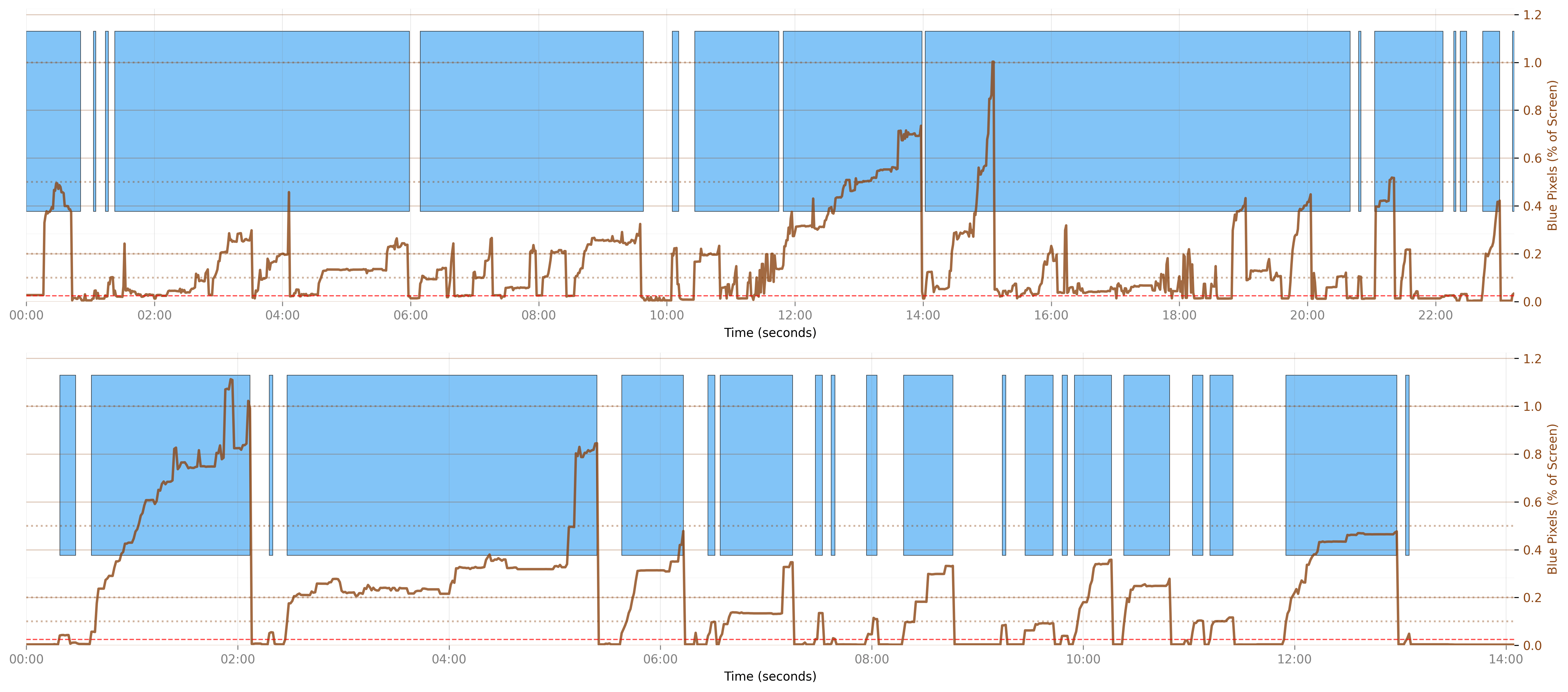}
\caption{Overlay slope-aware analysis for Participant P10 showing blue pixel density (brown line) and annotation activity periods (blue bars) across both lessons. The visualization reveals temporal patterns in visual attention and annotation behavior during software instruction.}
\label{fig:p10-overlay-slope-aware}
\Description{Two stacked line charts analyzing Participant P10's visual attention and annotation behavior across both lessons. Both charts share a common X-axis (Time in seconds) and Y-axis (Blue Pixels as percentage of screen, 0.0-1.2). The top chart represents Lesson 1 (22 minutes) and the bottom chart represents Lesson 2 (14 minutes). Each chart displays a brown line showing blue pixel density over time, indicating visual attention to blue interface elements, and blue bars showing annotation activity periods occurring approximately every 2 minutes in 1-minute segments. The brown line shows fluctuating patterns with several peaks throughout both lessons, often correlating with annotation activity periods. In Lesson 1, notable peaks occur around 14:30 and 15:00, while Lesson 2 shows peaks around 02:00 and sustained activity from 10:00-13:00. The correlation between increased blue pixel density and annotation activity periods suggests a relationship between visual attention and annotation behavior during the learning sessions.}
\end{figure*}
Figure~\ref{fig:p10-overlay-slope-aware} presents the overlay slope-aware analysis for Participant P10, showing the temporal relationship between blue pixel density and annotation activity across both instructional lessons. The visualization demonstrates the dynamic nature of visual attention patterns during software instruction, with the brown line representing continuous blue pixel density and blue bars indicating periods of high annotation activity.

\end{document}